\documentclass[11pt]{cernrep}
\usepackage{graphicx}
\usepackage{here}
\begin{document} 
\def \slhclum{$10^{35}\ \mathrm{cm}^{-2}\mathrm{s}^{-1}$}
\def \lhclum{$10^{34}\ \mathrm{cm}^{-2}\mathrm{s}^{-1}$}
\def \timeslhclum#1{$#1\times 10^{34}\ \mathrm{cm}^{-2}\mathrm{s}^{-1}$}
\def \ifb {fb$^{-1}$}
\def \ipb {pb$^{-1}$}
\def \gev {\mathrm{GeV}}
\def \tev {\mathrm{TeV}}
\def\d {$} 
\def\ed{$} 
\def \lambdahhh {\ifmmode \lambda_{\sss HHH}\else $\lambda_{\sss HHH}$\fi}
\def \lambdahhhsm {\ifmmode \lambda_{\sss HHH}^{\sss SM}\else 
$\lambda_{\sss HHH}^{\sss SM}$\fi}
\def \mh {\ifmmode m_H \else $m_H$\fi}
\def \pythia {{\small PYTHIA}}
\def    \nn             {\nonumber} 
\def    \=              {\;=\;} 
\def    \frac           #1#2{{#1 \over #2}} 
\def    \ret            {\\[\eqskip]} 
\def    \ie             {{i.e.} } 
\def    \etal             {{et al.} } 
\def    \eg             {{e.g.} } 
\def    \lsim {\raisebox{-3pt}{$\>\stackrel{<}{\scriptstyle\sim}\>$}}  
\def    \gsim {\raisebox{-3pt}{$\>\stackrel{>}{\scriptstyle\sim}\>$}}  
\def    \gtrsim {\raisebox{-3pt}{$\>\stackrel{>}{\scriptstyle\sim}\>$}}  
\def    \esim {\raisebox{-3pt}{$\>\stackrel{-}{\scriptstyle\sim}\>$}}  
\def\wup{{W^+}}
\def\wum{{W^-}}
\def\wupm{{W^\pm}}
\def\wump{{W^\mp}}
\def\abs#1{\left| #1\right|}
\def\sgn{\mathop{\rm sgn}}
\def\gtap{\raisebox{-.4ex}{\rlap{$\sim$}} \raisebox{.4ex}{$>$}}  
\def\fb{{\rm fb}}
\def\ltap{\raisebox{-.4ex}{\rlap{$\sim$}} \raisebox{.4ex}{$<$}}
\def\tG{{\tilde G}}
\def\ns{{\rm ns}}
\def\tell{{\tilde\ell}}
\def\ttau{{\tilde\tau}}
\def\fbi{{\rm fb}^{-1}}
\def\Meff{M_{\rm eff}}
\def\Msusy{M_{\rm SUSY}}
\def\lsp{{\tilde\chi_1^0}}
\def\ra{\rightarrow}
\def\GeV{{\rm GeV}}
\def\TeV{{\rm TeV}}
\def\mhalf{m_{1/2}}
\def\tchi{\tilde\chi}
\def\tg{\tilde g}
\def\tq{\tilde q}
\newcommand     \be     {\begin{equation}} 
\newcommand     \ee     {\end{equation}} 
\newcommand     \ba     {\begin{eqnarray}} 
\newcommand     \ea     {\end{eqnarray}} 
\newcommand     \sst            {\scriptstyle} 
\newcommand     \sss            {\scriptscriptstyle} 
\newcommand     \auno{a^{(1)}} 
\newcommand     \avg[1]         {\left\langle #1 \right\rangle} 
\newcommand     \ptmin     {\ifmmode p_{\scriptscriptstyle T}^{\sss min} \else 
                           $p_{\scriptscriptstyle T}^{\sss min}$ \fi} 
\def     \muf           {\mbox{$\mu_{\sss F}$}} 
\def     \mur            {\mbox{$\mu_{\sss R}$}} 
\def    \muo            {\mbox{$\mu_0$}} 
\newcommand\as{\alpha_{\sss \mathrm S}} 
\newcommand\astwo{\alpha_{\sss \mathrm S}^2} 
\newcommand\asthree{\alpha_{\sss \mathrm S}^3} 
\newcommand\asfour{\alpha_{\sss \mathrm S}^4} 
\newcommand\aem{\alpha_{\mathrm em}} 
\newcommand\cb{\overline{c}} 
\newcommand\bb{\overline{b}} 
\newcommand\tb{\overline{t}} 
\newcommand\Qb{\overline{Q}} 
\newcommand\qq{{\scriptscriptstyle Q\overline{Q}}} 
\def \asopi{\mbox{$\frac{\as}{\pi}$}} 
\def \oafour {\mbox{${\cal O}(\asfour)$}} 
\def \oacube {\mbox{${\cal O}(\asthree)$}} 
\def \oatwo {\mbox{${\cal O}(\astwo)$}} 
\def \oas   {\mbox{${\cal O}(\as)$}} 
\def \bbbar {\mbox{$b \bar b$}}
\def \ttbar {\mbox{$t \bar t$}}  
\def \ccbar {\mbox{$c \bar c$}} 
\def \pt   {\mbox{$p_{\scriptscriptstyle T}$}} 
\def \et   {\mbox{$E_{\scriptscriptstyle T}$}} 
\def \Emu  {\ifmmode{E_{\mu}
    }\else{$E_{\mu}$}\fi} 
\def \Enu  {\ifmmode{E_{\nu}
    }\else{$E_{\nu}$}\fi} 
\def \nudis {$\nu$DIS}
\def \nubar {\bar{\nu}}
\def \nue  {\ifmmode{\nu_e}\else{$\nu_e$}\fi} 
\def \numu  {\ifmmode{\nu_{\mu}}\else{$\nu_{\mu}$}\fi} 
\def \nufact {$\nu$-Factory}
\def \rap   {\mbox{$\eta$}} 
\def \deltar {\mbox{$\Delta R$}} 
\def \dphi {\mbox{$\Delta \phi$}} 
\def \to   {\mbox{$\rightarrow$}} 
\def    \mb             {\mbox{$m_b$}} 
\def    \mc             {\mbox{$m_c$}} 
\def    \mt             {\mbox{$m_t$}} 
\newcommand \jpsi{\ifmmode{J/\psi 
    }\else{$J/\psi$}\fi} 
\def\calF{{\cal F}} 
\def\calP{{\cal P}} 
\def\calM{{\cal M}} 
\def\calO{{\cal O}} 
\def        \mW         {\mbox{$m_W$}} 

\pagestyle{empty}
\begin{titlepage}
{\flushright{
        \begin{minipage}{5cm}
        CERN-TH/2002-078\\
        {\tt hep-ph/0204087}\\
        April 1, 2002\hfill  
        \end{minipage}        }
        
}
\vfill
\begin{center}
{\Large {\bf 
    PHYSICS POTENTIAL AND EXPERIMENTAL \\[0.3cm]
    CHALLENGES OF THE LHC LUMINOSITY UPGRADE}
}
\end{center}                                   
\vfill                                                       
{\bf Conveners}:
  F. Gianotti~$^{1}$, M.L. Mangano~$^{2}$, T. Virdee~$^{1,3}$ \\
{\bf Contributors}:
S.~Abdullin~$^{4}$, G.~Azuelos~$^{5}$, A. Ball~$^{1}$,
D.~Barberis~$^{6}$, A.~Belyaev~$^{7}$, P. Bloch~$^{1}$, M.
Bosman~$^{8}$, L. Casagrande~$^{1}$, D.~Cavalli~$^{9}$, P.
Chumney~$^{10}$, S. Cittolin~$^{1}$, S.Dasu~$^{10}$, A.~De
Roeck~$^{1}$, N. Ellis~$^{1}$, P. Farthouat~$^{1}$, D.
Fournier~$^{11}$, J.-B.~Hansen~$^{1}$, I.~Hinchliffe~$^{12}$,
M.~Hohlfeld~$^{13}$, M. Huhtinen~$^{1}$, K.~Jakobs~$^{13}$, C.
Joram~$^{1}$, F.~Mazzucato~$^{14}$, G.Mikenberg~$^{15}$,
A.~Miagkov$^{16}$, M.~Moretti$^{17}$, S.~Moretti~$^{2,18}$, T.
Niinikoski~$^{1}$, A.~Nikitenko$^{3,\dag}$, 
A. Nisati~$^{19}$, F.~Paige$^{20}$, S.
Palestini~$^{1}$, C.G.~Papadopoulos$^{21}$,
F.~Piccinini$^{2,\ddag}$, R.~Pittau$^{22}$, G.~Polesello~$^{23}$,
E.~Richter-Was$^{24}$, P. Sharp~$^{1}$, S.R.~Slabospitsky$^{16}$, W.H.
Smith~$^{10}$, S. Stapnes~$^{25}$, G. Tonelli~$^{26}$,
E.~Tsesmelis~$^{1}$, Z. Usubov$^{27,28}$, L.~Vacavant~$^{12}$,
J.~van~der~Bij$^{29}$, A. Watson~$^{30}$, M.~Wielers~$^{31}$
\\
{\small
  $^{1}$~Experimental Physics Division, CERN, Geneva, Switzerland\\
  $^{2}$~Theoretical Physics Division, CERN, Geneva, Switzerland\\
  $^{3}$~Imperial College, London, UK \\
  $^{4}$~University of Maryland, USA \\
  $^{5}$~Group of Particle Physics, University of Montreal,
              Montreal, Canada \\
  $^{6}$~Dipartimento di Fisica and INFN, Universit\`a di Genova, 
              Italy\\
  $^{7}$~Florida State University, Tallahassee, FL, USA \\
  $^{8}$~IFAE, Barcelona, Spain\\
  $^{9}$~INFN Milano, Italy \\
  $^{10}$~Univ. of Wisconsin, Madison, WI, USA.\\
  $^{11}$~LAL, Orsay, France\\
  $^{12}$~Lawrence Berkeley National Laboratory, Berkeley, CA, USA \\
  $^{13}$~Institut f\"ur Physik, Johannes-Gutenberg
  Universitaet, Mainz, Germany\\
  $^{14}$~Section de Physique, Universit\'e de Gen\`eve, Switzerland\\
  $^{15}$~Weizman Institute, Israel\\
  $^{16}$~Institute for High Energy Physics, Protvino, Russia \\
  $^{17}$~Dipartimento di Fisica and INFN, Universit\`a di Ferrara, Italy\\ 
  $^{18}$~Institute for Particle Physics Phenomenology, Durham
    University, UK\\
  $^{19}$~Dipartimento di Fisica and INFN, Universit\`{a} di Roma 1, Italy\\
  $^{20}$ Brookhaven National Laboratory, Upton, New York, USA. \\
  $^{21}$~Institute of Nuclear Physics NCSR "Demokritos", Athens, Greece \\
  $^{22}$~Dipartimento di Fisica Teorica and INFN, Universit\`a di Torino,
  Italy\\ 
  $^{23}$~INFN Pavia, Italy. \\
  $^{24}$~Institute of Computer Science, Jagellonian University;
  Institute of Nuclear Physics, Cracow, Poland \\
  $^{25}$~University of Oslo, Norway\\
  $^{26}$~Universit\'a di Pisa and INFN, Italy\\
  $^{27}$~Institute of Physics, Academy of Sciences of Azerbaijan,
  Baku, Azerbaijan \\ 
  $^{28}$~Joint Institute for Nuclear Research, Dubna, Russia \\
  $^{29}$~Fakult\"at f\"ur Physik, Albert-Ludwigs Universit\"at Freiburg,
  Germany\\
  $^{30}$~School of Physics and
  Astronomy, the University of Birmingham, UK \\
  $^{31}$~TRIUMF, Vancouver, Canada }\\
{\footnotesize
$^{\dag}$On leave of absence from ITEP, Moscow, Russia.\\
$^{\ddag}$On leave of absence from INFN, Sezione di
  Pavia, Italy.}
\newpage
\begin{abstract} 

 We discuss the physics potential and the experimental challenges of an
 upgraded LHC running at an instantaneous luminosity of \slhclum.  
 The detector R\&D needed to operate ATLAS and CMS in a very high
 radiation environment and the expected detector performance are discussed. 
   A few examples of the increased physics potential are given, ranging from
  precise measurements within the Standard Model (in particular
  in the Higgs sector) to the discovery reach for several New Physics
 processes.

\end{abstract}                                                
\tableofcontents

\end{titlepage}

\pagestyle{plain}

\section{INTRODUCTION}
\label{sec:intro}
 This note documents preliminary studies of the
physics potential and experimental challenges of a future
high-luminosity (\slhclum) upgrade of CERN's Large Hadron Collider.
Hereafter, we shall refer to this upgrade as the Super-LHC (SLHC).
It is impossible at this stage to give a conclusive judgement of what
will be the most interesting topics to study after the first few years
of LHC operation at the nominal \lhclum\ luminosity. We shall assume
however that the physics programme of the LHC will have been accomplished,
and in particular that the Higgs boson and Supersymmetry will have been
found, if they are in the mass ranges expected
today~\cite{ATLAS-tdr,CMSTP}. 
The physics potential of the SLHC can then be roughly
divided into the following main topics:

\begin{enumerate}
\item improvement of the accuracy in the determination of Standard
  Model (SM)
  parameters (e.g. triple (TGC) and quartic (QGC) gauge boson
  couplings, Higgs couplings);
\item improvement of the accuracy in the determination of 
  parameters of New Physics possibly discovered at the LHC
  (e.g. sparticle spectroscopy, $\tan\beta$ measurements);
\item extension of the discovery reach in the high-mass region
  (e.g. quark compositeness, new heavy gauge bosons,
  multi-TeV squarks and gluinos, Extra-dimensions);
\item extension of the sensitivity to rare processes (e.g.  
   FCNC top decays, Higgs-pair production, multi gauge boson
   production).
\end{enumerate}

The detector performance at high luminosity will have a different
impact on the physics output depending on the topic considered.  In
the case of searches at the high-mass frontier, in most cases
detection of multi-TeV objects should not be impaired by the high
luminosity environment. In contrast, accurate measurements of systems
in the few hundred GeV range could be significantly affected by the
large event pile-up, and reduced efficiencies or increased backgrounds
could spoil the advantage of the higher luminosity. Accurate
predictions will, in this case, depend on the actual detector
configuration, e.g. whether a fully functional tracker can be operated
at \slhclum. Given the rapid progress of technology and R\&D,
it is premature at this stage to attempt to select
a completely defined detector scheme. 
Therefore, in the studies that follow, we have worked in most
cases under the optimistic assumption that the main parameters of the
detector performance (e.g. the $b$-tagging efficiency, the jet energy
resolution) will remain the same as those expected at \lhclum. To
fully benefit from a tenfold increase in statistics, this is an almost
mandatory requirement.

The main goal of the studies presented here is to illustrate how the
SLHC could allow good progress to be made in the understanding of
fundamental interactions, at a moderate extra cost relative to the
overall initial LHC investment, given that the existing tunnel, accelerator and
detectors would be in large part reused. We shall mostly concentrate on physics
studies which are not feasible at the standard LHC. As a result, we
shall not cover $B$ physics.  While an increase in luminosity would in
principle improve the ability to study rare $B$ decays, we expect
a $B$ physics programme to be extremely unlikely at the SLHC given the
difficulties to reconstruct low momentum particles.

We shall not attempt here any exhaustive cross-comparison with the
potential of other machines.  A discussion of different accelerator
options relevant to the CERN future programme is given
in~\cite{DeRoeck:2001nz}.  A first study of the SLHC, including
comparisons with the potential of an LHC energy upgrade, was presented
in~\cite{atlas-note}.  The prospects of several options for future
hadron colliders have recently been reviewed in~\cite{Baur:2002ka}.

\section{THE MACHINE UPGRADE}
\label{sec:machine}

 A feasibility study for upgrading the LHC has been
launched at CERN~\cite{machine-rep}, which develops scenarios 
for increasing both, the luminosity 
in each of the two high-luminosity experiments and the beam energy.
 The study presents some baseline options and discusses a few alternative
solutions, identifying further investigations needed and
proposing an $\rm{R \& D}$ programme.

A staged upgrade of the LHC and its injectors has been considered,
compatible with established accelerator design criteria and
fundamental limitations of the hardware systems, aiming at a target
luminosity in proton operation of \slhclum\ in each
of the two high-luminosity experiments, and an upgrade of the centre
of mass energy to 28~TeV.
Three stages in the  upgrading process were identified:

\begin{itemize}

\item Phase 0: maximum performance without hardware changes to the LHC.

\item Phase 1: maximum performance while keeping the LHC arcs unchanged.

\item Phase 2: maximum performance with major hardware changes to the LHC.

\end{itemize}
The nominal LHC performance for a beam energy of 7~TeV corresponds to
a total beam-beam tune spread of 0.01, with $\rm{1.1\times 10^{11}}$
protons per bunch, yielding a luminosity of \lhclum\ 
in IP1 (ATLAS) and IP5 (CMS), halo
collisions in IP2 (ALICE) and low luminosity in IP8 (LHCb). Any
performance beyond these conditions will be considered as an LHC
upgrade.

The steps required to reach the maximum performance without hardware
changes to the accelerator (Phase~0) are:

\begin{itemize}

\item Collide beams only in IP1 and IP5.

\item Increase the bunch population up to the beam-beam limit 
of $\rm{1.7\times 10^{11}}$ protons per bunch, resulting
in a luminosity of \timeslhclum{2.3}\ at IP1 and IP5.

\item Increase the main dipole field to 9~T (ultimate field), resulting
in a maximum proton energy of 7.54~TeV. This ultimate dipole field corresponds
to a beam current limited by cryogenics and by beam dump considerations. 

\end{itemize}
  Increasing the LHC luminosity with hardware changes only in the
 LHC insertions and/or in the injector complex (Phase~1) includes the following
 steps:

\begin{itemize}
  
\item Modify the insertion quadrupoles and/or layout to yield a
  $\rm{\beta^*}$ of 0.25~m from the nominal 0.5~m. In addition,
  although this is not the favoured option, a possible modification to
  the layout is to include separation dipoles closer to the
  interaction point to reduce the effect of long-range beam-beam
  collisions.

\item Increase the crossing angle by $\rm{\sqrt{2}}$ to $\rm{424~\mu
    rad}$ from the nominal $\rm{300~\mu rad}$. The reason for
  increasing the nominal crossing angle by $\rm{\sqrt{2}}$ for half
  the nominal $\rm{\beta^*}$ is to keep the same small contribution of
  long-range collisions to the beam-beam footprint.
  
\item Increase the bunch population up to the ultimate intensity of
  $\rm{1.7\times 10^{11}}$ protons per bunch, resulting in a
  luminosity of \timeslhclum{3.3}\ at IP1 and
  IP5.

\item Upgrading the injectors to deliver beams with higher brilliance could
increase the luminosity without exceeding the beam-beam limit, by increasing
the product of crossing angle times bunch length. This option may yield a
luminosity of up to \timeslhclum{4}\ with a 
$\rm{\beta^*}$~=~0.5~m at IP1 and IP5.

\item Halving the bunch length with a new high-harmonic RF system would
increase the luminosity to \timeslhclum{4.7}\ 
at IP1 and IP5.

\end{itemize}
 However, there is an interesting alternative scheme to increase the LHC
luminosity based on very long `super-bunches'. This scheme would consist of the
following points:

\begin{itemize}

\item Modify the LHC insertion quadrupoles and/or layout to reach a 
$\rm{\beta^*}$ of 0.25 m.

\item Possibly increase the crossing angle to several mrad in order to
  pass each beam through separate final quadrupoles of reduced
  aperture.

\item Inject a bunched beam of 1~A and accelerate it to 7~TeV.

\item Use barrier buckets to form a single long super-bunch of 1~A current.

\end{itemize}
A 300-m long super-bunch in each of the LHC rings would be compatible
with the beam-beam limit, and the corresponding luminosity in ATLAS
and CMS (with alternating horizontal-vertical crossing planes) would
be about \timeslhclum{9}. The super-bunch option
is very interesting for large crossing angles. It can potentially
avoid electron cloud effects and minimize the cryogenic heat load.
However, the associated RF manipulations and beam parameters are
challenging and require further studies.  This scheme requires
upgrades to the detectors to achieve an effective length of between
20~m and 30~m.
In the following studies of physics performance and detector R\&D
we shall not analyze the impact of this super-bunch option, and
assume for the SLHC a 12.5~ns bunch crossing interval.

Finally, possible steps to increase the LHC performance with major hardware
changes in the LHC arcs and/or in the injectors (Phase~2) include:

\begin{itemize}

\item Modify the injectors to significantly increase the brilliance to beyond
its ultimate value (in conjunction with beam-beam compensation schemes).

\item Equip the SPS with superconducting magnets and inject into the LHC at 
1~TeV. This also implies a corresponding upgrade to the transfer lines. For 
given mechanical and dynamic apertures at injection, this option can increase
the LHC luminosity by about a factor of two. This would
also be the natural first step in view of an LHC energy upgrade, since the
corresponding energy swing would be reduced by a factor of two. 

\item Install new superconducting dipoles in the LHC arcs to reach a
  beam energy of 14~TeV. Magnets with a nominal dipole field of
  between 16 and 16.5~T, providing a safety margin of 1--2~T, can be
  considered a reasonable target for 2010 and could be operated by
  2015. This requires a vigorous $\rm{R \& D}$ programme
  on new superconducting materials.

\end{itemize}
  More details can be found in~\cite{machine-rep}. 
  As stated above, energy upgrades of the LHC, although being
  considered and being indeed most interesting from the physics point
  of view, require full replacement of the machine and a major $\rm{R
    \& D}$ activity to develop the needed dipole technology.
  Therefore the physics potential studies presented here were limited
  to the more realistic luminosity upgrade. In a few cases, however,
  the expected performance of a $pp$ machine running at a
  centre-of-mass energy of 28~TeV is given for comparison. More
  results for this option can be found in Ref.~\cite{atlas-note}.

\section{THE EXPECTED DETECTOR PERFORMANCE}
\label{sec:det-performance}

The expected ATLAS and CMS performance at \slhclum, and the
assumptions adopted for the physics studies discussed here, are
summarised below for the most relevant issues.

\subsection{Tracking and $\mathbf{b}$-tagging}
\label{sec:tracking}

It has been assumed that, provided that a large part of the inner
detectors of both experiments can be replaced with more radiation hard
and granular devices resulting from the R\&D described in
Section~\ref{sec:experiments}, reconstruction of isolated high-\pt
charged particles (e.g. muons and electrons with $\pt>20$~GeV) will be
possible with similar efficiency and momentum resolution as with the
present detectors operating at design luminosity. Obviously, the
information from the electromagnetic calorimeter and from the external
Muon Spectrometer will  be used to improve the performance.

The impact of the higher luminosity on b-tagging has initially been 
evaluated by
assuming that the new pixel detectors will have the
same two-track resolution as the current silicon systems.  In this way, the
probability of confusion in the pattern recognition remains low, and
the extra (fake) b-tags are given by real tracks from the minimum bias
events which are produced near the main event primary vertex and
within the jet cone. The results are shown in Table~\ref{tab:btag} as
a function of \pt. It can be seen that for a fixed b-tagging
efficiency of 50\%, the rejection against light-quark jets is
deteriorated by a factor of about eight at low \pt\ ($\pt<50$~GeV)
and by a factor of less than three above 100~GeV.
    
\begin{table}
\begin{center}
\caption{Rejection against u-jets (R$_u$) for a b-tagging efficiency
 of 50\% and in various \pt bins, as expected in ATLAS at the LHC
 design luminosity and with the upgraded luminosity.} 
\label{tab:btag}
\vspace*{0.1cm}  
\begin{tabular}{|c|c|c|}\hline
 \pt\ (GeV)         & R$_u$ at   \lhclum\  & R$_u$ at \slhclum \cr
\hline
 30-45   &  33  & 3.7 \\
 45-60   & 140  & 23  \\
 60-100  & 190  & 27  \\
 100-200 & 300  & 113 \\
 200-350 & 90   & 42  \\
\hline
\end{tabular}
\end{center}
\end{table}
      
Given that these results were obtained using the same 
response time as in the current detectors, which is a conservative
assumption in view of the discussion of Section~\ref{sec:IT}, 
we shall assume in most of the studies presented here that the
b-tagging performance in terms of efficiency and fake rate will be the
same at the SLHC as at the standard LHC.

\subsection{Electron identification and measurement}
\label{sec:calo}

  An increase of a factor of ten in luminosity, and therefore in the
  number of pile-up minimum-bias events, increases the contribution
 of the pile-up noise to the calorimeter energy resolution
 by about a factor of three. 
 The energy resolution of the ATLAS electromagnetic calorimeter at 
  \slhclum\ has been studied with
 electrons of \et=30~GeV and a full GEANT simulation which
  included the
 expected pile-up.  The energy resolution obtained by using also the
 information of the tracker is 3.6\%, to be compared to 2.5\% at the
 LHC design luminosity. The deterioration is expected to be
 smaller at higher electron
 energies, because the  contribution of the pile-up noise
 to the energy resolution decreases with the particle energy as $1/E$.
 
 Electron identification  at \slhclum\ has also
 been studied with a full simulation of ATLAS.  Table~\ref{tab:eID}
 compares the achieved rejection against jets faking electrons to the
 performance expected at the LHC design luminosity, for electrons with
 \et=40~GeV. For a fixed
 electron efficiency of about 80\%, the jet rejection
 decreases at the SLHC by about 50\% compared to the standard LHC.  This is
 mainly due to the fact that isolation cuts and shower shape criteria
 in the calorimeter are less powerful in the presence of a larger
 pile-up. As in the case of the energy resolution, the loss in jet rejection
 power is expected to decrease at higher electron energies.
\begin{table}
\begin{center}
\caption{Electron identification efficiency and rejection against
  jets, for \et=40~GeV, as expected in ATLAS at the LHC design
  luminosity and at the upgraded luminosity.}
\label{tab:eID}
\vspace*{0.1cm}  
\begin{tabular}{|c|c|c|}\hline
 L (cm$^{-2}$s$^{-1}$) & Electron efficiency  & Jet rejection \cr
\hline
 10$^{34}$ & 81\%   & 10600$\pm$2200 \\ 
 10$^{35}$ & 78\%   & 6800$\pm$1130 \\
\hline
\end{tabular}
\end{center}
\end{table}

\subsection{Muon identification and measurement}
\label{sec:muons}

 If enough shielding can be installed in the forward regions
to protect the external Muon Spectrometers from the increased
radiation background, the muon reconstruction efficiency and momentum
resolution provided by the muon chambers
are not expected to be seriously deteriorated when running
at the SLHC. The additional shielding tanslates into a reduced
rapidity acceptance, which will most likely be limited to the region
$|\eta|<$2. This is however not a big penalty, given the centrality of
very high-\pt\ final states.

\subsection{Forward-jet tagging and central-jet veto}
\label{sec:jettag}
    
The presence of two jets emitted in the forward and backward regions
of the detector is a distinctive signature of processes arising from
$WW$ or $ZZ$ fusion, where the vector bosons are radiated by the
incoming quarks. This is for instance one of the main Higgs production
mechanisms at the LHC.  Tagging these forward/backward jets (``forward
jet tagging") is an essential handle to increase the
signal-to-background ratio for these processes.  Similarly, the
relatively low jet activity at central rapidities in electroweak
processes (e.g. Higgs production) can be used to reject QCD
backgrounds (e.g. $t\overline{t}$). This is done by vetoing the
presence of additional jets.
 
As the luminosity increases, both these handles are expected to become
less powerful, mainly because the increased pile-up can give rise to
additional jets in the detector. Additional central jets overlapping
purely electroweak processes spoil the efficiency of the jet veto.
Similarly, additional forward jets from pile-up can mimic the typical
signature of vector-boson fusion processes.
  
The performance of the forward jet tag and central jet veto at the
SLHC was estimated in a preliminary way by using a full simulation of
the ATLAS detector.  The pile-up was generated by summing minimum-bias
events over the correct number of bunch crossings, and by taking into
account the shape of the electronic response of the various
calorimeters.  Jets were then found using a jet finder cone algoritm
with cone sizes $\Delta R=0.4$ and $\Delta R =0.2$, and were assigned
to ranges of rapidity:

\begin{itemize}
\item forward: $\eta>2.0$
\item backward: $\eta<-2.0$
\item central:  $\abs{\eta}<2.$
\end{itemize}

 A single tag is defined as an event with either a forward or backward
jet; a double tag has both.
 The probability of an event consisting only of minimum-bias
interactions having either a single or double 
jet tag is shown in Fig.~\ref{fig:jettag} as a function of the jet energy.
The  probability of an event having an additional central jet is shown
in Fig.~\ref{fig:jetveto} as a function of $p_T$. The numbers should
be compared to typical forward-jet tagging efficiencies of $\ge 80\%$.

\begin{figure}
\begin{center}
\includegraphics[width=0.45\textwidth,clip]{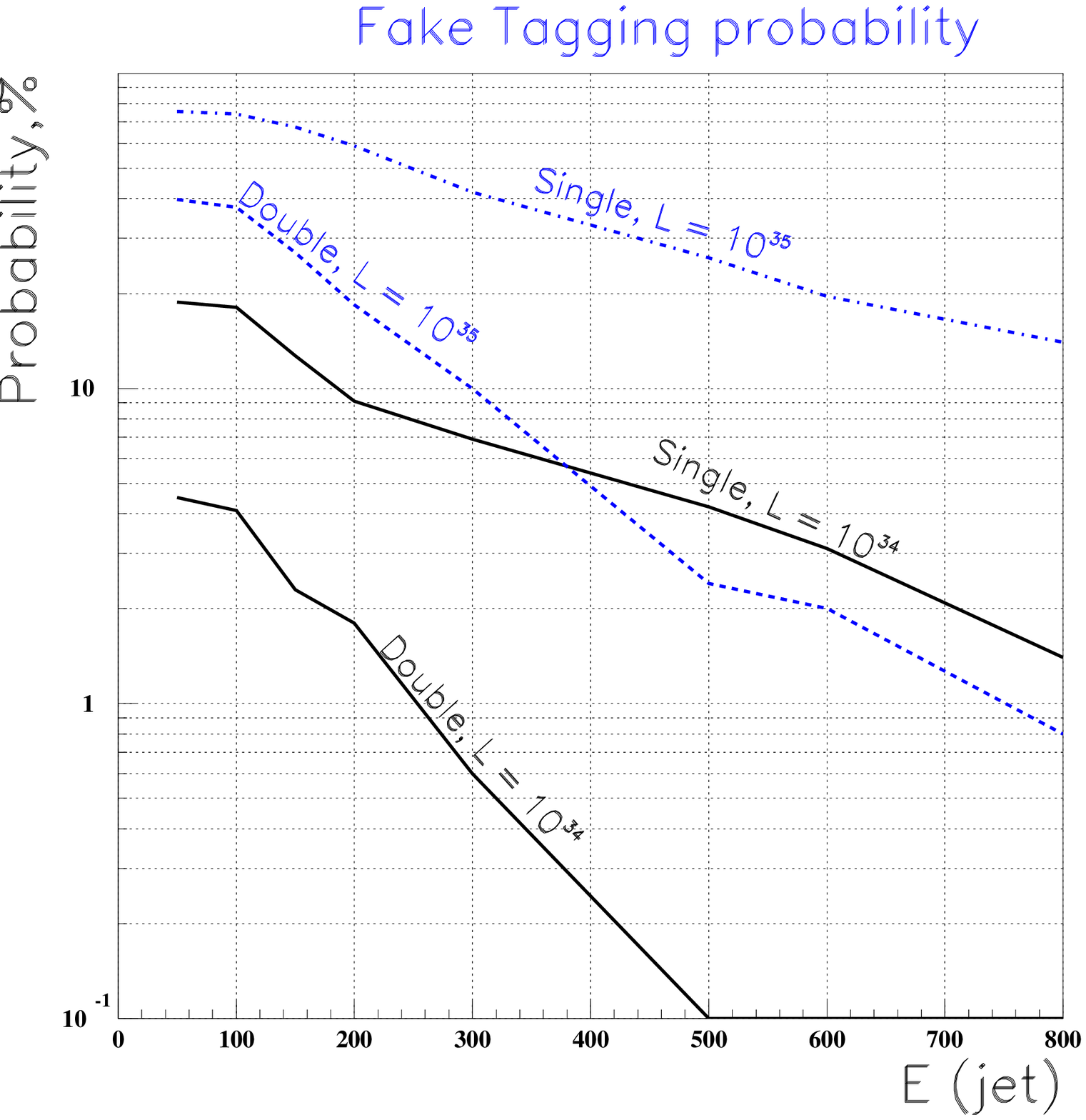} 
\includegraphics[width=0.45\textwidth,clip]{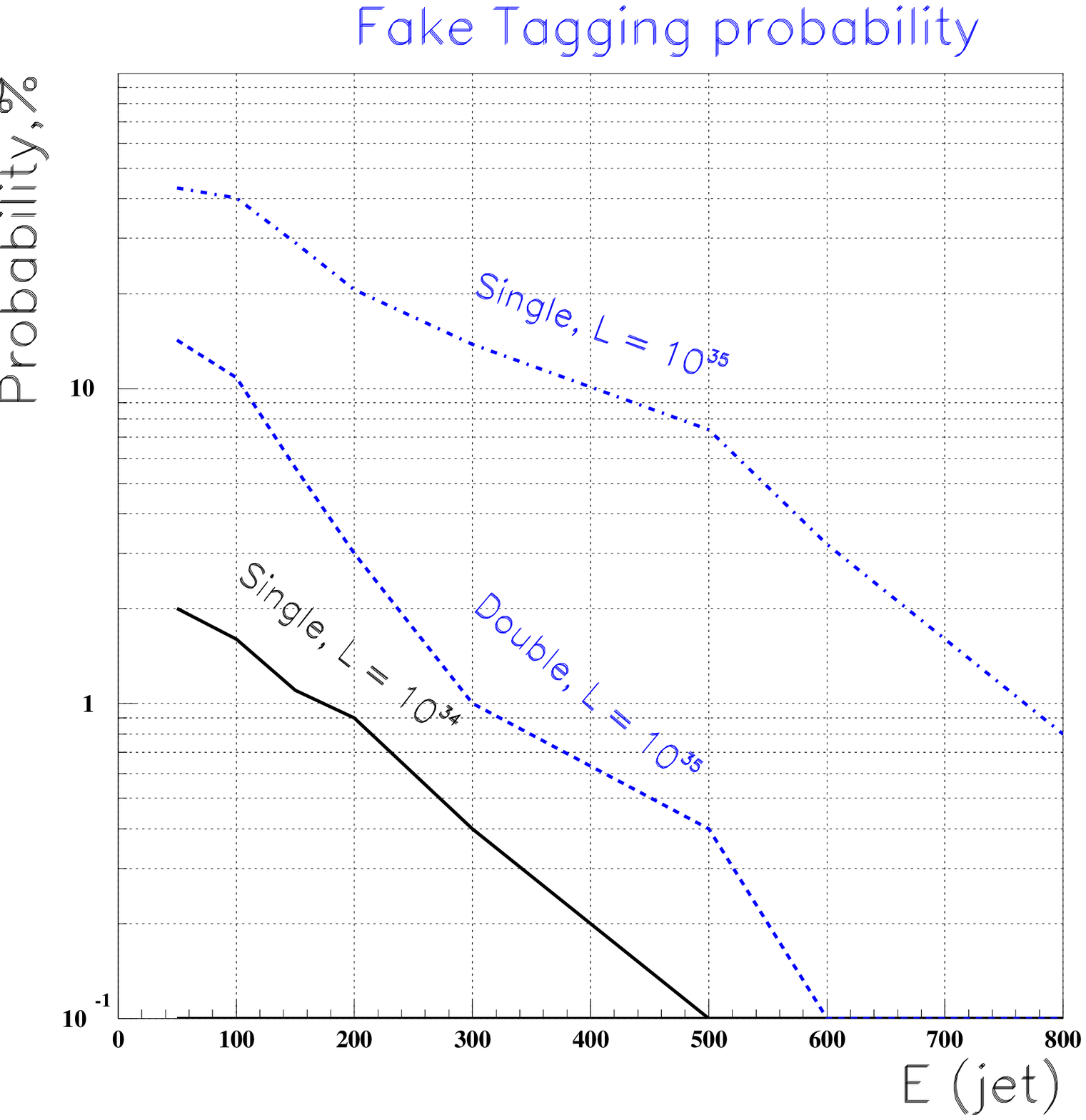} 
\end{center}
\vskip -0.4cm 
\caption{Estimates of probabilities of single and double forward jet tagging 
    from pile-up at the LHC, for the
    nominal and the upgraded luminosities,  as a function of the
    jet energy, and for jet cone sizes 
    $\Delta R = 0.4$ (left) and $\Delta R = 0.2$ (right).
\label{fig:jettag}}
\end{figure}

\begin{figure}
\begin{center}
\includegraphics[width=0.45\textwidth,clip]{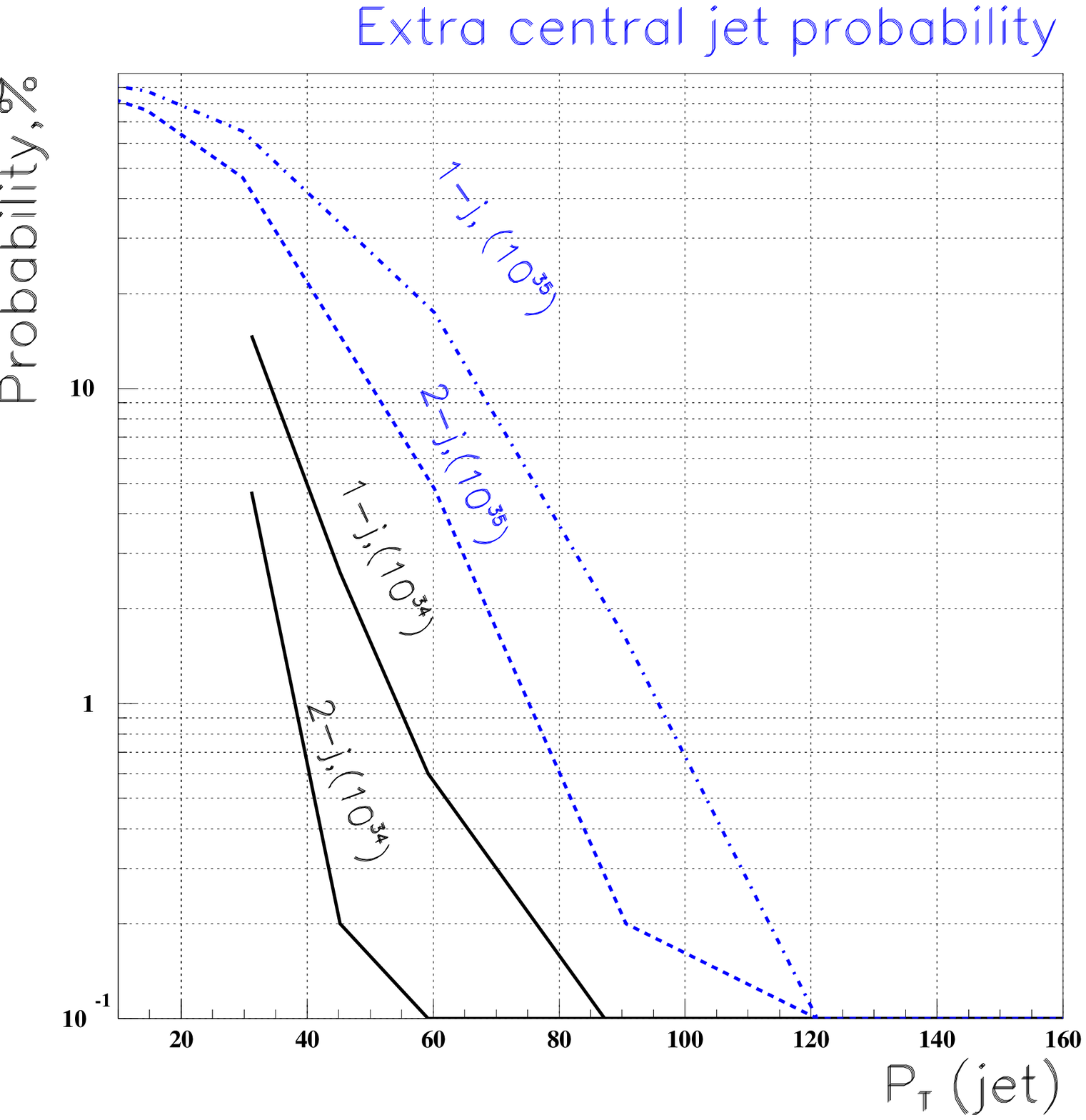} 
\includegraphics[width=0.45\textwidth,clip]{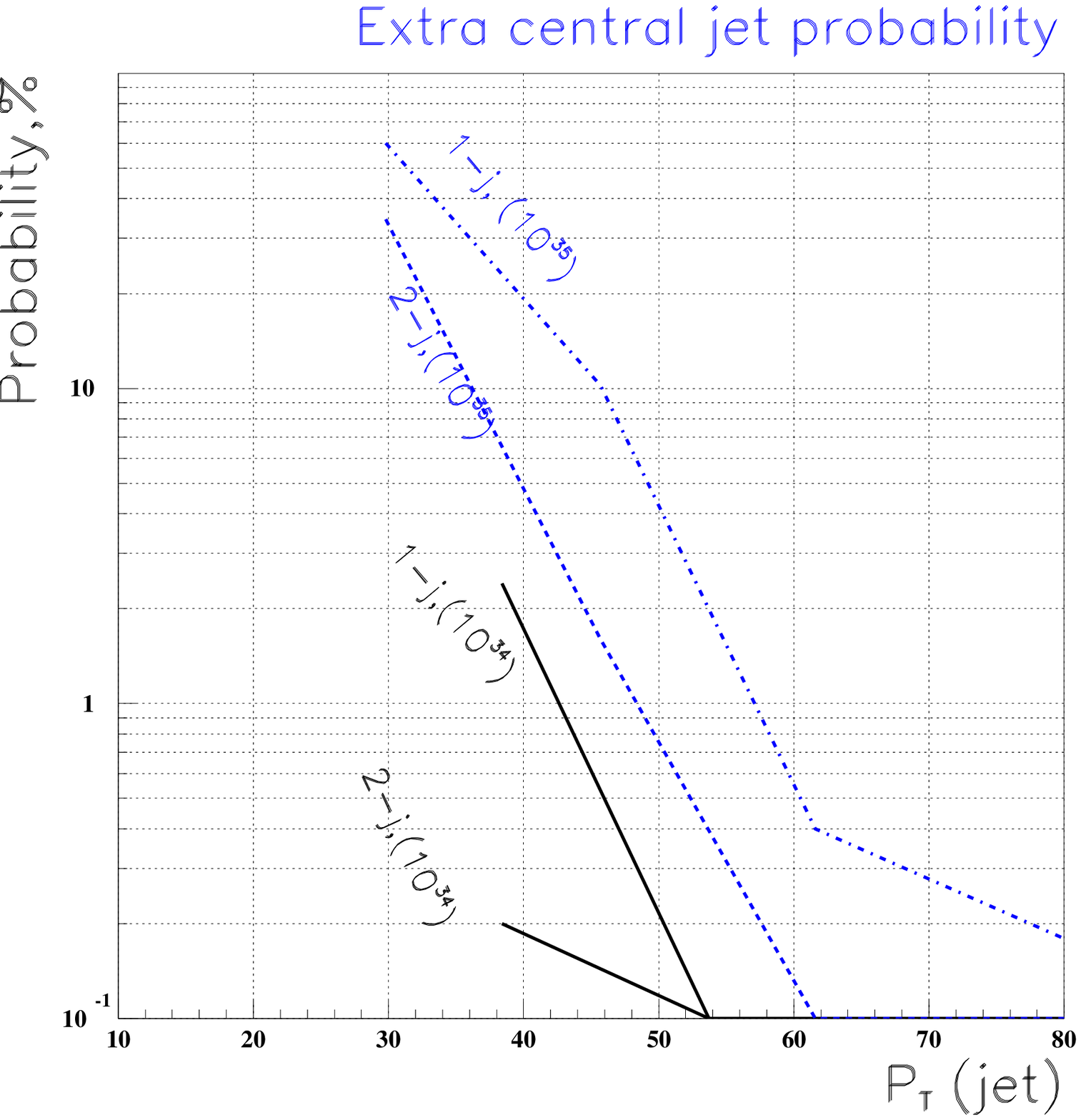} 
\end{center}
\vskip -0.4cm 
\caption{Estimates of 
probabilities of one or two extra central jets from pile-up 
 at the LHC, for the
    nominal and upgraded luminosities, as a function of the
    jet $p_T$ threshold, and for jet cone sizes $\Delta R = 0.4$ (left) 
    and $\Delta R = 0.2$ (right).
\label{fig:jetveto}}
\end{figure}

The main conclusion is that at a luminosity of \slhclum\ the forward
jet tag and central jet veto performances are significantly degraded
compared to design luminosity. However, if sufficiently small cone
sizes are used to limit pile-up effects ($\Delta R \leq 0.2$), these
two strategies can still be effective.  For instance, with a jet cone
$\Delta R = 0.2$ the probability of double tag is only 2\% for forward
jets with energy above 300~GeV. For the same cone size, an additional
central jet with $p_T>$50~GeV is found in 10\% of the cases.  We
stress that these results are preliminary. In particular, the fake
probabilities reported in Figs.~\ref{fig:jettag} and~\ref{fig:jetveto}
depend significantly on the jet energy calibration and reconstruction
conditions. Furthermore, the performance reported here may be too
pessimistic as it will be possible to increase the calorimenter
and trigger granularity in the forward regions, and to 
reduce the pile-up noise in the
calorimeters by using optimal filtering techniques. Finally, possible
algorithms to distinguish between incoherent pile-up of energy and QCD
jets (e.g. based on the longitudinal shower profiles) have not been
investigated.


\subsection{Trigger}
The strategy and expected performance of the trigger 
at the SLHC are discussed in Section~\ref{sec:TDAQ}.
Here we anticipate briefly
 that a robust trigger menu for physics can be
obtained, for an acceptable rate, by adopting the following philosophy:
setting high thresholds for the inclusive triggers aiming at selecting
very high-\pt\ final states; pre-scaling inclusive and semi-inclusive
low-\pt\ triggers aiming at selecting well-known control/calibration
samples (e.g. $Z\to\ell\ell$); using a set of exclusive menus
aiming at selecting specific final states (e.g. a given decay
channel of a low-mass Higgs already observed at  design
luminosity).

\subsection{Summary of the assumptions used in this work}
\label{sec:detector}

For the physics studies presented in this document, we have worked
mostly under the assumption that the detector performance at \slhclum\ 
is comparable to that at \lhclum. The material presented in
Section~\ref{sec:experiments}, as well as the considerations made
above, show that indeed the performance degradation is not expected
 to be dramatic in most cases.
 For example, for forward-jet tagging and central-jet veto one can
 recover to some extent the signal purity and background
 rejection by increasing the jet thresholds.  We shall explicitly
 mention the cases in which a more conservative scenario, with
 degraded detector
 performance, has been adopted.
 
 The detector simulations have been performed with different levels of
 detail.  In some cases, fast simulations of fully showered
 final states have been employed (e.g.
 {\small{ATLFAST}}~\cite{ATLFAST}). These fast simulations smear the
 momenta, energies and positions of the final-state particles
 according to the detector resolutions, as obtained from GEANT-based
 studies and from test beams.
 In other cases, the analyses are simply based on parton-level
 studies, with acceptances and efficiencies estimated as a function of
 the partons kinematics, but with no smearing for resolution effects.
 The precise assumptions will be
 listed for each individual study.

 We have assumed integrated luminosities corresponding to the standard
 data taking time of $10^{7}$ s/year.  This gives rise to total
 integrated luminosities for each experiment of $10^3$ ($3\times
 10^3$) \ifb\ for 1 (3) year(s) of running at \slhclum, and $10^2$
 ($3\times 10^2$) \ifb\ for 1 (3) year(s) at \lhclum.

\section{THE PHYSICS POTENTIAL}
\label{sec:physics}

In this Section we discuss examples of physics areas where the SLHC is
expected to improve on the LHC potential, either because of the
extended mass reach or because of the improved measurement precision.
It should be noticed that the quality and the depth of the studies
presented here are by far not comparable to those of previous studies
performed by ATLAS and CMS for the standard LHC and documented in
Technical Proposals and Technical Design Reports.  These results
should therefore be considered preliminary and illustrative only.

\subsection{Electroweak physics}
\label{sec:ewk}

The high precision studies performed at LEP have clearly indicated the
essential role played by precision determinations of the electroweak
(EW) parameters~\cite{Abbaneo:2001ix}, also as a tool to look
indirectly for New Physics.  We show in this Section a few examples of
measurements where the SLHC should improve significantly on the LHC
accuracy. In addition, new channels which are rate-limited at the LHC
should become accessible at the upgraded machine.

\subsubsection{Multiple gauge boson production}
\label{sec:multiw}

Production of multiple
($n_V\ge 3$) gauge bosons provides an important test of the
high energy behaviour of weak interactions. The cleanest  final states
are those where all $W$'s and $Z$'s decay leptonically, but these are 
compromised by  small branching ratios (BRs). 
As a rule of thumb, each
additional gauge boson emitted in the hard process costs a factor of
approximately 1000 in rate. 
 As a result, luminosity limits the number of possible channels
 accessible at the LHC. 

 Table~\ref{tab:nW} shows the
expectation at the SLHC. To obtain these results, we have assumed a 90\%
identification efficiency for each lepton, and applied the following
set of acceptance cuts:
\be \label{eq:lcuts}
\vert \eta_\ell \vert < 2.5 \; , \quad p_T^\ell > 20~\gev \ee
All rates were evaluated at LO, using the parton distribution set
CTEQ5M and renormalization and factorization scales
$\mur=\muf=\sum_{i=1,n_W+n_Z} \; m_{V_i}$.
NLO predictions are available
in the case of di-boson production~\cite{Dixon:1999di}, 
leading to $K$ factors of the order
of 1.3, 1.5 and 1.8 for $ZZ$, $W^+W^-$ and $W^{\pm}Z$,
respectively~\cite{Haywood:1999qg}.
Similarly,  $K$ factors of the order of 1.5 are expected for the higher
multiplicities. \\
From Table~\ref{tab:nW} it can be seen that,
  although with limited statistics, 
the channels $W^{\pm}ZZ\to \ 5\ell$, 
$ZZZ\to \ 6\ell$ and even the four-gauge-boson final state $W^+W^-W^+W^-$
become accessible at the SLHC. As a result, 
 the first limits on 5-ple gauge boson
vertices (expected to vanish in the SM) could be set.
 The standard LHC luminosity would not allow these 
channels to be oberved. 
 All rates are increased in presence of a
Higgs boson above threshold for the decay into boson pairs (see the
row corresponding to \mh=200~GeV in Table~\ref{tab:nW}).

\begin{table}
\begin{center}
\caption{Expected numbers
 of events in fully leptonic final states from multiple
 gauge boson production, for an integrated luminosity of 6000~\ifb 
 and after cuts.}
\label{tab:nW}
\vspace*{0.1cm}
\begin{tabular}{|l|cccccc|} \hline
Process        & $WWW$ & $WWZ$ &$ZZW$& $ZZZ$ & $WWWW$
                    & $WWWZ$ 
\\
N(\mh=120~GeV) & 2600          &  1100      &  36       & 7  &  5 
                   &  0.8    \\ \hline
N(\mh=200~GeV) & 7100          &  2000      &  130       & 33 &  20
                   &  1.6     \\ \hline
\end{tabular}
\end{center}
\end{table}

In addition to the direct production of multiple gauge bosons, the
SLHC luminosity should allow also an accurate measurement of triple boson
production in boson-boson fusion (allowing, again,  a test of the
quintuple couplings). For example, for the  $WZ$ fusion processes
\be
ZW^{\pm} \to W^+W^-W^{\pm}
\ee
 870 fully leptonic events are expected for \mh=120~GeV
 and 6000~\ifb. In the case of a
200~GeV Higgs, including the resonant production of $W$ pairs leads to
2700 events. 

\subsubsection{Triple gauge boson couplings}
\label{sec:TGC}
 As discussed in~\cite{Haywood:1999qg}, the LHC will
significantly improve the precision of the measurements of the
Triple Gauge boson Couplings (TGC) compared to the LEP and
Tevatron results.
In the SM, the TGC's are
uniquely fixed by gauge invariance and renormalizability. Extensions
to the SM, in which for example the gauge bosons are not elementary
but are bound states of more fundamental particles, generically
lead to deviations from the SM prediction for the TGCs. The larger the
available statistics, the higher the sensitivity to these deviations.
  In the case of a positive indication of non-SM TGC's at
the LHC, increased statistics at the SLHC should allow a deeper
understanding of which specific realization of New Physics is
responsible for these deviations.  The latter are in fact parameterized by
effective interactions which, in order to preserve unitarity
 at high energy, require the inclusion of form factors. The
mass scale $\Lambda$ which is needed to define such form factors is
typically associated to the scale at which New Physics 
manifests itself. A measurement of the energy dependence of the TGC's
will probe the shape of the form factor~\cite{Haywood:1999qg}, 
and therefore allow extraction of the
value of the scale $\Lambda$. As a result, there is no limit a priori
to the accuracy one would like to achieve in the determination of
TGCs: the larger the precision, the stronger the reach in searches for
 phenomena beyond the Standard Model.

  Assuming electromagnetic gauge invariance and C- and P-conservation, 
 five parameters can be used to describe the triple-gauge vertices:
 $g_1^Z$, $\Delta\kappa_{\gamma}$, $\Delta\kappa_{Z}$, $\lambda_{\gamma}$, 
 and $\lambda_{Z}$~\cite{Haywood:1999qg}. 
  The SM values of these parameters are one for
 $g_1^Z$ and zero for the others, at the tree level. 
 
 At the LHC two processes can be studied to access these TGCs:
 $W\gamma\to\ \ell\nu\gamma$ production, which probes the couplings
 $\lambda_{\gamma}$ and $\Delta\kappa_{\gamma}$, and $WZ\to\ 
 \ell\nu\ell\ell$ production, which probes the couplings
 $\lambda_{Z}$, $\Delta\kappa_{Z}$ and $g_1^Z$.  After simple
 kinematic cuts and lepton and photon identification, a few thousands
 events are expected for an integrated luminosity of 100~\ifb\ in one
 experiment, with a background contamination of 20-30\%.  In contrast,
 the $WW\to\ \ell\nu\ell\nu$ process, which also proceeds through
 triple-gauge interactions in the $s$-channel, suffers from a large
 $\ttbar$ background and therefore has not been considered here.
 
  The experimental sensitivity to anomalous TGC's arises from the increase
 of the production cross-section and from 
 alteration of differential distributions. 
 The $\lambda$-type couplings have a strong $\sqrt{s}$ dependence, being
 enhanced at high centre-of-mass energy. Therefore they can be
 constrained by  measuring the total cross-section for the above-mentioned
 processes and by looking for an excess of gauge boson pairs with
 high $p_T$ compared to the SM expectation.  
  The $\kappa$-type couplings, on the other hand, 
 modify mainly the angular distributions of the final state particles. 
   
 For a luminosity of \slhclum, the analysis
 reported here uses conservatively only final states containing muons
 and photons, because these final states do not necessitate
 significant detector upgrades.  This choice entails a loss of 50\%
 (75\%) of the $W\gamma$ ($WZ$) effective rate. In addition, only
 transverse momentum distributions have been used  to constrain
 TGC's, therefore these results are pessimistic in the case of the
 $\kappa$-type couplings.
 
 The expected sensitivity is summarised in Table~\ref{tab:TGC} and
 Fig.~\ref{fig:TGC} for different
 luminosity scenarios.  For comparison, the performance of
 a $pp$ machine running at $\sqrt{s}$~=~28~TeV is also shown.  
 Only statistical errors have been taken
 into account. The dominant systematic uncertainty is expected to come
 from higher-order QCD corrections.  Their contribution, which has not
 been evaluated for the studies presented here, depends on the extent
 to which they can be constrained by using theory and data and
 controlled by applying a jet veto at \slhclum.
   
 It can be seen that a tenfold luminosity increase should
  extend the sensitivity for the $\lambda$-type and $g_1^Z$
 parameters into the range ($\sim$ 0.001) expected from radiative
 corrections in the Standard Model, and should therefore allow a
 meaningful test of these corrections and others that arise for
 example in Supersymmetric models.  It should also be noted that, even
 in the pessimistic approach adopted here and with only one
 experiment, the precision on the $\lambda$-type and $g_1^Z$
  parameters is equal to or better
 than that expected at a Linear Collider with $\sqrt{s}=$~500~GeV and
 an integrated luminosity of 500~\ifb~\cite{TESLA}. 
   On the other hand even the
 SLHC is not competitive with a Linear Collider for the
 measurement of the $\kappa$-type parameters, which do not exhibit a
 strong energy dependence and which are optimally constrained by
 angular measurements in the clean environment of an $e^+e^-$ machine.

\begin{table}
\begin{center}
\caption{Expected 95\%~C.L. constraints on Triple Gauge Couplings
 in ATLAS for various luminosity/energy scenarios ($\Lambda$~=10~TeV).
 Only one coupling is allowed to vary at the time, while the others are fixed
 at their SM values. The last column shows the expectation for a Linear Collider
 with $\sqrt{s}$=500~GeV and 500~\ifb~\cite{TESLA}.}
\label{tab:TGC}
\vspace*{0.1cm}
\begin{tabular}{|c|c|c|c|c|c|}\hline
Coupling&14 TeV& 14 TeV &28 TeV& 28 TeV & LC \cr
&100 fb$^{-1}$&1000 fb$^{-1}$&100 fb$^{-1}$&1000 fb$^{-1}$ & 500~\ifb, 500~GeV\cr\hline
$\lambda_\gamma$&0.0014&0.0006&0.0008&0.0002& 0.0014\cr\hline
$\lambda_Z$&0.0028&0.0018&0.0023&0.009 & 0.0013\cr\hline
$\Delta \kappa_{\gamma}$&0.034&0.020&0.027&0.013 & 0.0010\cr\hline
$\Delta \kappa_Z$&0.040&0.034&0.036&0.013 & 0.0016\cr\hline
$g_1^Z$&0.0038&0.0024&0.0023&0.0007 & 0.0050\cr\hline
\end{tabular}
\end{center}
\end{table}

\begin{figure}
\begin{center}
\includegraphics[width=0.60\textwidth,clip]{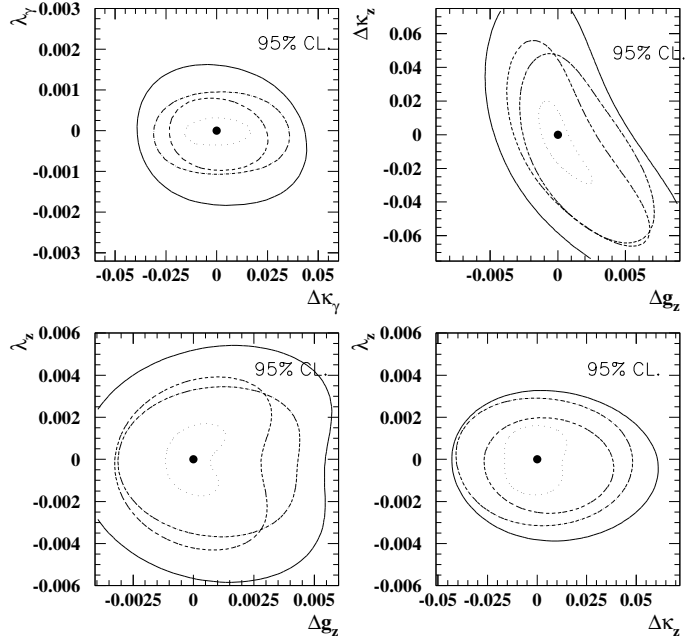} 
\end{center}
\vskip -0.4cm 
\caption{Expected 95\%~C.L. constraints on Triple Gauge Couplings 
 in ATLAS,  resulting from two-parameter fits ($\Lambda$~=10~TeV). 
  The contours correspond to
 14~TeV and 100~\ifb\  (solid), 28~TeV and 100~\ifb\  (dot dash), 14~TeV 
 and 1000~\ifb\  (dash) and 28~TeV and 1000~\ifb\  (dotted).\label{fig:TGC}}
\end{figure}

\subsubsection{Quartic gauge boson couplings}
\label{sec:qgc}
 Quartic boson couplings (QGC) are an essential component of the EW
theory. Similarly to the TGCs, they are required by gauge invariance
and their values are
uniquely determined within the SM by the value of the EW gauge
coupling. As in the case of TGC's, possible deviations from the SM
prediction are parametrised in terms of effective terms in the
Lagrangian. 

The results presented here are based on the
work of Ref.~\cite{Belyaev:1998ih}, where the following operators
leading to genuine quartic vertices are considered:
\begin{eqnarray}
{\cal L}_4 &=& \alpha_4\left[{\rm{Tr}}
\left(V_{\mu}V_{\nu}\right)\right]^2
\label{eq:Leff4}
\;, \\
{\cal L}_5 &=& \alpha_5\left[{\rm{Tr}}
\left(V_{\mu}V^{\mu}\right)\right]^2
\label{eq:Leff5}
\;, \\
{\cal L}_6 &=& \alpha_6 \; {\rm{Tr}}\left(V_{\mu}V_{\nu}\right)
{\rm{Tr}}
\left(TV^{\mu}\right){\rm{Tr}}\left(TV^{\nu}\right) \;, \\
{\cal L}_7 &=& \alpha_7\;{\rm{Tr}}\left(V_{\mu}V^{\mu}\right)
\left[{\rm{Tr}}\left(TV^{\nu}\right)\right]^2
\;, \\
{\cal L}_{10} &=& \frac{\alpha_{10}}{2}
\left[{\rm{Tr}}\left(TV_{\mu}\right)
\;{\rm{Tr}}\left(TV_{\nu}\right)\right]^2
\; .
\label{eff:10}
\end{eqnarray}
In the unitary gauge, there
are new  anomalous contributions to the $ZZZZ$ vertex coming from
all five operators, to the $W^+W^-ZZ$ vertex from all operators
except ${\cal L}_{10}$, and to the $W^+ W^- W^+ W^-$ vertex
  from ${\cal L}_4$ and ${\cal
L}_5$.  
A possible way to probe these couplings is via the 
scattering of gauge bosons in
reactions like $pp \rightarrow q q V V \rightarrow V V j j$
\cite{bagger0,bagger,dobado}, with $V = W^\pm$ or $Z$. 

Table~\ref{tab:QGClimits} shows the limits on the couplings $\alpha_i$
($i=4,5,6,7,10$) expected at the LHC, as a function of integrated
luminosity, compared  to current indirect limits from
Ref.~\cite{loops}. Fully leptonic final states were required and
the cuts applied are those of Eq.~(\ref{eq:lcuts}). It can be seen that
  in few cases the improvement obtained with the luminosity
upgrade goes beyond the simple statistical scaling. This is due
to the fact that almost no events are expected in the $ZZ$
final state at \lhclum. 
The interplay among the various channels, and the correlations among
different parameters $\alpha_i$, are shown in 
Figs.~\ref{fig:QGC45}--\ref{fig:QGC10}. 
\begin{table}
\caption{1$\sigma$ limits on the anomalous quartic couplings $\alpha_i$
  at LHC and SLHC (95\% C.L. limits are also given in this case), 
   as well as the present indirect
 bounds from Ref.\ \protect\cite{loops}.} 
\label{tab:QGClimits}
\begin{tabular}{|c | c | c |c |c |}
\hline
         & Indirect Limits   & LHC, 100 fb$^{-1}$ & LHC, 6000 fb$^{-1}$ & LHC, 6000 fb$^{-1}$\\
 Coupling &    $(1\sigma)$   & $(1\sigma)$       &  $(1\sigma)$       &  $95\%$ C.L.         \\
          & ($\times 10^{-3}$)& ($\times 10^{-3}$) & ($\times 10^{-3}$)  & ($\times 10^{-3}$) \\
\hline
$\alpha_4$ &$-120. \leq \alpha_4 \leq 11. $
           &$-1.1  \leq \alpha_4 \leq 11. $
           &$-0.67  \leq \alpha_4 \leq 0.74 $
           &$-0.92  \leq \alpha_4 \leq 1.1 $
\\ \hline
$\alpha_5$ &$-300. \leq \alpha_5 \leq 28. $
           &$-2.2  \leq \alpha_5 \leq 7.7 $
           &$-1.2  \leq \alpha_5 \leq 1.2 $
           &$-1.7  \leq \alpha_5 \leq 1.7 $
\\ \hline
$\alpha_6$ &$-20. \leq \alpha_6 \leq 1.8 $
           &$-9.6 \leq \alpha_6 \leq 9.1 $
           &$-3.5 \leq \alpha_6 \leq 3.2 $
           &$-4.3 \leq \alpha_6 \leq 3.9 $
\\ \hline
$\alpha_7$ &$-19. \leq \alpha_7 \leq 1.8 $
           &$-10. \leq \alpha_7 \leq 7.4 $
           &$-4.4 \leq \alpha_7 \leq 2.2$
           &$-5.4 \leq \alpha_7 \leq 2.8$
\\ \hline
$\alpha_{10}$&$-21. \leq \alpha_{10} \leq 1.9$
             &$-24. \leq \alpha_{10} \leq 24.$
             &$ -4.1\leq \alpha_{10} \leq 4.1$
             &$ -4.8\leq \alpha_{10} \leq 4.8$\\
\hline       
\end{tabular}
\end{table}

\begin{figure}
\begin{center}
\includegraphics[width=0.45\textwidth,clip]{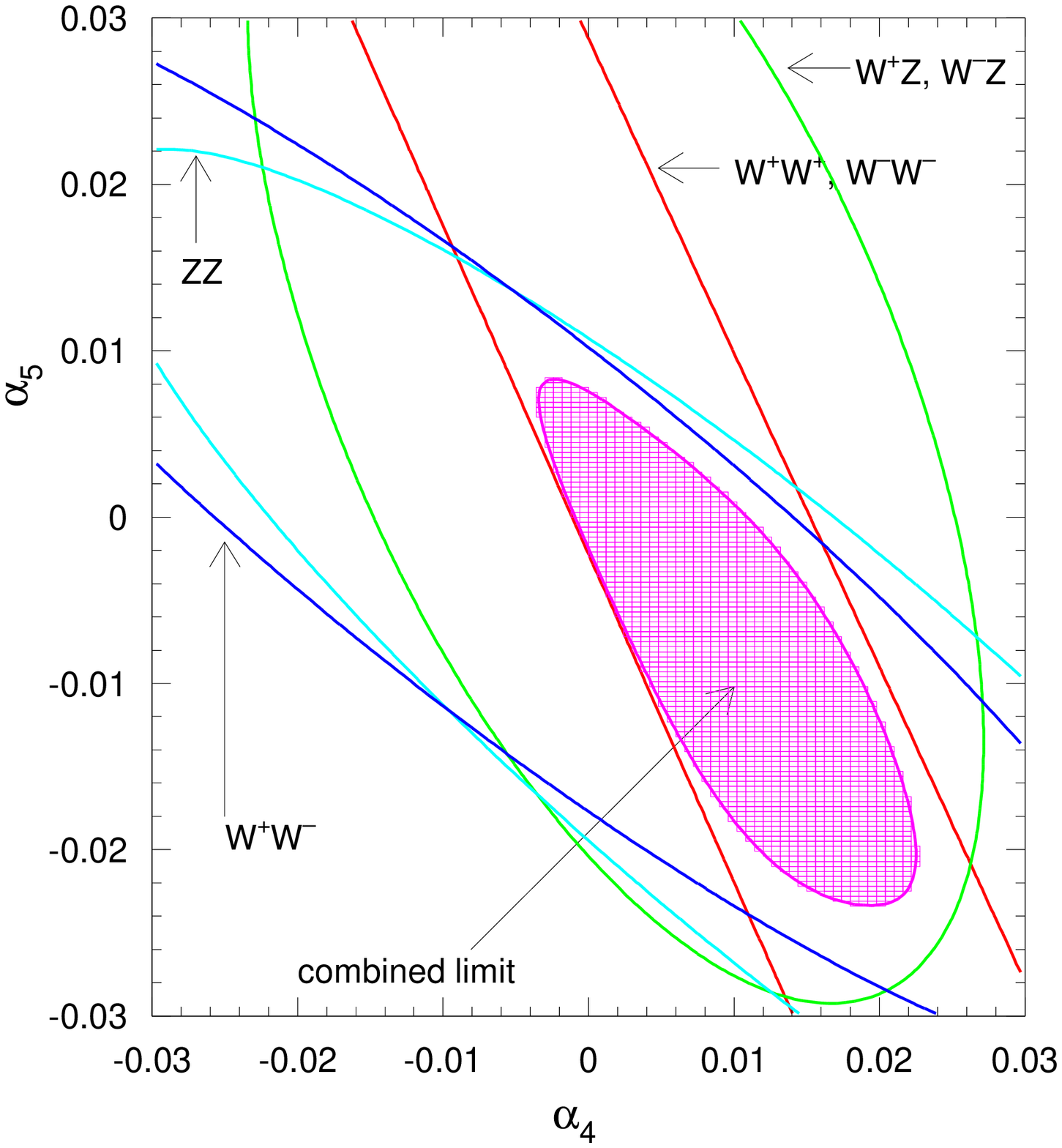} \hfill
\includegraphics[width=0.45\textwidth,clip]{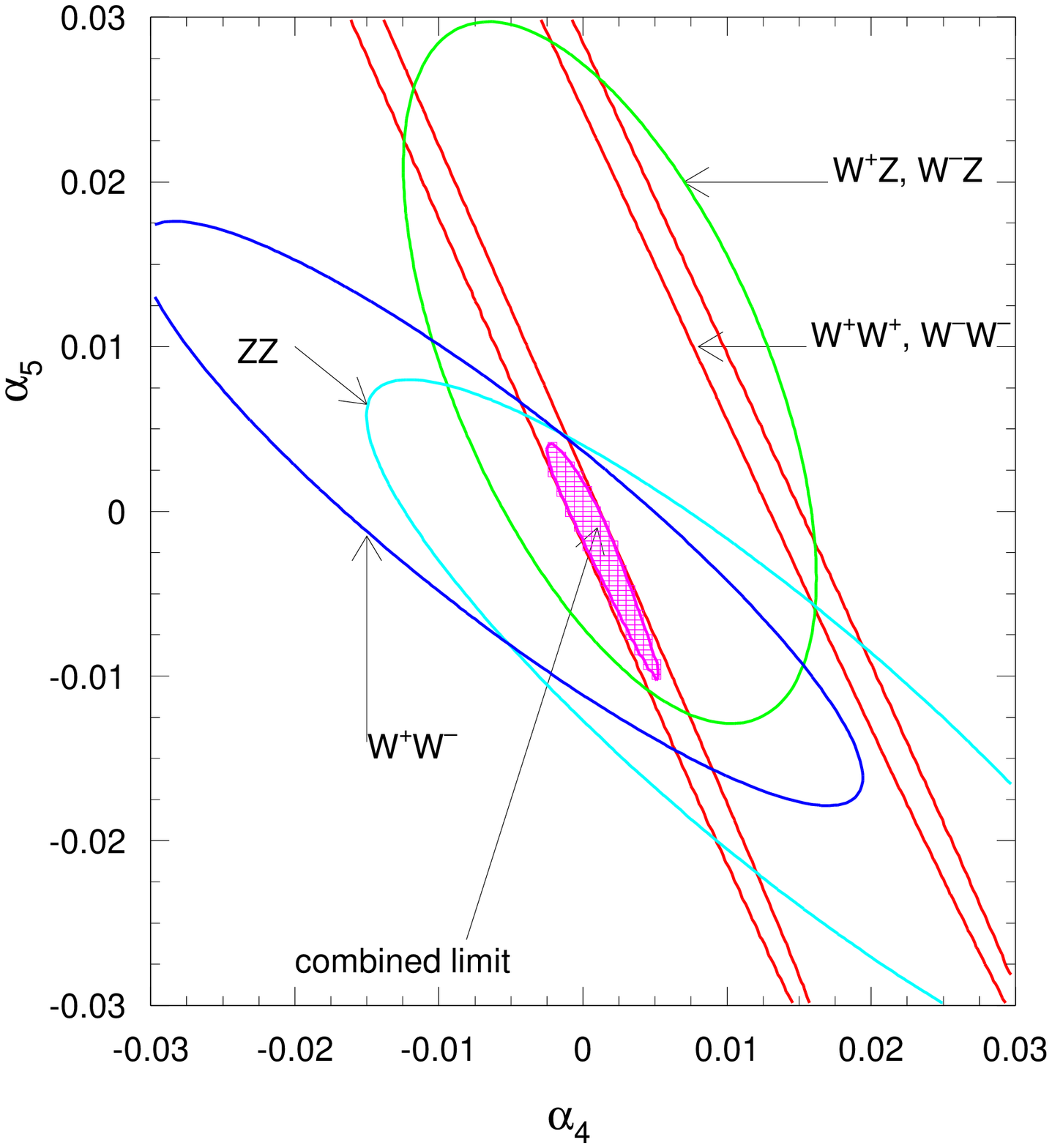} 
\end{center}
\vskip -0.4cm 
\caption{1-$\sigma$ contours in the $\alpha_4 -\alpha_5$
plane for $W^+W^-$, $W^\pm W^\pm$, $W^\pm Z$ and $ZZ$ production for
an integrated
luminosity of 100 fb$^{-1}$ (left plot)  and 6000 fb$^{-1}$ (right plot).
 We have assumed $\alpha_6 = \alpha_7 = \alpha_{10} = 0$. 
\label{fig:QGC45}}
\end{figure}


\begin{figure}
\begin{center}
\includegraphics[width=0.45\textwidth,clip]{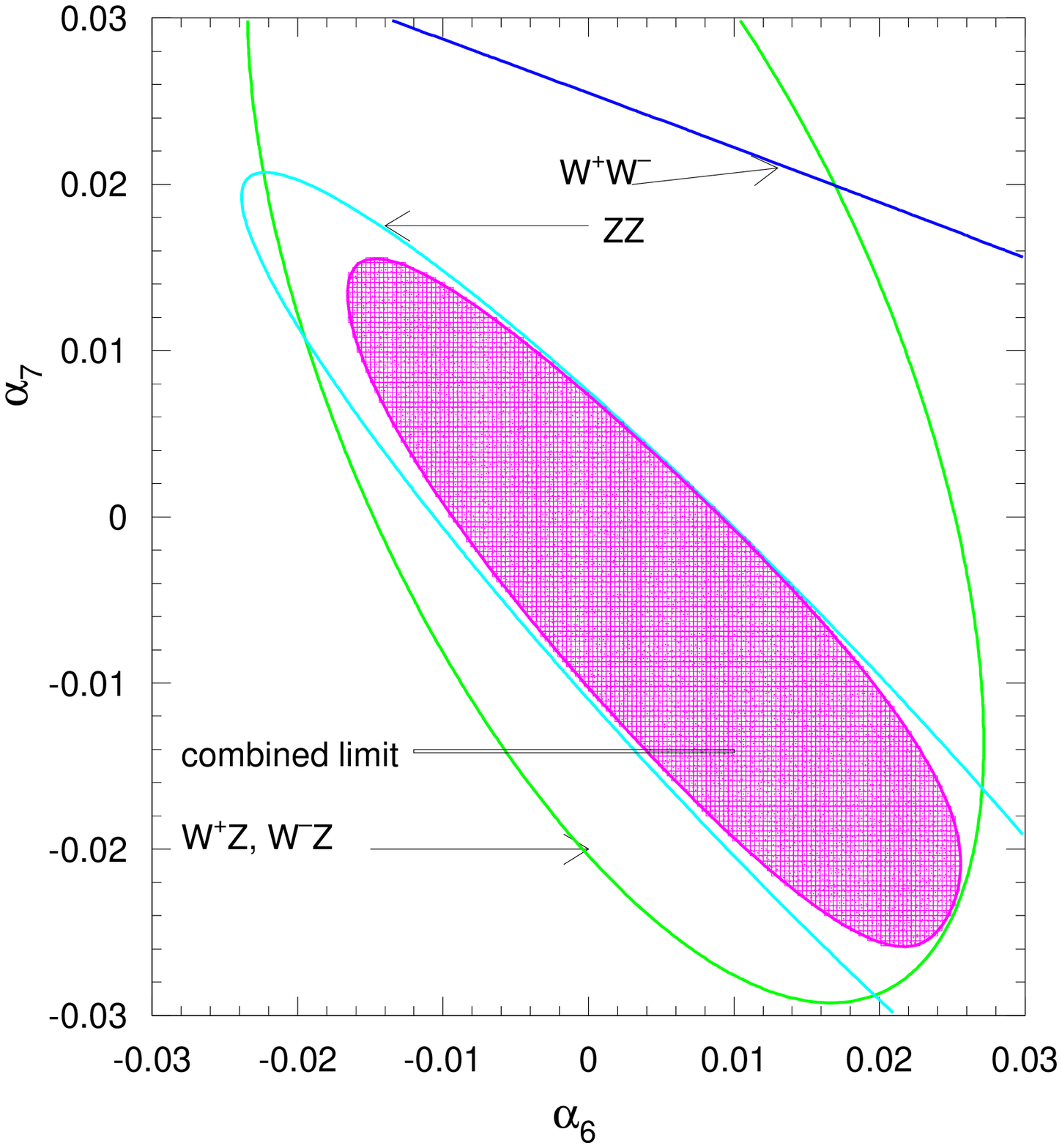} \hfill
\includegraphics[width=0.45\textwidth,clip]{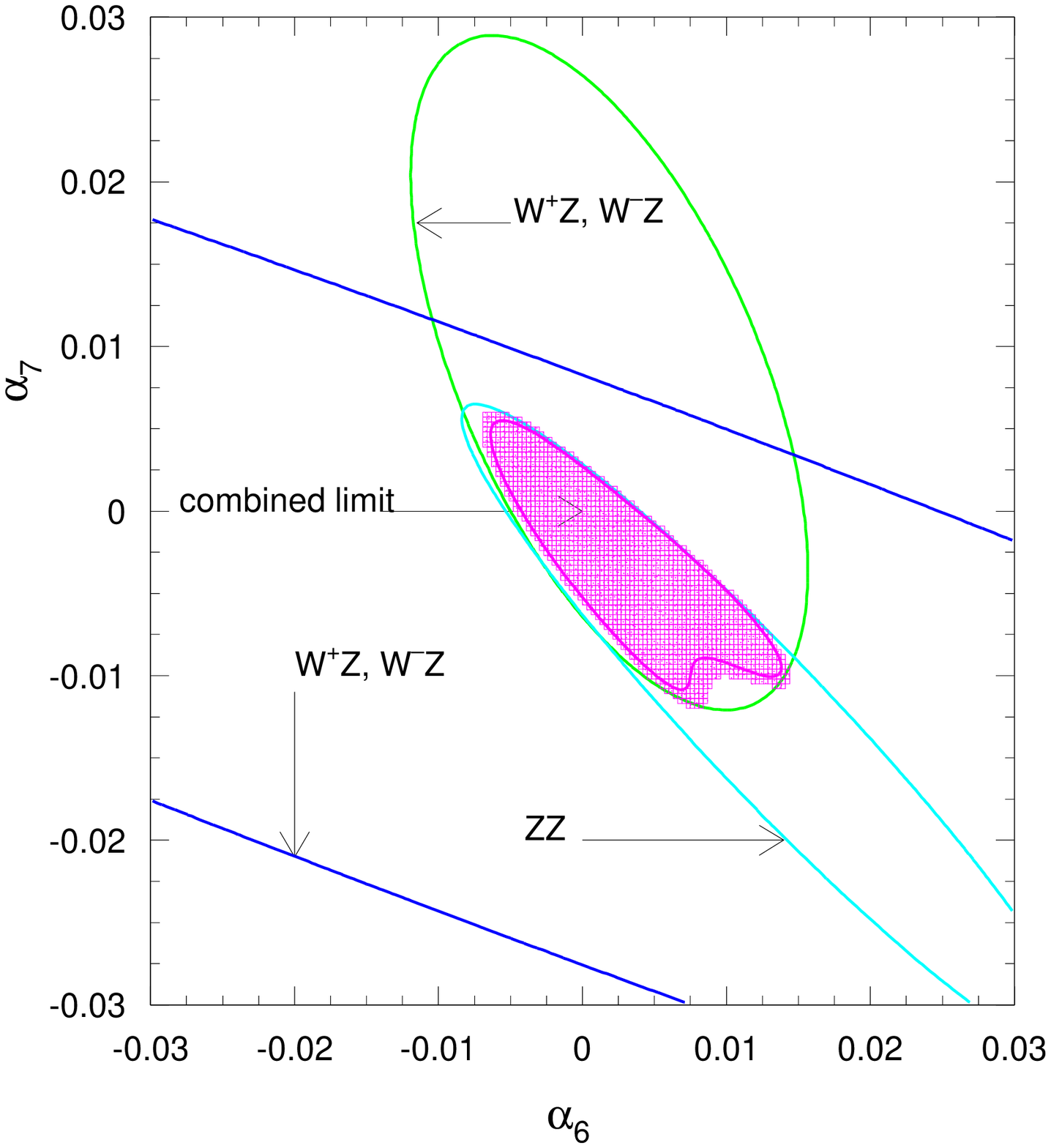} 
\end{center}
\vskip -0.4cm 
\caption{1-$\sigma$ contours in the $\alpha_6 - \alpha_7$ plane 
 from $W^+W^-$, $W^\pm Z$ and and $ZZ$ production for an integrated 
 luminosity of 100~fb$^{-1}$(left plot) and 6000~fb$^{-1}$ (right plot).   
 We have assumed $\alpha_4 = \alpha_5 = \alpha_{10} = 0$. 
\label{fig:QGC67}}
\end{figure}

\begin{figure}
\begin{center}
\includegraphics[width=0.45\textwidth,clip]{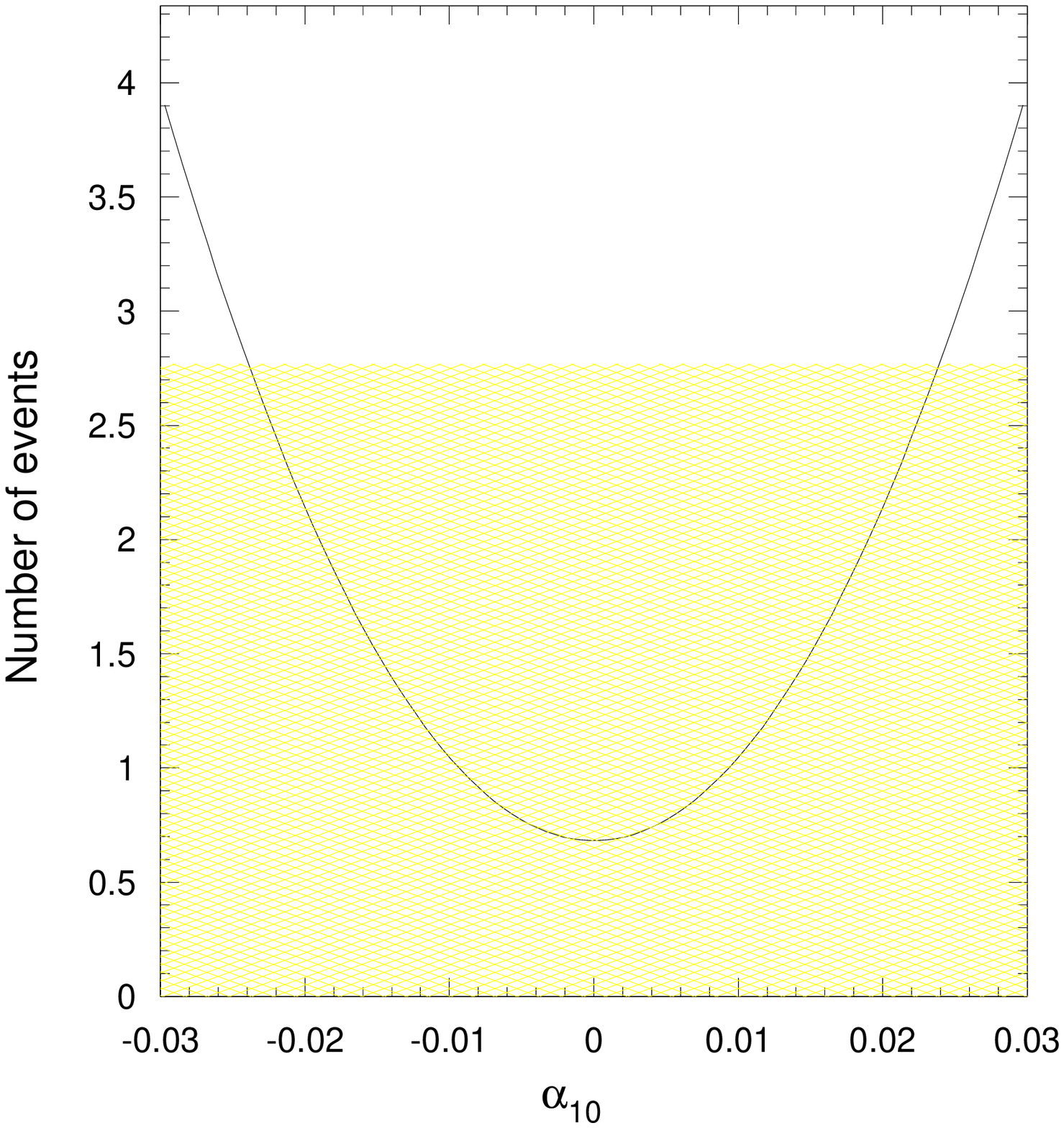} \hfill
\includegraphics[width=0.45\textwidth,clip]{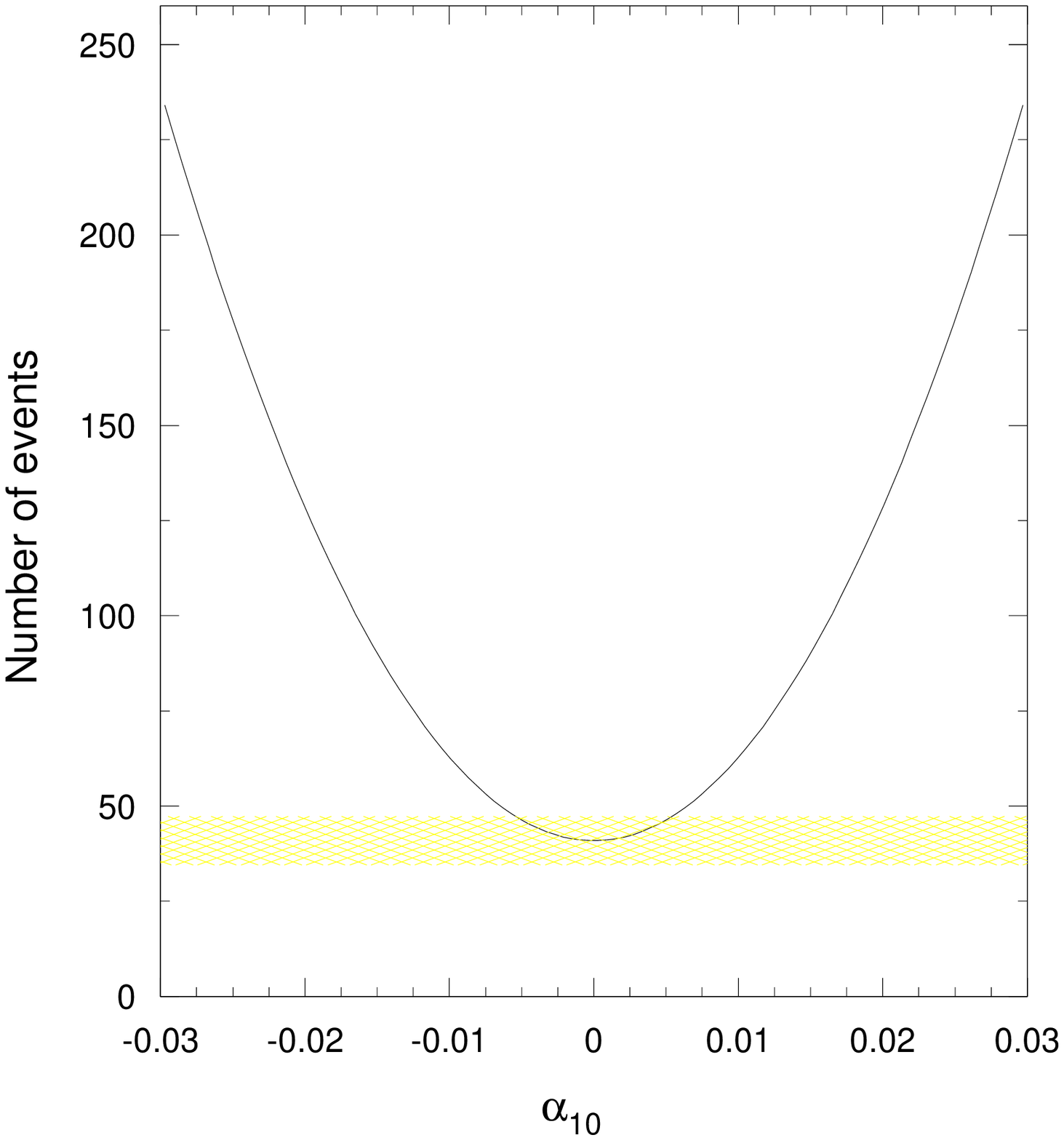} 
\end{center}
\vskip -0.4cm 
\caption{Expected number of events from $ZZ$ production as a function of
  $\alpha_{10}$ for an integrated luminosity
  of 100~fb$^{-1}$(left plot) and 6000~fb$^{-1}$ (right plot).
  The horizontal line represents a 64\% C.L. effect.  We
  have assumed  $\alpha_4 = \alpha_5 = \alpha_6 = \alpha_7 =0 $.  
\label{fig:QGC10}}
\end{figure}

In addition to the vector boson scattering processes, an alternative
probe of quartic couplings is given by the production of three 
gauge bosons via the off-resonance production of a W or Z decaying
into a system of three gauge bosons ($V^* \to VVV$).  In this case,
a different kinematical configuration is probed. For vector boson
scattering, two of the bosons are space-like, with virtualities of the
order of $M_W$; for triple gauge boson production, one is
off-shell but  is time-like and with large virtuality. The
observation of anomalies in the two channels would therefore provide
complementary information, and would also be sensitive to different
combinations of the QGC parameters. For triple gauge boson production
 we updated the studies presented in~\cite{Haywood:1999qg}, assuming a
total integrated luminosity of 6000~\ifb. Given the number of events
quoted in Table~\ref{tab:nW}, and using just the $ZZZ$ final state,
the limit $\vert\alpha_4+\alpha_5\vert < 0.025$ was obtained at 95\% C.L.
 for $\Lambda=2~\tev$. This is to be compared with 0.09 with
100~\ifb. 


\subsection{Higgs physics}

The Higgs search programme, including the discovery of a SM Higgs
over the full allowed mass range or of at least one SUSY Higgs boson,
will be largely accomplished at the standard LHC. There are
issues where the LHC potential is limited by statistics and where a
luminosity upgrade could therefore have a significant impact.
 
Four examples are discussed in this Section. The possibility of
observing a Higgs boson in decay channels which are rate-limited at
the LHC; the measurements of Higgs couplings to fermions and bosons
with improved precision; the possibility of observing the production
of Higgs pairs and measuring Higgs self-couplings; the increased
discovery potential for heavy MSSM Higgs bosons in the difficult
region of the decoupling limit.
 It should be stressed that Higgs physics requires fully functional
detectors, providing powerful b-tagging and electron identification
and allowing precise measurements of particles with moderate energies.
Major detector upgrades are therefore needed in this case for the SLHC
phase.

\subsubsection{Rare decay modes}
Two examples of channels which are accessible with difficulty at the
LHC because of their tiny rates are discussed here. Assuming that a
Higgs boson will have been previously discovered in higher-rate final
states, these rare decay modes can be used to extend the information
available on the Higgs couplings to fermions and bosons.
   
The possible decay $H\to Z\gamma$ of a SM Higgs is relevant only in
the mass region 100-160~GeV and has a branching ratio of a few per
mill. In contrast to the $H\to\gamma\gamma$ channel, which has a
similar branching ratio, an additional suppression in the $Z\gamma$
case comes from the fact that only decays of the $Z$ into electron or
muon pairs lead to observable final states above the background at the
LHC. Taking this into account, the production cross-section times
branching ratio for $H\rightarrow Z\gamma\rightarrow\ell\ell\gamma$ is
only $\sim$2.5~fb, too small to be observed at the LHC.  Indeed, the
expected significance for 600~fb$^{-1}$ (300~fb$^{-1}$ per experiment)
is $\sim3.5\sigma$. A factor of ten in luminosity, i.e.
6000~fb$^{-1}$, would allow the observation of a signal at the
$\sim11\sigma$ level.  This additional channel would provide an
additional measurement of the Higgs couplings to bosons (see
Section~\ref{sec:hcoupl}).
  
  Another interesting decay mode which may become accessible at the
  SLHC is $H \rightarrow \mu^+\mu^-$. This channel, which
  in the SM has a BR of order $10^{-4}$, has recently been
  studied for the LHC design luminosity in~\cite{Plehn:2001qg} using
  production via gauge boson fusion, and in~\cite{Han:2002gp} using
  production via $gg$ fusion. The analyses were performed at the
  generator level, taking into account the experimental acceptance and
  muon momentum resolution.  The large background from $Z \rightarrow
  \mu\mu$ was reduced by constructing a signal likelihood
  function based on the $\mu\mu$ system kinematics and the event $E_T$
  distribution. It should also be noticed that, since a Higgs signal
  will have been previously observed in higher-rate decay modes, the
  Higgs mass will be known with an accuracy of $\sim$0.1\%. Therefore
  the Z background can be precisely measured in the signal region by
  using a sample of $Z\to ee$ decays.  Due to the small 
  branching ratio, this channel cannot be
  observed to better than $3.5\sigma$ at the LHC design luminosity,
  even combining both production channels.
   Extrapolation to the SLHC gives, for
  the $gg$-fusion channel alone, the
 results of Table~\ref{tab:hmm}. In the 
  mass range 120--140~GeV, a $5\sigma$ evidence or larger
  can be obtained, and
  the square root of the production cross-section times branching ratio (which is
  directly proportional to the muon Yukawa coupling $g_{H\mu\mu}$) could be
  measured with statistical accuracies of better than 10\%. These results are
  comparable to those obtained for a 200~TeV Very Large Hadron Collider (VLHC)
   with an integrated
  luminosity of 300~fb$^{-1}$. The possibility of adding the contribution of
  the gauge-boson fusion production channel  rests on the
  viability of the forward jet tagging, and has so far not been
  explored in detail for the SLHC.
\begin{table}
\begin{center}
\caption{Expected signal significance of a SM $gg\to\ H\to\ \mu\mu$ signal
 for various  mass values, as obtained by combining ATLAS and
 CMS and for an integrated luminosity of 3000~\ifb per
 experiment~\cite{Han:2002gp}. 
  The expected
 statistical accuracy on the measurement of the product of
 cross-section times BR is also given.}
\label{tab:hmm}
\vspace*{0.1cm}  
\begin{tabular}{|l|c|c|}\hline
$m_H$ (GeV) & $S/\sqrt{B}$ &
$\frac{\delta\sigma \times {\mathrm{BR}}(H \rightarrow \mu\mu)}{\sigma \times 
{\mathrm{BR}}}$ \\
\hline 
120~GeV & 7.9 & 0.13\\
130~GeV & 7.1 & 0.14\\
140~GeV & 5.1 & 0.20\\
150~GeV & 2.8 & 0.36 \\
 \hline
\end{tabular}
\end{center}
\end{table}

\subsubsection{Higgs couplings to fermions and bosons}
\label{sec:hcoupl}

Assuming that a SM Higgs boson will have been discovered at the LHC,
measurements of Higgs couplings to fermions and bosons should be
possible~\cite{Zeppenfeld:2000td}, but in most cases the precision
will be limited by the available statistics~\cite{hohlfeld}. A
luminosity upgrade should therefore be useful for this physics.
  
In principle, the Higgs coupling for instance to a given fermion
family $f$ could be obtained from the following relation:
      
\begin{equation}
R(H\to ff)=\int{L dt}\cdot \sigma(pp\to H)\cdot {\Gamma_f \over \Gamma}
\end{equation}
where $R(H\to ff)$ is the Higgs production rate in a given final
state, which can be measured experimentally, $\int {L dt}$ is the
integrated luminosity, $\sigma(pp\to H)$ is the Higgs production
cross-section, and $\Gamma$ and $\Gamma_f$ are the total and partial
Higgs widths respectively.  Hence, a measurement of the Higgs
production rate in a given channel allows extraction of the partial
width for that channel, and therefore of the Higgs coupling $g_{f}$ to
the involved decay particles ($\Gamma_f\sim g_f^2$), provided that the
Higgs production cross-section and the total Higgs width are known
from theory.

Model-independent measurements are only possible if one
considers ratios of couplings, which are experimentally accessible
through the measurements of ratios of rates for two different final
states, because in the ratio the total Higgs cross-section,  width
and luminosity cancel.  Examples are shown in
Fig.~\ref{fig:coupl}. The left plot gives the expected precision on
the ratio of the Higgs widths for the decays into $WW$ and $ZZ$. For
masses larger than approximately 150 GeV a comparison of the $H\to
ZZ\to \ 4\ell$ and $H\to WW\to \ \ell\nu\ell\nu$ rates provides a
direct measurement of $\Gamma_W/\Gamma_Z$. At smaller
masses the process $H\to WW\to \ \ell\nu\ell\nu$ has too low a rate but
one can use the measured rate of $H\to\ \gamma\gamma$ to extract
$\Gamma_W$, at the prize of introducing some theoretical assumptions
(indirect measurement).  The coupling $H\to\ \gamma\gamma$ is
dominated by a loop graph with an intermediate $W$ and hence the rate of
$H\to\ \gamma\gamma$ can be related to the $HWW$ coupling. 
 Similarly,
the right plot in Fig.~\ref{fig:coupl} shows the expected precisions
on the measurements of ratios of Higgs couplings to fermions and
bosons.  The coupling $H\ttbar$ can be probed in the
mass region below 150~GeV by comparing the $WH\to \ 
\ell\nu\gamma\gamma$ rate and the $H\to\ \gamma\gamma$ rate.  The
latter production rate is determined by the coupling of the Higgs to
gluon pairs (since $gg\to \ H$ is the production mechanism) and the
dominant contribution to this coupling is from a top quark loop.  An
indirect measurement of the ratio of couplings $HWW$ and
$Ht\overline{t}$ can therefore be performed. In the mass region above
150~GeV, $\Gamma_W/\Gamma_t$ can be obtained in a similar way by using
the $WH\to \ WWW$ and the $H\to WW$ channels.  
Since in this mass range the dominant systematics is the theoretical
uncertainty on the absolute cross-sections for the two independent
production channels, the higher luminosity leads to a minor
improvement. Progress in the understanding of the theoretical
systematics will however allow to take full benefit of the higher
statistics. 
The processes
$t\overline{t}H\ (\to\ \gamma\gamma)$ and $t\overline{t}H\ (\to\ 
b\overline{b})$ can be combined to give the ratio of widths to $WW$
and to $b\overline{b}$ in an indirect way.
Finally, measurements of the $H\to\ \tau\tau$ and 
$H\to \ WW\to\ell\nu\ell\nu$ rates
in events with tagged forward jets, which arise from the fusion
process $qq\to qqH$, can be combined to directly obtain
$\Gamma_W/\Gamma_{\tau}$. At this time, the impact of the \slhclum\
environment on the combined request of tau identification, missing
\et\ and forward jets has not been evaluated, and we do not have an
estimate of the improvement possible at the SLHC for this channel.
   
It can be seen that at the SLHC ratios of Higgs couplings to fermions
and bosons should be measured with precisions of 10\% or better in
most cases. In some cases, this represents an improvement by up to a
factor of two on the ultimate precision expected at the standard LHC.
Progress for other channels can be anticipated as a result of improved
theoretical understanding of the Higgs production mechanisms, and of
the impact of the experimental environment on the detector
performance.

\begin{figure}
\begin{center}
\includegraphics[width=0.45\textwidth,clip]{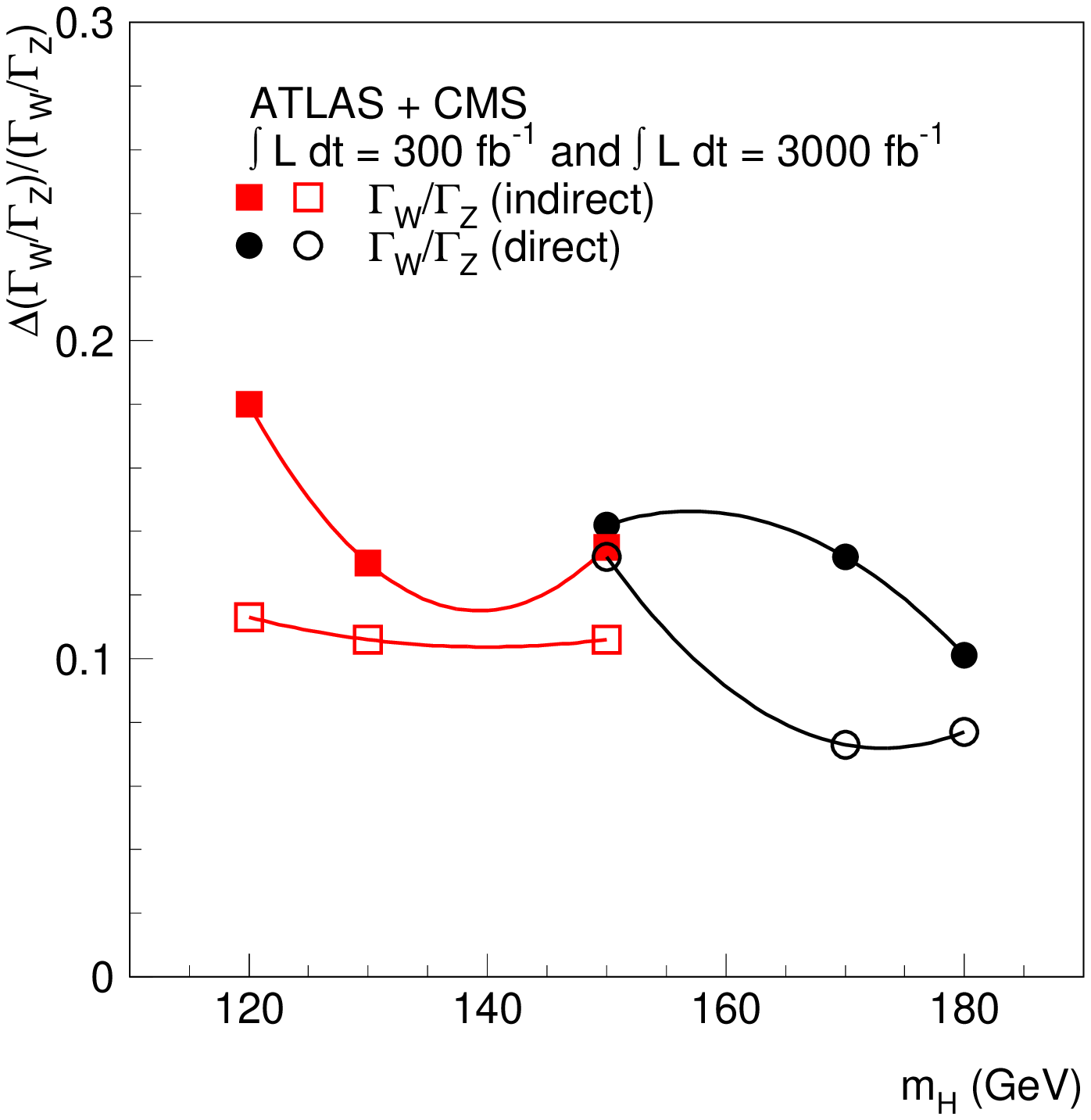} \hfil
\includegraphics[width=0.45\textwidth,clip]{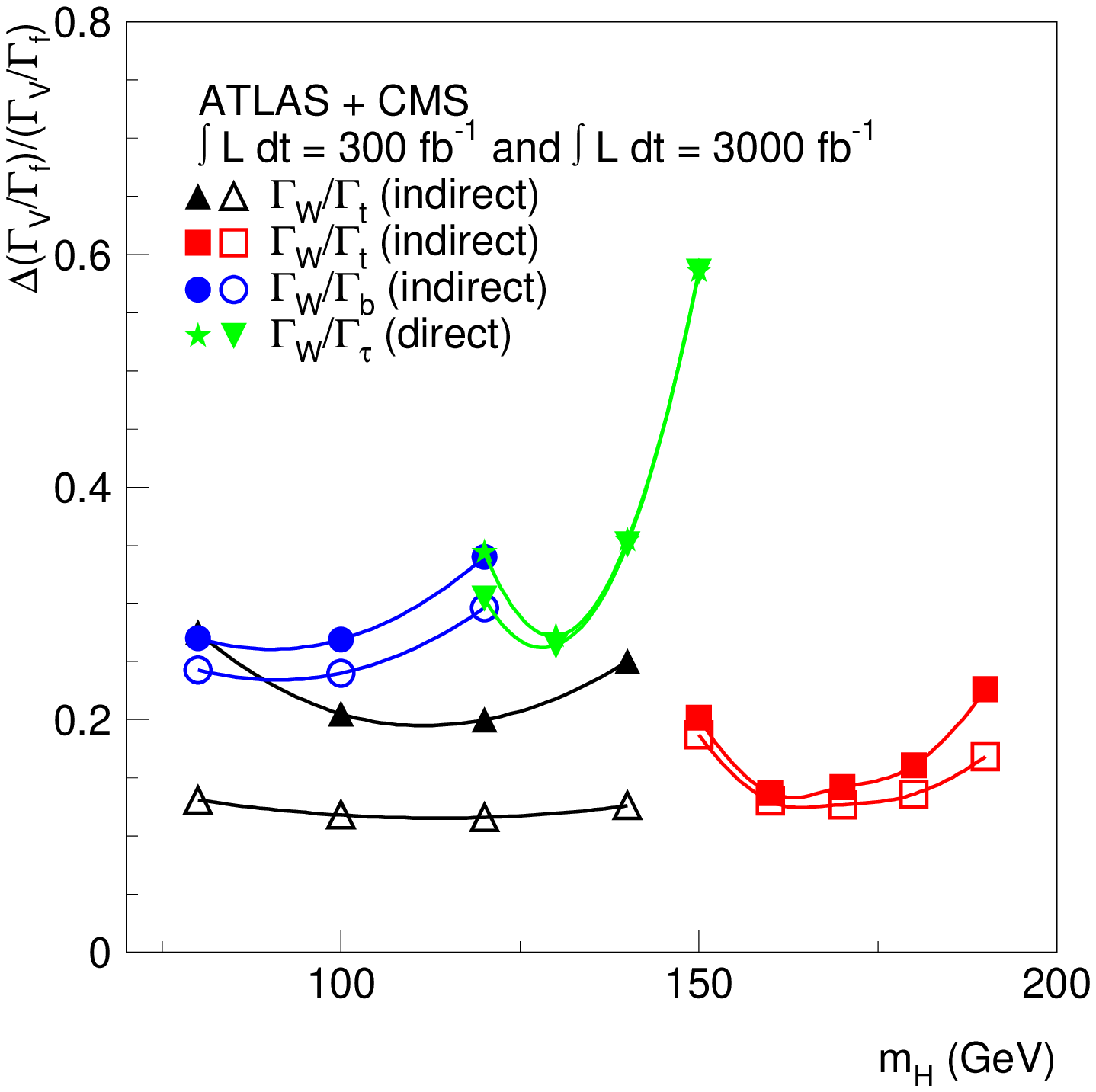} 
\end{center}
\vskip -1.0cm 
\caption{Expected uncertainties on the measured ratios of the 
 Higgs widths to final states involving  bosons only (left) and 
 bosons and fermions (right), 
 as a function of the Higgs mass. Closed symbols:  
 two experiments and 300~\ifb\  per experiment 
 (standard LHC); open symbols: two experiments and 
 3000~\ifb\  per experiment (SLHC). Direct and indirect 
 measurements have been included (see text).\label{fig:coupl}}
\end{figure}

\subsubsection{Higgs self-couplings}
\label{sec:hself}
A complete determination of the parameters of the Standard Model 
requires the measurements of the Higgs selfcouplings. These include a
trilinear and a quartic interaction. In the SM, the corresponding couplings
are fixed at LO in terms of the Higgs mass and vacuum expectation value $v$:
\be
\lambdahhhsm=3\, \frac{m_H^2}{v} \; , \quad \lambda_{\sss HHHH}^{\sss SM}=3 \,
\frac{m_H^2}{v^2} 
\ee
A direct measurement of \lambdahhh\ can be obtained via the detection
of Higgs pair production, where a contribution is expected from the
production of a single off-shell Higgs which decays into
a pair of Higgs. This contribution will always
be accompanied by diagrams where the two Higgs bosons are radiated
independently, with couplings proportional to the Yukawa couplings or
to the gauge couplings. As a result, different production mechanisms
will lead to different sensitivities of the $HH$ rate to the
value of \lambdahhh. The production mechanisms which have been considered in
the literature in the context of hadron-hadron collisions 
include~\cite{Djouadi:1999rc}:
\begin{enumerate}
\item inclusive $HH$ production, dominated by the partonic process
$gg\to HH$~\cite{Glover:1987nx,Dawson:1998py}
\item vector boson fusion~\cite{Dobrovolskaya:1990kx}: $q q \to qq
  V^*V^*$, followed by $V^*V^* \to HH$ (where possible different quark
  flavours are understood in both initial and final state)
\item associated production with $W$ or $Z$ bosons~\cite{Barger:1988jk}: 
$q\bar{q} \to V HH$
\item associated production with top quark pairs: $gg/q\bar{q} \to
  t\bar{t}HH$
\end{enumerate}
In theories beyond the SM, alternative production channels may exist. 
For example, when several Higgs multiplets exist, as is typical of
Supersymmetry, pairs of lighter Higgs bosons can be produced in the
resonant decay of a heavier one. In this document we concentrate on
the SM case.

\begin{figure}
\begin{center}
\includegraphics[width=0.75\textwidth,clip]{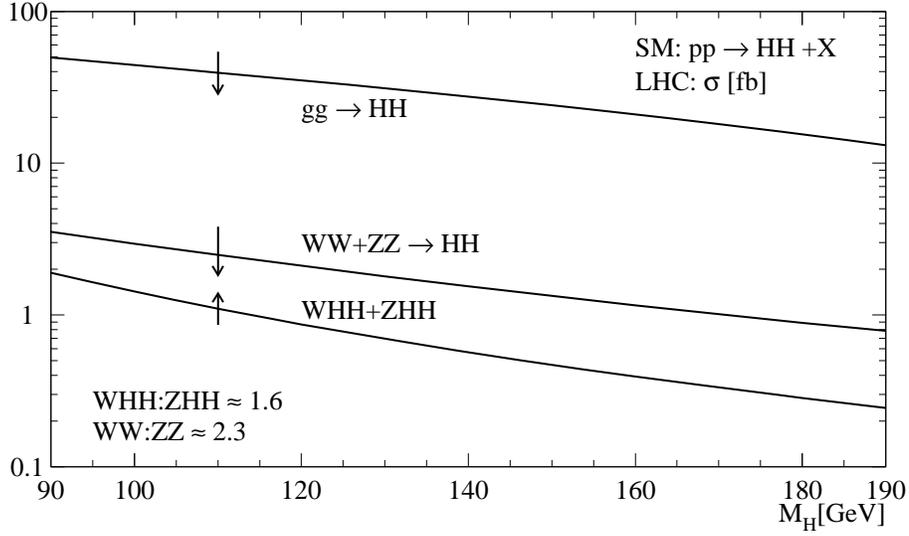} 
\end{center}
\vskip -0.4cm 
\caption{Cross-sections for the production of
  Higgs boson pairs from $gg$ fusion, $WW/ZZ$ fusion and double
  Higgs-strahlung (from~\cite{Djouadi:1999rc}). The vertical arrows
  correspond to a variation of $\lambdahhh$ from 1/2 (arrow's tail) to
  3/2 (arrow's tip) of the SM value.\label{fig:sigHH}}
\end{figure}

The $HH$ production rates are shown in Fig.~\ref{fig:sigHH} for the
first three channels~\cite{Djouadi:1999rc}, and in
Table~\ref{tab:ttHHrate} for the $t\bar{t}HH$ case\footnote{C.G.
  Papadopoulos, unpublished, using results of~\cite{phegas}.}. 
 The arrows indicate the variation in
rate expected when changing the self-coupling in the range
$\lambdahhhsm/2 < \lambdahhh < 3/2\lambdahhhsm$.
\begin{table}
\begin{center}
\caption{Total production cross-section (fb)
  fb for $ pp \to t\bar{t} HH $,
  as a function of \mh\ and of \lambdahhh\ (given in units of the SM
  value $ \lambdahhhsm $).}
\label{tab:ttHHrate}
\vspace*{0.1cm}
\begin{tabular}{|l|cccc|} \hline
\mh(GeV) & $ \sigma(\lambdahhh=1) $ & $ \sigma(\lambdahhh=1/2) $ 
& $ \sigma(\lambdahhh=2) $ & $ \sigma(\lambdahhh=0) $
\\
\hline
120      & 1.0                  & 0.9                  & 1.3 & 0.8
\\
140      & 0.54                 & 0.48                   & 0.73 & 0.42
\\
160      & 0.32                 & 0.32                  & 0.47 & 0.24 \\
\hline
\end{tabular}
\end{center}
\end{table}
 Depending on the value of \mh, different decay channels
dominate. For $\mh\lsim 140\ \gev$ 
$H\to b\bar{b}$ decays dominate, for $ 170 \lsim \mh \lsim 190\ \gev$ 
 $H\to WW$ dominates, and for larger masses
   $WW$ and $ZZ$ final states give the largest rate. 

Given the extremely low production rates, and the potentially large
backgrounds associated to the final states with the largest signal rates,
naive arguments lead to the conclusion
that detection of SM Higgs pairs at \lhclum\ is 
not feasible. As a result, no complete study of the backgrounds is
present in the literature. We present here the first preliminary
results of studies performed specifically for the \slhclum\ option. We
analysed the cases of production via gluon-gluon fusion, vector boson
fusion, and associated production with top quark pairs. In the latter
two cases, we concluded that the large
level of backgrounds make it very hard, if not impossible, to extract
meaningful signals. These findings will be documented in more detail
elsewhere. In the case of the gluon fusion channel, 
we shall now show that the extra factor of ten in luminosity, if accompanied
by a detector performance  comparable to that expected at the
LHC, allows extraction of a signal,and provides the first measurements of
\lambdahhh,  if the Higgs lies within the mass range
$170\ \gev < \mh < 200\ \gev$.
\\[0.3cm]
\noindent
$\mathbf{  gg \to HH}$\\
The $gg$ fusion process has a strong dependence on the value of
\lambdahhh. This is shown  in  
Fig.~\ref{fig:ggHH170}~\cite{Dawson:1998py}.
\begin{figure}
\begin{center}
\includegraphics[width=0.45\textwidth,clip]{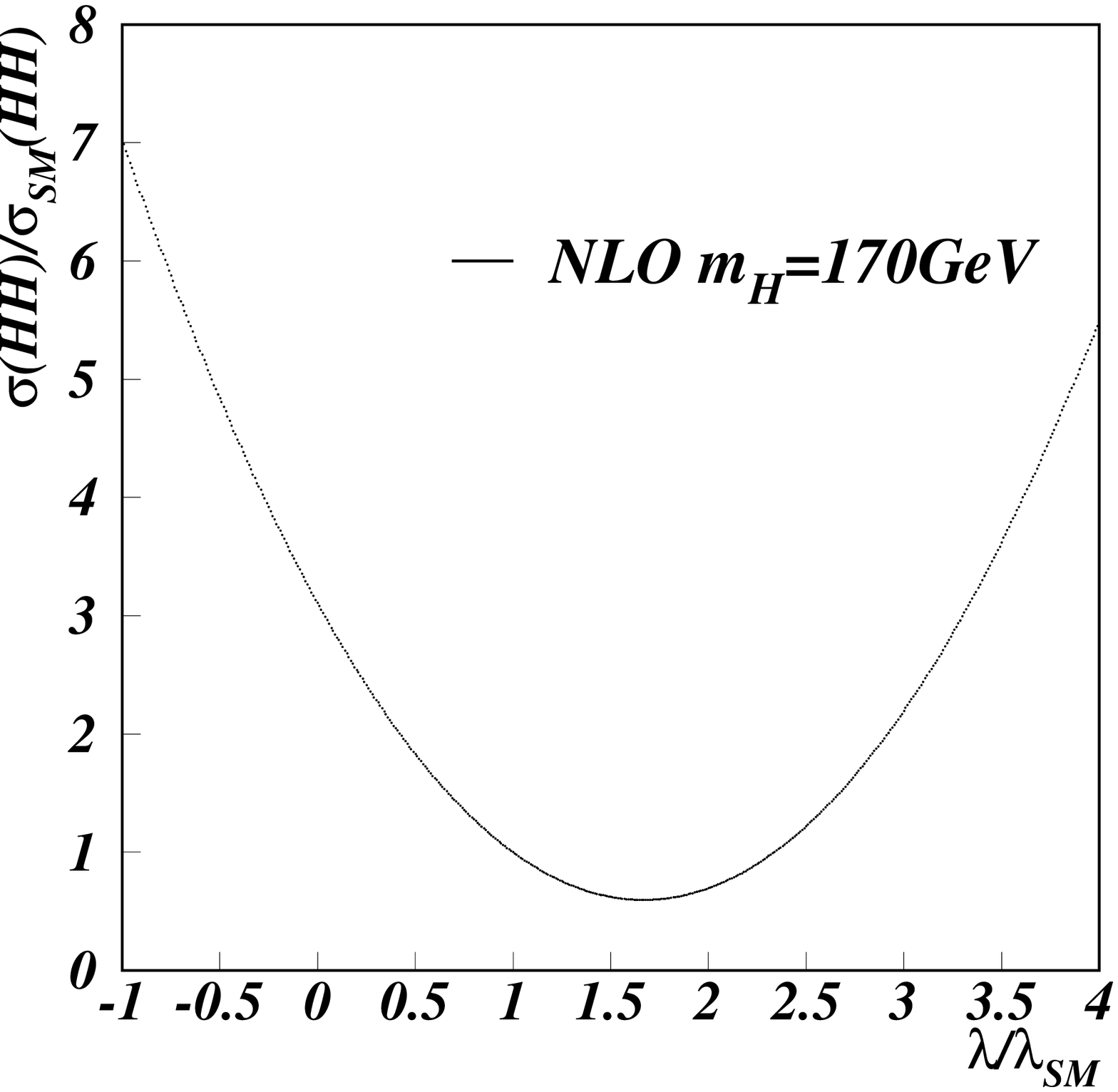} 
\hfil
\includegraphics[width=0.45\textwidth,clip]{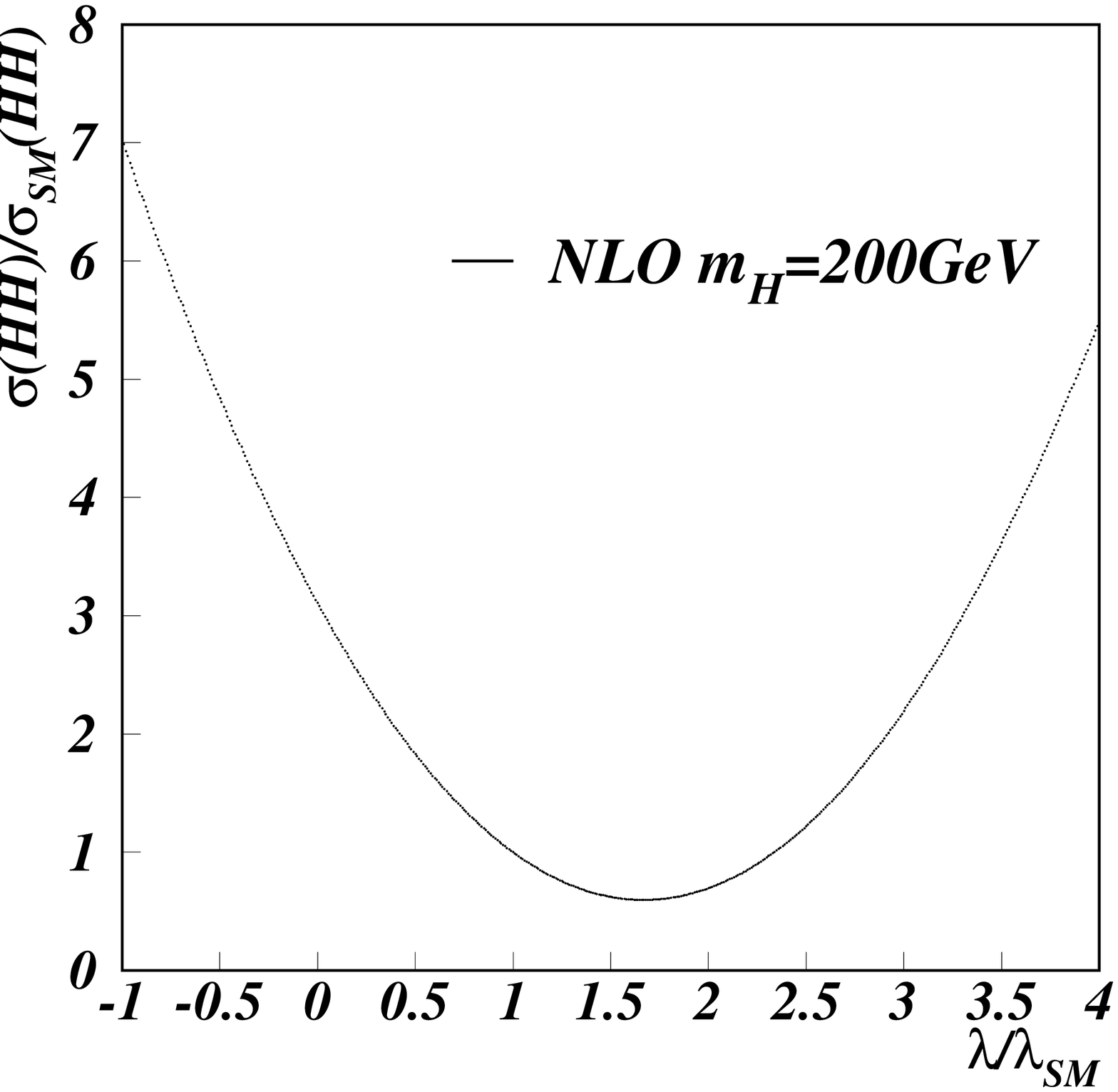} 
\end{center}
\vskip -0.4cm 
\caption{\lambdahhh\ dependence of the $gg \to HH$
  rate~\cite{Dawson:1998py}, for $ \mh=170$~GeV (left) and 
200~GeV (right).\label{fig:ggHH170}}
\end{figure}
 For small \mh,  4$b$ final states dominate. 
 In spite of the double resonance, the QCD bacgkround is
immense and limits any chance to observe a signal. Among all channels
we considered, the most interesting one turned out to be 
\be
gg\to HH \to W^+W^- \, W^+W^- \to \ell^{\pm} \nu jj \ell^{\pm} \nu jj
\ee
which has a good overall BR for  $\mh\gsim 170
$ GeV. The like-sign lepton requirement is essential to reduce the 
high-rate opposite-sign lepton final states from Drell Yan (DY) 
and $t\bar{t}$.
Potential backgrounds to this signal arise from: 
\begin{enumerate}
\item $t\bar{t}$+jets, where the second lepton comes from $b$ decays
\item $WZ $+jets, where one of the  leptons from the $Z$ is not identified
\item $t\bar{t}W$
\item $WWWjj$, including the resonant channel $
  W(H\to WW) jj$
\item $t\bar{t} t\bar{t} $
\end{enumerate}
All these backgrounds are in principle reducible, since they share one
or more of the
following features not present in the signal: presence of $b$-jets;
 presence of additional hard and isolated leptons;
 jet-jet invariant masses not consistent with $ W $ decays.

 To reduce the top-quark backgrounds, all events with one (or more)
 $b$-tags are rejected (this is accomplished by reweigthing each event
 with a factor of 0.5 for each $b$ jet with $\et >30\ \gev $ and $
 \vert \eta \vert < 2.4 $). To select hadronic $ W $ decays we require
 the presence of at least two jet pairs with invariant masses $ 50 <
 m_{jj} < 110$~GeV.  To reduce the contribution of leptons from
 untagged $ b $ decays we apply an isolation cut. We assumed a 10\%
 identification inefficiency for either leptons, to estimate the
 residual DY contamination from non-reconstructed DY pairs.
 In addition, the following cuts are applied:
\be
\pt^{\ell}>\ 20\gev \; , \quad \vert \eta_\ell \vert < 2.4 \; \quad
\forall \; \ell
\ee
\be
\ge 4~\mbox{jets with}~\et>20~\gev~\mbox{and}~\vert \eta \vert <
2.4 \; \mbox{, two of which  with }~  \et>30~\gev
\ee
To reduce the high jet multiplicity 4$t$~processes, we finally ask
that there be at most six jets with $ \et>30 $~GeV in the event. 
Backgrounds 1-3 were
generated with \pythia, and processed through ATLFAST;
  backgrounds 3-5 were evaluated using parton-level simulations
based on exact multi-particle matrix elements,
 following~\cite{Mangano:2001xp}. 
We used the
channel $ t\bar{t}W$, for which simulations were carried out using
both approaches, to cross-check our overall results. We also verified
that a \pythia+ATLFAST simulation of the $ q\bar q' \to WH $
contribution to the $ WWWjj $ background  is consistent with 
a partonic simulation of the  $ q\bar q' \to WH $ subset
of the full set of contributing processes.

 The resulting event numbers 
after all cuts are shown in Table~\ref{tab:ggHH}.
\begin{table}
\begin{center}
\caption{Expected numbers
 of signal and background events after all cuts for the $ gg\to HH\to
  \ 4W \to \ \ell^+\ell '^{+} 4j $ final state, for
  $\int {\cal L}=6000$ \ifb.}
\label{tab:ggHH}
\vspace*{0.1cm}
\begin{tabular}{|l|ccccccc|} 
\hline
\mh & Signal & $ t\bar{t} $ & $ W^{\pm} Z $ & 
 $ W^{\pm}W^+W^- $ & $ t\bar{t} W^{\pm} $ & $ t\bar{t} t\bar{t} $ 
& $S/\sqrt{B}$\\
\hline
170~GeV &   350                &  90              & 60             &
   2400              &  1600            & 30   
& 5.4
                          \\ 
200~GeV & 220               & 90             & 60              &
   1500              & 1600           & 30
& 3.8 \\
\hline                          
\end{tabular}
\end{center}
\end{table}
In spite of the signal being smaller than the backgrounds, the number
of events is large enough 
 to provide a statistical excess of 5.3 (3.8) $\sigma$ 
  for \mh=170 (200)~GeV. 
 It is important to remark that the precise size of the
backgrounds is subject today to large theoretical uncertainties, and
the above significance values should be taken as indicative only.
However, once the data will be available these uncertainties
can be determined experimentally by using control samples
or by other tools. For example, the largest backgrounds ($ t\bar{t}
W^\pm $ and $ W^{\pm} W^+W^- jj $) have a potentially measurable charge
asymmetry. The ratios $ \sigma(XW^+)-\sigma(XW^-)/
\sigma(XW^+)+\sigma(XW^-) $ (with $ X= t\bar{t}, \, W^+W^-jj $)
are quite insensitive to theoretical uncertainties (scales, parton
densities, etc), thereby 
 allowing a determination of the background normalization. 
  Additional handles for an accurate
estimate of the backgrounds come from  counting  events where
some of the cuts are relaxed (e.g. the $b$-veto, the
lepton misidentification). 
We therefore assume that the uncertainty on the background
subtraction will be dominated by statistics. In the case of SM Higgs
production, this leads to a determination of the total production
cross-section with a statistical uncertainty of $ \pm 26\% $ ($ \pm 20\% $ )
for \mh=200~GeV (\mh=170~GeV). This allows a measurement of
\lambdahhh\ with statistical errors of 25\% (19\%).

%
%
%
%
%
%
%
%

\subsubsection{The heavy Higgs bosons of the MSSM}
   The LHC discovery potential for MSSM Higgs bosons decaying into
  SM particles is summarised in
  Fig.~\ref{fig:mssm} in the usual tan$\beta$ vs $m_A$ plane.  
\begin{figure}
\begin{center}
\includegraphics[width=0.6\textwidth,clip]{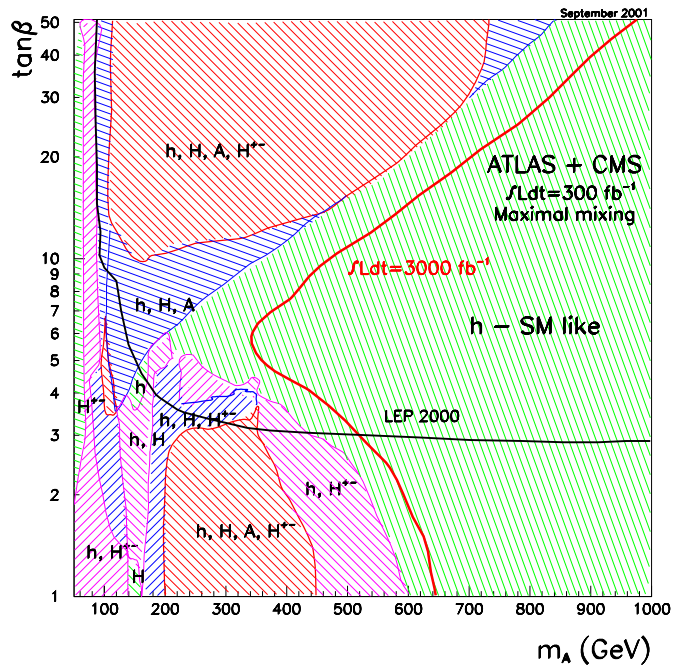} 
\end{center}
\vskip -0.4cm 
\caption{Regions of the MSSM parameter space where the various
  Higgs bosons can be discovered at $\geq5\sigma$ at the LHC 
  (for 300~\ifb\ per experiment and both experiments combined) 
   through their decays into SM
  particles. In the dashed regions at least two Higgs bosons can be
  discovered, whereas in the dotted region only $h$ can be discovered
  at the LHC. In the region to the left of the rightmost contour at
  least two Higgs bosons can be discovered at the SLHC (for
  3000~\ifb\ per experiment and both experiments combined).\label{fig:mssm}}
\end{figure} 
  This plot shows that over a good part of the parameter
 space (note the logaritmic scale) the LHC should be able to discover 
 two or more SUSY Higgs bosons, except in the region at large 
 $m_A$ (the so-called ``decoupling limit"). In this region, only the
 lightest Higgs boson $h$ can be observed, 
  unless the heavier Higgs bosons ($H,\ A,\ H^{\pm}$) have detectable
 decay modes into SUSY particles. This means that the LHC cannot
  promise  a complete and model-independent observation of the heavy
  part of the MSSM Higgs spectrum, although the observation 
  of sparticles (e.g. squarks and gluinos) will clearly indicate 
  that additional Higgs bosons should exist. 
  
  In the region of the decoupling limit, the heavy Higgs bosons are
  not accessible at a future Linear Collider like TESLA either. In
  this case, however, very precise measurements of the $h$ parameters
  should be able to demonstrate indirectly (i.e. through radiative
  corrections), and limited to the region $m_A~<~$500~GeV, the
  existence of heavier Higgs bosons~\cite{TESLA}.
  Figure~\ref{fig:mssm} also shows that the SLHC should be able to
  extend significantly the region over which at least one heavy Higgs
  boson can be discovered at $\geq~5\sigma$ in addition to $h$
  (rightmost contour in the plot), covering in particular almost the
  full part of the parameter space where TESLA should be able to
  constrain (at the 95\% C.L.)  the heavy part of the SUSY Higgs
  spectrum through precise measurements.

\subsection{Strongly-coupled vector boson system}
\label{sec:strongWW}

 If there is no light Higgs boson, then general
arguments~\cite{arguments} imply that
scattering of electroweak gauge bosons at high energy will show
structure beyond that expected in the Standard Model. In order to
explore such signals it is necessary to measure final states 
containing pairs
of gauge bosons with large invariant mass.

\subsubsection{$W_L Z_L \to W_L Z_L$}

Estimates of the production of a $\rho$-like vector resonance of 
$W_LZ_L$ can be obtained
from the Chiral Lagrangian model, with the inverse amplitude method of
unitarization~\cite{Haywood:1999qg,Dobado}. The cross-section depends, in
next to leading order, on a linear
combination $a_4-2a_5$ of two quadrilinear coupling parameters.
The model was implemented in PYTHIA. Only the channel
$W_L Z_L \to W_L Z_L \to \ell \nu \ell^+ \ell^-$ is
considered here, although the resonance can
be produced in the $q \bar q$ fusion channel at higher rate.
Forward jet tagging is here an essential ingredient to reduce the
background.

 The irreducible Standard Model background $ q q \to q q W Z$,
with transverse gauge bosons in the final state, was generated with 
COMPHEP~\cite{COMPHEP} with cuts $p_T (q,W,Z) > 15$ GeV, and
$m_{WZ}~>~500$~GeV, with CTEQ5L structure functions and $Q=m_Z$. 
The process includes electroweak and QCD
diagrams, as well as the quadrilinear gauge boson couplings. The
Higgs mass was set at the low value of 100 GeV, and the signal is
then defined, as in~\cite{Bagger95}, as the enhancement of the
SM prediction over the 100~GeV Higgs. Other backgrounds considered were
$Z b \bar b$ and $Z t \bar t$, also generated with COMPHEP, with
cuts $p_T(b,t)>15$ GeV and $p_T(Z) > 50$ GeV, and 
SM production of $WZ, ~ZZ$, generated with PYTHIA. Table~\ref{tab:WZback}
gives the cross-sections for the different backgrounds.

\begin{table}
\begin{center}
\caption{Cross-sections for backgrounds to the $W_L Z_L \to W_L Z_L$ process.}
\label{tab:WZback}
\vspace*{0.1cm}  
\begin{tabular}{|c|c|c|c|c|}\hline
  Process         &  $\sigma$ (pb) \\ \hline
 $qq \to qqWZ$    &  1.45          \\
 $Z b \bar b$     &  141           \\
 $Z t \bar t$     &  2.23          \\
 $q q \to WZ$     &  3.00          \\
 $q q \to ZZ$     &  0.81          \\  
\hline
\end{tabular}
\end{center}
\end{table}

 The selection criteria are based on leptonic cuts:
\begin{eqnarray}
 p_T(\ell_1) > 150 \ \mbox{GeV}, ~~~~p_T(\ell_2) > 100\ \mbox{GeV}, ~~~~p_T(\ell_3) > 50\  \mbox{GeV} \nonumber \\
 |m(\ell_1\ell_2) - m_Z| < 10 \ \mbox{GeV} \nonumber \\
 E_T^{miss} > 75 \ \mbox{GeV} \nonumber
\end{eqnarray}
and forward jet tagging, i.e. the presence of one forward and one
backward jet ($|\eta| > 2$) with energy greater than 300~(400)~GeV at
LHC (SLHC) luminosity.  In addition, events with jets with transverse
momenta greater than 50~(70)~GeV at LHC (SLHC) luminosity in the
central region ($|\eta| < 2$) were rejected.  The degraded jet tag and
jet veto performances discussed in Section~\ref{sec:jettag} were used
for the SLHC case.
 
 Figure~\ref{fig:WZ} shows the expected signal for a 1.5 TeV
 resonance, corresponding to the choice of the Chiral Lagrangian
 parameters of eqs.~(\ref{eq:Leff4})-(\ref{eq:Leff5})
$\alpha_4-2\alpha_5=0.006$, at the LHC and at the SLHC.
 The resonance is at the limit of the observation at LHC, with
 6.6~events expected over a background of about 2.2~events around the
 region of the peak, but at the SLHC the signal has a significance of
 $S/\sqrt{B} \sim 10$.

\begin{figure}
\begin{center}
\includegraphics[width=0.4\textwidth,clip]{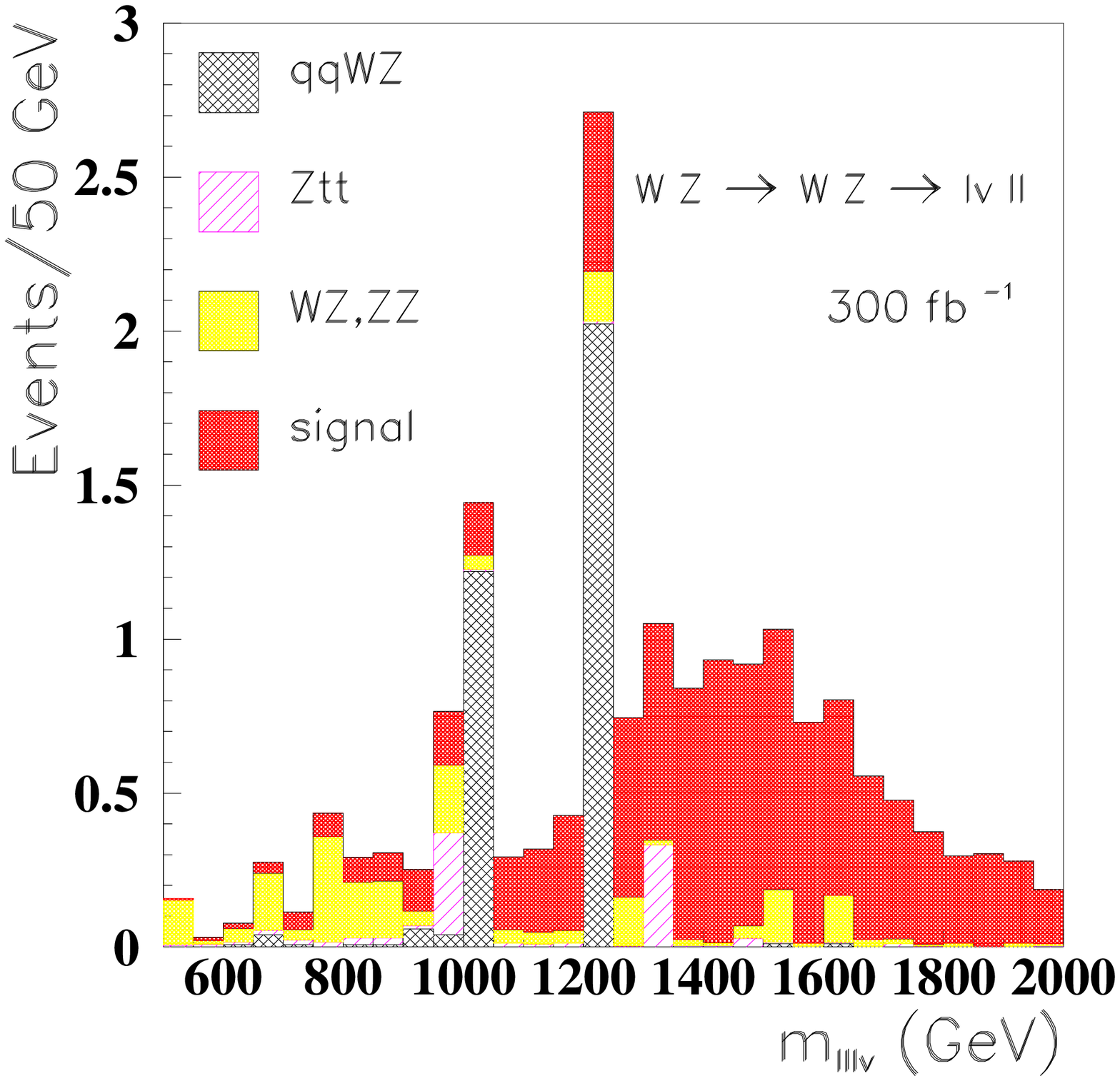}
\includegraphics[width=0.4\textwidth,clip]{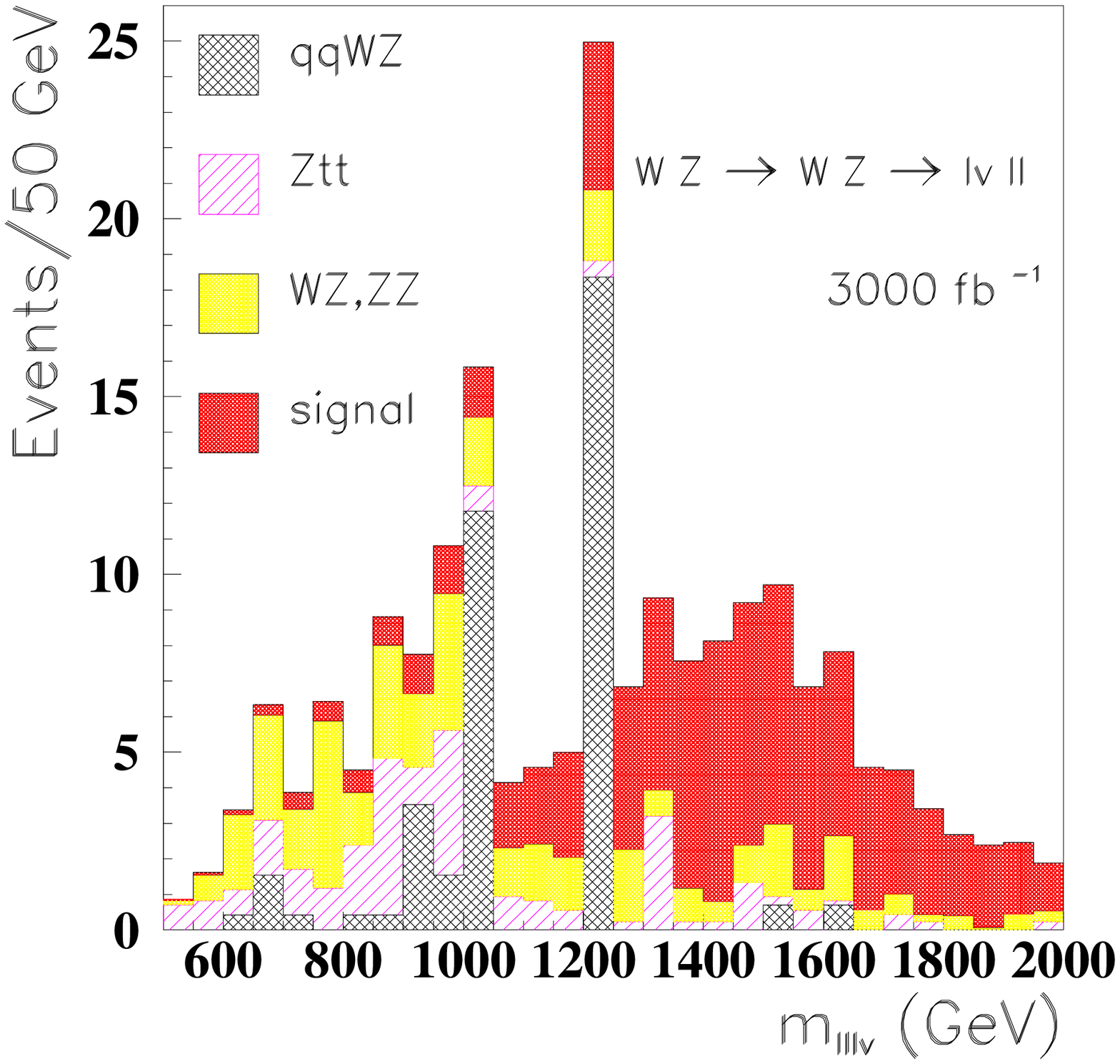}  
\end{center}
\vskip -0.4cm 
\caption{Expected signal and background for a 1.5 TeV WZ resonance 
in the leptonic decay channel for
300 fb$^{-1}$ (left) and 3000 fb$^{-1}$ (right).
\label{fig:WZ}}
\end{figure}

\subsubsection{$Z_L Z_L$ scalar resonance}

 Production of a scalar $Z_LZ_L$ resonance decaying via the gold-plated
channel $ZZ \to \ 4 \ell$
is a rare process, well suited to the SLHC. As in the case of  the $WZ$ 
resonance discussed above, the Chiral Lagrangian
model with regularization by the inverse amplitude method
was used.  The cross-section depends on a
linear combination, $7\alpha_4+11\alpha_5$, of the same parameters as those
of the vector resonance, and therefore observation of the scalar resonance
will resolve  $\alpha_4$ and $\alpha_5$ unambiguously.

 Production occurs through the scattering processes
$W^+_LW^-_L \to Z_LZ_L$ and $Z_LZ_L \to Z_LZ_L$. Standard Model
backgrounds leading to $qqZZ$ in the final state have been generated
with COMPHEP, with cuts $p_T (q,Z) > 15$ GeV, 
$m_{ZZ} > 500$ GeV, $m_{qq}>200$ GeV, with CTEQ5L  structure
 functions and $Q=m_Z$. The process was
implemented in PYTHIA as an external process. 
 The Higgs mass was set at 100~GeV, so that the contribution from
longitudinal vector boson scattering was negligible in this background. 
With these conditions, the SM cross-section is 69.4~fb. Other backgrounds
considered were $qq \to ZZ$, with cut $m_{ZZ} > 500$ GeV
and cross-section 8.66~fb. The background $gg \to ZZ$ was not included,
but is expected to contribute about one third of the $q\bar q$ fusion
process~\cite{Dobado}.

 The analysis requires the presence of four isolated leptons with transverse
momenta greater than 30 GeV, and with two-lepton invariant masses
compatible with coming from $Z$ bosons. 
  Forward jet tagging was applied by requiring the presence of one
forward and one backward jet  with energies greater than 
400 GeV. No central jet veto was imposed, as this is not needed
to reject the main backgrounds.  
 The expected signal and background for a resonance 
 of mass 750 GeV, corresponding to $7\alpha_4+11\alpha_5 = 0.063$, is shown 
in Fig.~\ref{fig:ZZ} for an integrated luminosity of 3000~fb$^{-1}$.
 Such a process would not be observable at the nominal LHC.

\begin{figure}
\begin{center}
\includegraphics[width=0.65\textwidth,clip]{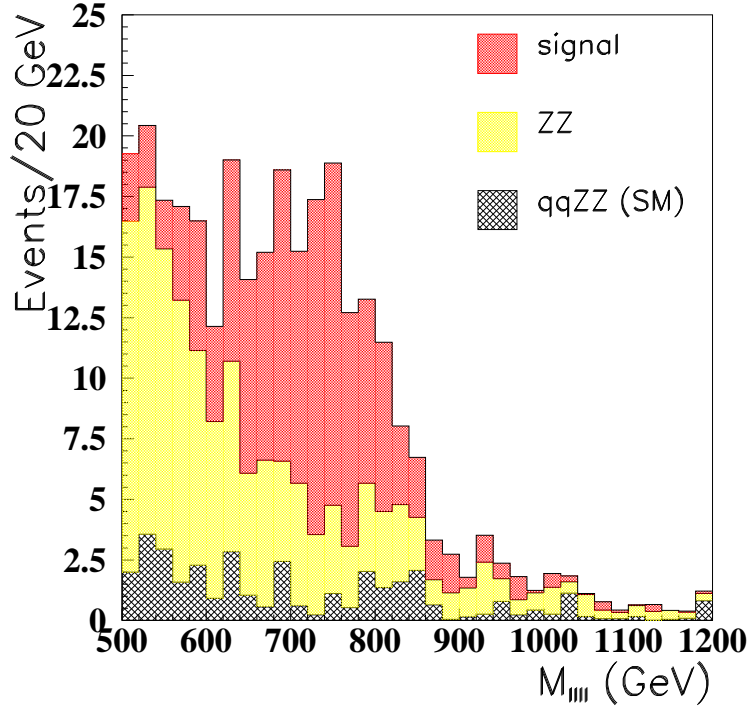} 
\end{center}
\vskip -0.4cm 
\caption{Expected signal and background at the SLHC (3000~fb$^{-1}$) 
 for a scalar resonance of mass 750~GeV decaying into four leptons.\label{fig:ZZ}}
\end{figure}

\subsubsection{$W^+_LW^+_L$}

Non-resonant production of vector boson pairs at high mass represents a
challenge at LHC because the background needs to be very
well understood.

 The production of $W^+W^+$ pairs has no contribution from $q\overline{q}$ fusion.
 Two models have been considered:
 
\begin{itemize}
\item A Higgs boson of mass 1 TeV (as a reference point).
\item WW production with K-matrix unitarization.
\end{itemize}

 Backgrounds arise from gauge boson pairs produced by electroweak and gluon
exchange diagrams~\cite{ATLAS-tdr}. Here only 
the most promising $\ell^+\ell^+\nu\nu$ final
state, arising from same-sign $W^+W^+$ production, is discussed. 
  Selection cuts were applied on the transversa momenta of both leptons,
  on their azimuthal angle and on the event missing transverse 
  energy~\cite{ATLAS-tdr}.  
  Two forward tagging jets with energies above 400 (600)~GeV 
 were required for the LHC (SLHC) scenario, and  events  
 were rejected if there was a
 central jet with transverse momentum greater than 40~(60)~GeV. 

  Table~\ref{tab:wwstrong} shows the number of signal and background
 events expected,  and the signal significance, for the LHC, the SLHC and
  (for comparison) for a machine with $\sqrt{s}=$28~TeV.

\begin{table}
\begin{center}
\caption{Expected numbers of reconstructed events above an invariant
 mass of 600 GeV (for $\sqrt{s}$=14~TeV) 
 and 800 GeV (for $\sqrt{s}$=28~TeV) for models with a strongly-coupled
  Higgs sector and for the background.
  The significance was computed as $S/\sqrt{S+B}$.}
\label{tab:wwstrong}  
\vspace*{0.1cm}  
\begin{tabular}{|c|c|c|c|c|}\hline
 & 300 fb$^{-1}$ & 3000 fb$^{-1}$
 & 300 fb$^{-1}$ & 3000 fb$^{-1}$\cr
 
Model &14 TeV&14 TeV& 28 TeV&
28 TeV\cr\hline
Background& 7.9& 44& 20& 180\cr
\hline
K-matrix Unitarization &14& 87& 57& 490\cr
Significance& 3.0& 7.6& 6.5& 18.9\cr
\hline
Higgs, 1 TeV & 7.2& 42& 18& 147\cr
Significance& 1.8& 4.5& 2.9& 8.1\cr\hline
\end{tabular}
\end{center}
\end{table}
  In spite of the increase 
in statistical significance at the SLHC compared
to the standard LHC, it should be noticed that extraction of a convincing 
signal in this channel will not be easy, because the shapes of the background 
and of the signal are similar.

\subsection{Top-quark physics}
\label{sec:top}
 Given the large top quark cross-section, most of the top
physics programme should be completed during the first few years
of LHC operation~\cite{Beneke:2000hk}. 
In particular, the $t\bar{t}$ and the single-top
production cross-sections should be measured more precisely
 than the expected theoretical uncertainties, and the determination
of the top mass should reach an uncertainty (dominated by systematics) of
$\sim$~1~GeV, beyond which more data offer no obvious improvement. 

There is however one issue in top physics, namely rare decays, that 
the LHC can only address with limited statistics. 
  While most of the rare decays expected in the SM are  
  beyond any possible reach, there is a large class of
theories beyond the SM where branching fractions for decays of
top quarks induced by flavour-changing
neutral currents (FCNC)  be as large as $ 10^{-5}-10^{-6}$. Studies
documented in~\cite{Beneke:2000hk} indicate that the data 
which can be collected with a luminosity of \lhclum\ are not
sufficient to explore these models. 

Three possible  FCNC decays have been investigated:
\begin{eqnarray}
 && t \to q \, \gamma, \quad  q=u \,\, {\rm or } \,\, c \\
 && t \to q \, Z,      \quad  q=u \,\, {\rm or } \,\, c \\
 && t \to q \, g,      \quad  q=u \,\, {\rm or } \,\, c
\end{eqnarray}
For each channel the number of signal events was evaluated for the
reference value of
\[ {\rm BR_{def}}(t \to (u \, + \, c) \, V) = 1.0 \times 10^{-3} \; ,
\quad V = \gamma, Z, g \]
The ``reachable'' branching ratio for $t \to q V$ decay was estimated 
as follows~\cite{bityukov}:
\begin{eqnarray*}
\frac{S}{\sqrt{S+B}+\sqrt{B}} \geq \frac{3}{2}\sigma , \quad ({\rm CL} = 99\%)
\end{eqnarray*}
where $S$ and $B$ stand for the numbers of signal and background events, respectively.
The considered background processes  include:
\begin{itemize}
\item  $t \bar t$ ($\sigma=830$~pb) 
\item  $ W(\to \ e,\mu) + $~jets
 ($\sigma\sim 7500$~pb for $p_{T,W}>20$~GeV )
\item  $WW+WZ+ZZ$ (\d \sigma=110\d~pb)
\item  $W \, \gamma$ (\d \sigma=17.3\d~pb)
\item  single top (generated with {\sf
    TopRex~\cite{Slabospitsky:2002ag}}) (\d \sigma=240\d~pb)
\end{itemize}
All $b$-tagged jets should have $|\eta|<2.5$.  We considered 
three cases for the $b$-tagging efficiency:
\begin{itemize}

\item An ideal case, where jets from $b$, $c$ and light quarks are 
 identified and distinguished with 100\% efficiency. 

\item A realistic case, based on a CMS simulation valid for a
 luminosity of \lhclum\, where the $b$-tagging  
 efficiency is $\epsilon_b\approx $ 60\%, the mistagging probability
 for $c$-jets $\approx $ 10\%, and the mistagging probability 
 for light-quark and gluon jets $\approx (1 - 2)$~\%. 

\item A pessimistic case, in which only semileptonic muon decays of the 
  $b$-quarks can be used. 
   In particular, we require  $p_{T}(\mu) \geq 20$~GeV for
  non-isolated muons carrying at least 60\% of the jet energy, and
  having a \pt\ relative to the jet axis larger than 700~MeV. This
  algorithm leads to a b-tagging efficiency \d \epsilon_{b}=6.4\% \d.
  
\end{itemize}

\begin{table} 
\begin{center}
  \caption{\label{tab:tqgamma} For $t \to q \gamma$ decays, the 
   achievable branching ratio (in units of $10^{-5}$) at the LHC and SLHC
    for different $b$-tagging hypotheses (see text).}
\vspace*{0.1cm} 
\begin{tabular}{|l||c|c|c|} \hline
$b$-tagging  & ideal   
                   & real. & $\mu$-tag \\ \hline
600~\ifb &  $0.48$  
                   & $0.88$  & $3.76$ \\ \hline
6000~\ifb &  $0.14$ 
                   & $0.26$  & $0.97$ \\ \hline
\end{tabular}
\end{center}
\end{table} 
\begin{table} 
\begin{center}
  \caption{\label{tab:tqg} For $t \to q g$ decays, 
the achievable branching ratio (in units of $10^{-5}$) at the LHC and SLHC
 for different $b$-tagging hypotheses (see text).}
\vspace*{0.1cm} 
\begin{tabular}{|l||c|c|c|} \hline
$b$-tagging  & ideal   
                   & real. & $\mu$-tag \\ \hline
600~\ifb & $22.3$  
                   & $60.8$  & $210.$ \\ \hline
6000~\ifb & $7.04$  
                   & $19.2$  & $66.2$ \\ \hline
\end{tabular}
\end{center}
\end{table} 
\begin{table} 
\begin{center}
  \caption{\label{tab:tqz} For $t \to q Z$ decays, 
the achievable branching ratio (in units of $10^{-5}$) at the LHC and SLHC
 for different $b$-tagging hypotheses (see text).}
\vspace*{0.1cm} 
\begin{tabular}{|l||c|c|c|} \hline
$b$-tagging & ideal 
                & real. & $\mu$-tag \\ \hline
600~\ifb & $0.46$ 
                 & $1.1$ & $83.3$  \\ \hline
6000~\ifb& $0.05$ 
                 & $0.11$ & $8.3$  \\ \hline
\end{tabular}
\end{center}
\end{table} 
\noindent 
$\mathbf{t \to q \, \gamma}$
\\
 We consider $(\gamma + \ell^{\pm} + \geq 2$~jets) final states, with
 the following cuts: 
 
\begin{itemize}
\item One isolated photon, with
 $E_{T} \geq 75$~GeV and $|\eta_{\gamma}| \leq 2.5$. 
\item One isolated lepton, with
 $p_{T} \geq 20$~GeV and $|\eta_{\ell}| \leq 2.5$. 
\item two or more jets with $E_{T \, j} \geq 30$~GeV and $|\eta_{j}|
  \leq 4.5$. No third jet with $E_{T \, j} \geq 50$~GeV. One of the
  jets should be $b$-tagged. 
\item One pairing of the jets, the $\gamma$ and the reconstructed
  semileptonic top such that:
\[ |M(j_1+\gamma)-m_t| \leq 15\ \gev\; \quad 
   |M({\mathrm b-jet}+W)-m_t| \leq 25\ \gev
\]
\end{itemize}

The results are given in Table~\ref{tab:tqgamma}.

\noindent $\mathbf{ t \to q \, g}$ \\

We consider $(\ell^{\pm} + \geq 3$~jets) final states, with the following cuts:
\begin{itemize}
\item One isolated lepton, with
 $p_{T} \geq 20$~GeV and  $|\eta_{\ell}| \leq 2.5$.
\item $N_{{\rm jets}} = 3$, with 
 $E_{T \, j} \geq 50$~GeV and $|\eta_{j}| \leq 4.5$. Only   
 one of the three jets should be $b$-tagged. 
\item At least one combination such that 
 $|M({\mathrm b-jet}\,+\,W)-m_t| \leq 25$~GeV and $|M(j_1 + j_2)-m_t| \leq 25$~GeV. 
\end{itemize}

The results are given in Table~\ref{tab:tqg}.

\noindent $\mathbf {t \to q \, Z}$ \\
We consider $(3\ \ell^{\pm} + \geq 2$~jets) final states, with the following cuts:

\begin{itemize}
\item Three isolated leptons with 
 $p_{T} \geq 20$~GeV and $|\eta_{\ell}| \leq 2.5$. 
\item $N_{{\rm jets}} \geq 2$, with
 $E_{T \, j} \geq 50$~GeV and $|\eta_{j}| \leq 4.5$. Only one of the jets
 should be $b$-tagged.
\item At least one combination with 
$|M({\mathrm b-jet}\,+\,W)-m_t| \leq 25$~GeV and
$|M(Z + j)-m_t| \leq 25$~GeV. 
\end{itemize}
The results are given in Table~\ref{tab:tqz}.

With a detector performance comparable to that expected
at \lhclum, the SLHC should enhance by a large factor the sensitivity
to FCNC top decays. In the case of $t\to Zq$, in particular, the
improvement is almost linear with the luminosity, thanks to the very
low background level. Branching ratios of order $10^{-6}$ are
achievable, which are of interest for some theories beyond the
Standard Model, as discussed in~\cite{Beneke:2000hk}. Loss of the
ability to tag $b$-quarks with a secondary vertex technique would
however downgrade the sensitivity to such a level that no gain could
be obtained from SLHC data compared to the standard LHC.

\subsection{Supersymmetry}
\label{sec:susy}

 If Supersymmetry is connected to the hierarchy problem, it is
 expected  that sparticles will be sufficiently light
 that at least some of them will be observed at the  LHC. 
  However it is not possible to set a rigorous bound on the sparticle
 masses, and it may well be that the heaviest part of the SUSY spectrum 
 (usually squarks and gluinos) is missed at the standard LHC. 
  
  The LHC discovery potential for squarks and gluinos, in several
  energy and luminosity 
 scenarios, is summarised in Fig.~\ref{fig:susy}. The various
 contours were derived within the
 framework of minimal Supergravity models (mSUGRA), and are shown as a function of the   
 universal scalar mass $m_0$ and of the universal gaugino mass $m_{1/2}$.
   They were obtained by looking for events with many high-\pt\ jets
  and large missing transverse energy. This is the most typical and most 
  model-independent signature for SUSY if R-parity is conserved.

\begin{figure}
\begin{center}
\includegraphics[width=0.65\textwidth,clip]{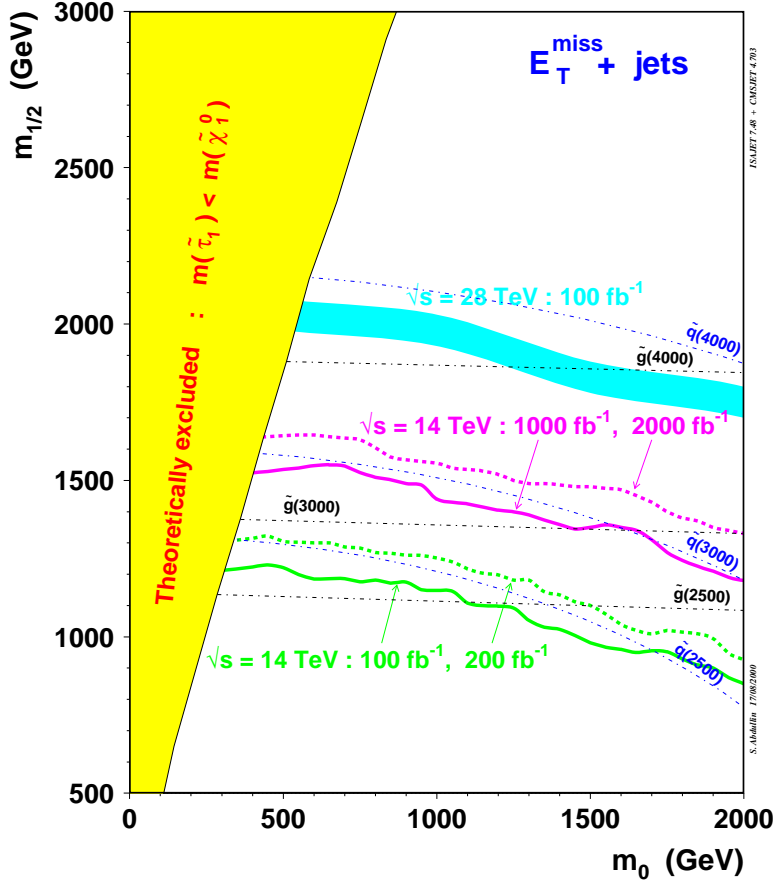} 
\end{center}
\vskip -0.4cm 
\caption{Expected $5\sigma$ discovery contours in the mSUGRA plane $m_0$ versus
 $m_{1/2}$ for $A_0=0$, tan$\beta$=10 and $\mu<0$. The various curves show
 the potential of the CMS experiment at the standard LHC (for luminosities of
 100~\ifb and 200~\ifb), at the SLHC 
 (for 1000~\ifb and 2000~\ifb), and (for comparison) at a machine with a 
 centre-of-mass energy of 28~TeV.\label{fig:susy}}
\end{figure}
 
It can be seen that a luminosity upgrade would extend the mass reach
for squarks and gluinos from about 2.5~TeV (standard LHC) to about
3~TeV (SLHC).  This performance does not require major detector
upgrades because these inclusive searches are based mainly on calorimetric
measurements of high-\pt\  jets and large missing transverse energy.
On the other hand, reconstruction of more exclusive decay chains which
may be rate-limited at the LHC, such as some cascade decays of
heavy gauginos, could become possible at the SLHC provided the full
detector functionality is preserved.  To illustrate this case, two
points (here called K and H) of the mSUGRA parameter space have been
studied in some detail~\cite{frank_ian}. 
 These points were taken from a
recently-proposed set of benchmark points~\cite{Battaglia}, that
satisfy existing bounds including dark matter constraints and results
from direct searches at LEP.  Three of these points (F, H, K), where
the squark and gluino masses exceed 2~TeV, might benefit from a
luminosity upgrade~\footnote{Point~M is beyond the sensitivity even of
  a SLHC, given that the squark and gluino masses are above 3~TeV.},
because the expected event rates are small at the LHC and
therefore detailed SUSY studies will not be possible.
  
Point K has gluino and squark masses slightly above
2~TeV. Squark pair production dominates, and is followed by the decays
$\tilde{q}_L \to \tilde{\chi}_1^\pm q, \tilde{\chi}_2^0 q$ and $\tilde
q_R \to \lsp q$.  The signal can inclusively be observed on top of the
background by using for instance the distribution of the effective
mass, defined as
  $$
  M_{eff} = E_T^{miss} + \sum_{jets}E_{T,jet}
  +\sum_{leptons}E_{T,lepton} $$
  where the sum runs over all jets with
  $E_T> 50 $ GeV and $\abs{\eta}<5.0$ and isolated leptons with $E_T>
  15 $ GeV and $\abs{\eta}<2.5$.  Events were selected with at least
  two jets with $p_T > 0.1\Meff$, $E_T^{miss} > 0.3\Meff$,
  $\Delta\phi(j_1, E_T^{mis}) < \pi-0.2$, and
  $\Delta\phi(j_1,j_2)<2\pi/3$.  The distributions in $ M_{eff}$ for
  signal and background are shown in Fig.~\ref{fig:pointk}.  The
  signal emerges from the background at large values of $M_{eff}$.
  For an integrated luminosity of 3000~fb$^{-1}$, a signal of 500
  events should be observed on top of a background of 100 events for
  $\Meff>4000~$GeV.  These rates are sufficiently large that a
  discovery could be made already at the LHC.

\begin{figure}
\begin{center}
\includegraphics[width=0.4\textwidth,clip]{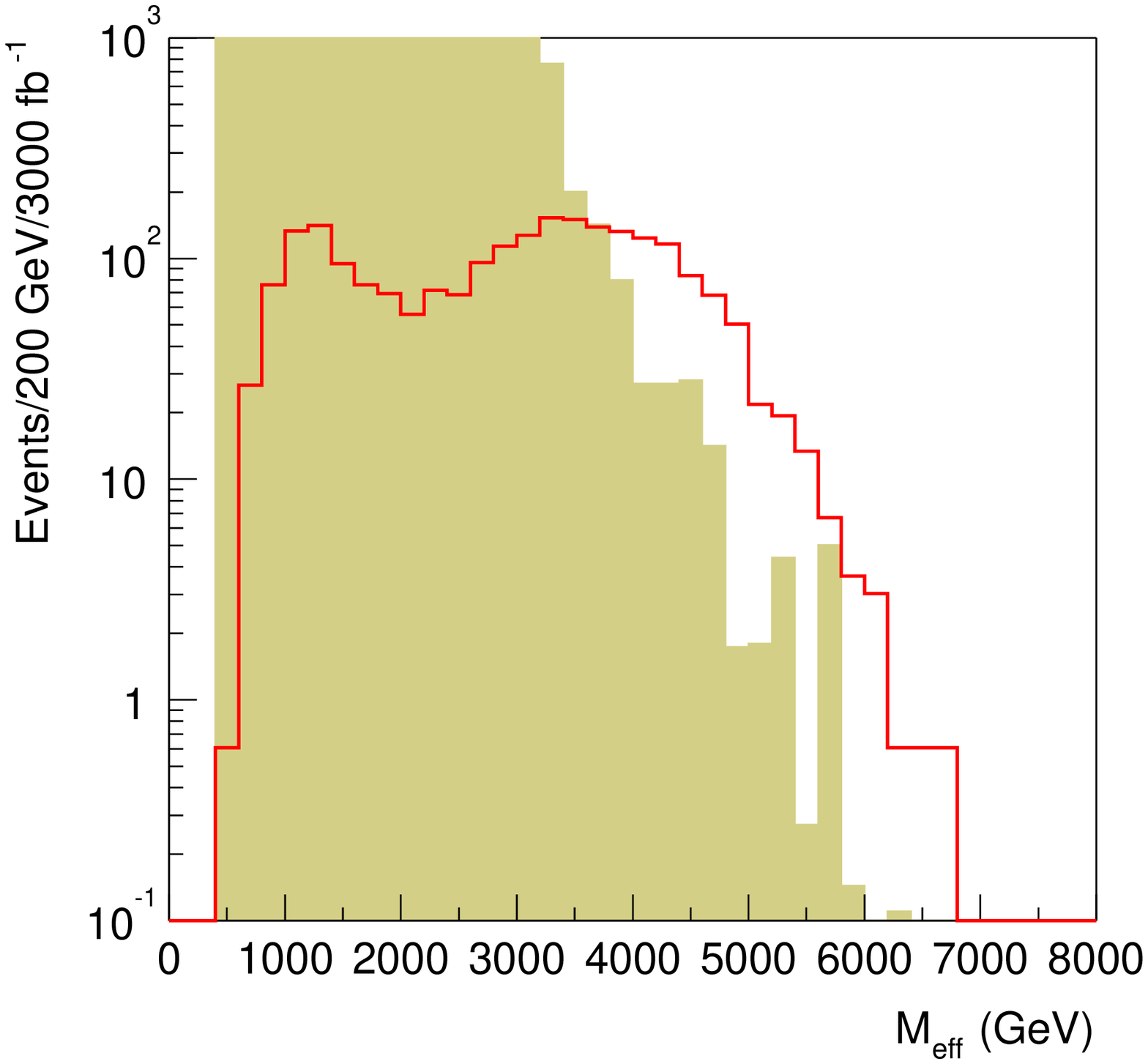} 
\includegraphics[width=0.4\textwidth,clip]{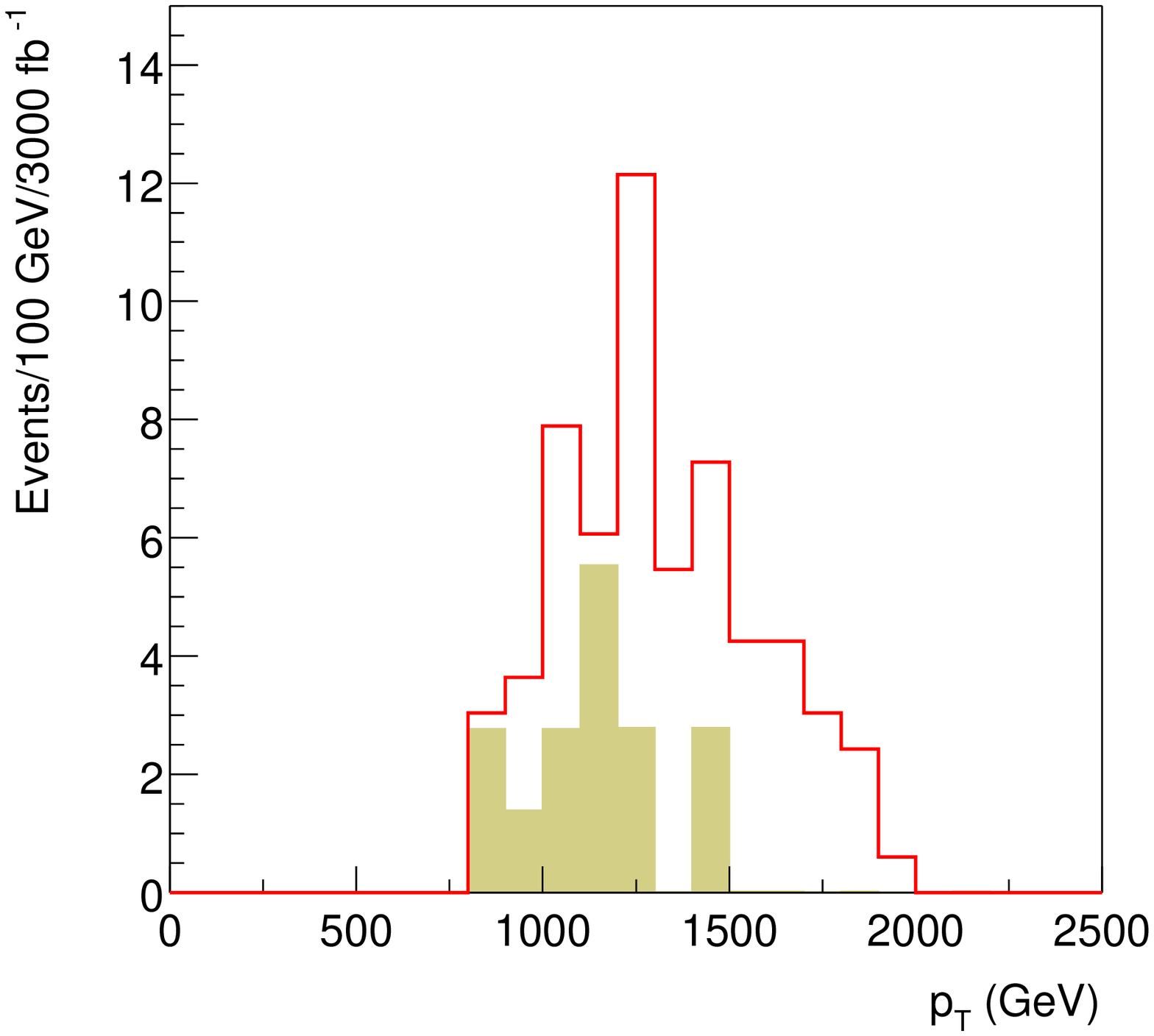}
\end{center}
\vskip -0.4cm 
\caption{For Point~K, the distributions of the effective mass (left) and of
 the $p_T$ of the hardest jet from a sample of
 2~jets + $E_T^{miss}$ events (right).   
 Solid: signal. Shaded: SM background.
\label{fig:pointk}}
\end{figure}

Production of $\tq_R\tq_R$ followed by the decay of each squark to
$q\lsp$ gives a di-jet signal accompanied by missing $E_T$. 
 In order to extract this signature from the Standard Model  background, 
hard cuts on the jets and $E_T^{miss}$ are needed.
 Events were required
to have two jets with $p_T>700$~GeV, $E_T^{miss}>$~600~GeV, and
$\Delta\phi(j_1,j_2)<0.8$. The resulting distribution of the
$p_T$ of the hardest jet is shown in
Fig.~\ref{fig:pointk}. Only a few events survive  with
3000~fb$^{-1}$, hence this exclusive channel is not observable
at the standard LHC. The transverse momentum of the hardest jet is 
 sensitive to the
$\tq_R$ mass~\cite{ATLAS-tdr}. The mass determination will be
limited by the available statistics.

 Since the decay $\tchi_2^0 \to \lsp h$ is dominant at Point~K, 
  Higgs particles
should be found  in the decay  $\tq_L\to \tchi_2^0 q$, followed by
$ \tchi_2^0 \to
\lsp h$. The Higgs signal should be observed as a peak in the
$b\overline{b}$ invariant mass distribution. It is therefore
essential that $b-$jets can be tagged with good efficiency and
excellent rejection against light-quark jets. 
 There is a large
background from $t \bar t$ production that must be overcome using topological cuts.
 Events were selected to have at least three
jets with $p_T > 600,300,100$~GeV, $E_T^{miss}>400$~GeV,
$\Meff>2500$~GeV, $\Delta\phi(j_1, E_T^{miss})<0.9$, and
$\Delta\phi(j_1,j_2)<0.6$. The $b\overline{b}$ invariant mass 
distribution is shown in
Fig.~\ref{fig:pointkh} assuming a $b$-tagging efficiency of 60\% 
and a rejection of $\sim 100$ against light-quark jets.

\begin{figure}
\begin{center}
\includegraphics[width=0.4\textwidth,clip]{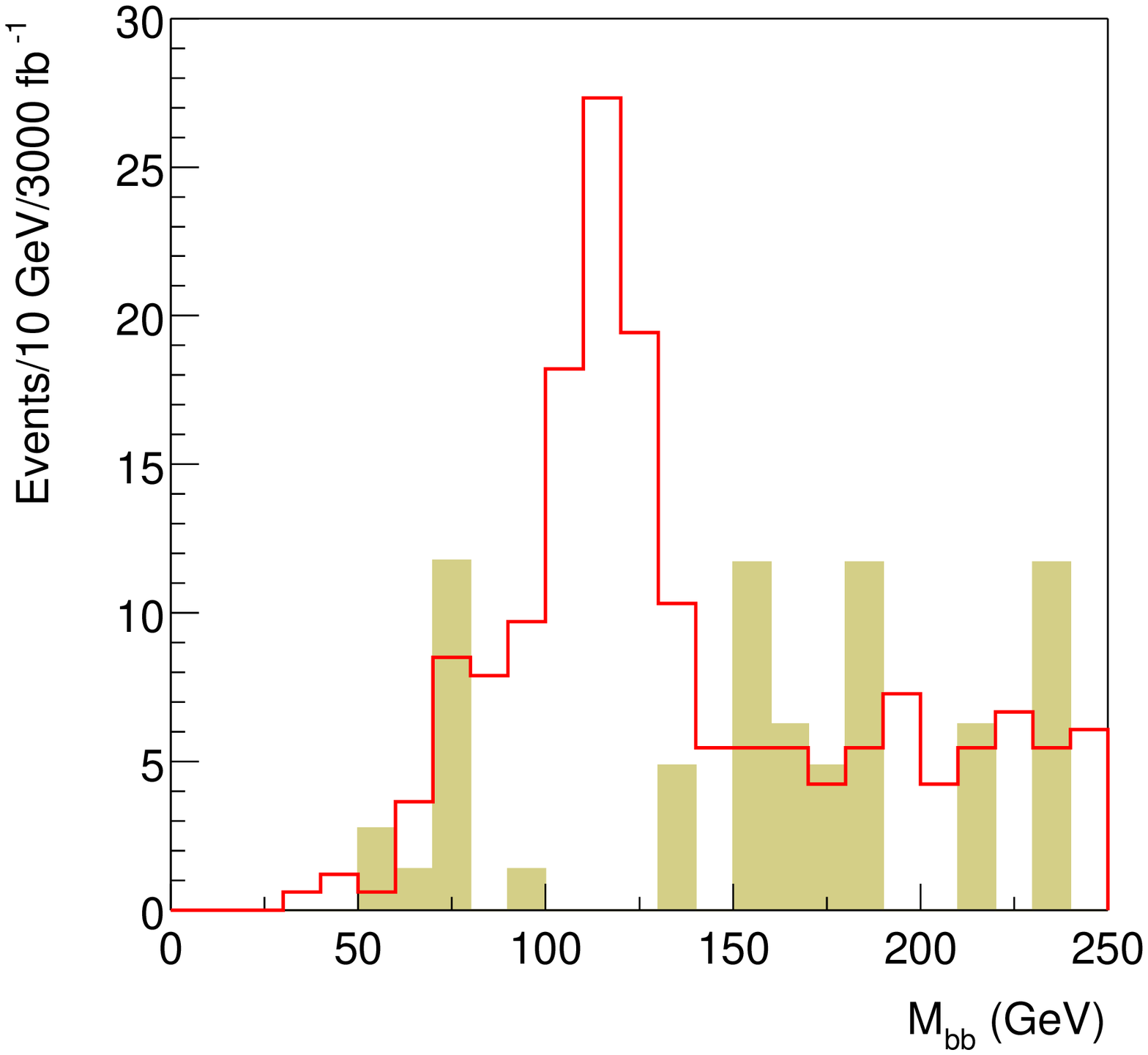}
\includegraphics[width=0.4\textwidth,clip]{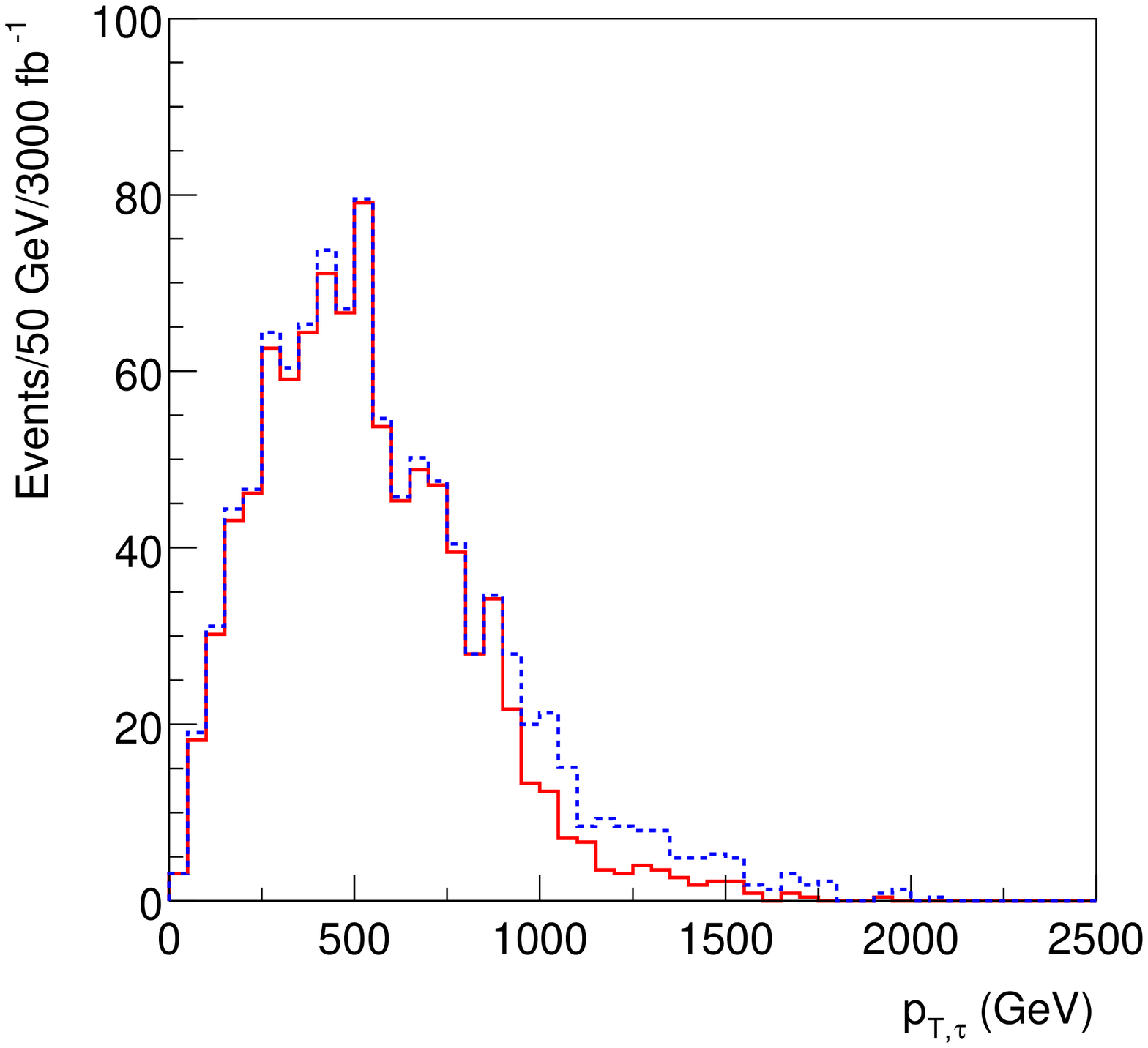}  
\end{center}
\vskip -0.4cm 
\caption{Left: the $b\overline{b}$ invariant mass distribution for Point~K 
 (solid: signal; shaded: SM background).  
 Right: the $p_T$ distribution of $\ttau_1$ for Point~H 
 (dashed: all $\ttau_1$; solid: 
 $\ttau_1$ reaching the ATLAS Muon Spectrometer 
  with a delay $\Delta t > 7{\,\rm ns}$).
\label{fig:pointkh}}
\end{figure}

 Point H is characterised by squark and gluino masses above 2.5~TeV 
  and by almost degenerate $\ttau_1$ and  $\lsp$ masses. In this
particular case,  $\ttau_1
\not{\!\!\!\to} 
\lsp \tau$, so $\ttau_1$  must decay by second order weak processes,
$\ttau_1 \to \lsp e\bar\nu_e\nu_\tau$, with a long lifetime. The
dominant SUSY rates arise from the strong production of squark pairs, with
$\tq_L \to \tchi_1^\pm q, \tchi_2^0 q$ and $\tilde q_R \to \ q\lsp$. The
staus which are produced from cascade decays of the gauginos
traverse the detector with a signal similar to a ``heavy muon".
 The $p_T$ spectrum of these quasi-stable $\ttau_1$  is
shown in Fig.~\ref{fig:pointkh}. The ATLAS 
 Muon Spectrometer~\cite{ATLAS-tdr} has a 
resolution of about 0.7~ns for the time of flight measurement. 
 The spectrum for staus reaching the muon chambers with a time delay 
 $\Delta t > 7\ {\rm ns}\ (10\sigma)$ is also shown in Fig.~\ref{fig:pointkh}. 
  This signal could be observed with $\sim 300$~fb$^{-1}$. 
  The mass of the stable stau can be determined
 by combining a momentum measurement  with
 a time of flight measurement in the Muon Spectrometer.  
Studies of such quasi-stable particles at somewhat smaller masses
carried out with simulations of the ATLAS detector showed a mass resolution of
approximately 3\% given sufficient statistics~\cite{ATLAS-tdr}.
 A precision of this order should be achieved for Point~H 
 with 3000~fb$^{-1}$.

 The stable $\ttau_1$ signature is somewhat exceptional. Therefore
other signatures that would be present if the
stau decayed inside the detector were examined. 
 Events were selected by requiring
 at least two jets
with $p_T > 0.1\Meff$, $E_T^{miss} > 0.3\Meff$, $\Delta\phi(j_1, E_T^{miss}) <
\pi-0.2$, and $\Delta\phi(j_1,j_2)<2\pi/3$. The $\Meff$ distributions
after these cuts show that the number of events in
the region where $S/B>1$ is of order 30 for 3000~fb$^{-1}$.
 Di-leptons arise from the cascade decay $\tq_L\to q \tchi_2^0\to
q\ell^+\ell^- \lsp$. The di-lepton invariant 
 mass distributions should have a 
kinematic end-point
corresponding to this decay. Figure~\ref{fig:pointh} shows the
distribution for same-flavor and different-flavor lepton pairs.
 Events were required to have $\Meff>3000$~GeV,
 $ E_T^{miss}>0.2\Meff$, and two isolated opposite-sign leptons
with $E_T>15$ GeV and $\abs{\eta}<2.5$.
 The end-point structure may be observable, although it should be noted
that the background has large errors as only three generated events passed
the cuts. The edge comes mainly from $\tchi_2^0 \to
\tell_L^\pm \ell^\mp$, which has  a branching ratio of 15\% per flavor. This
gives an end-point at
$$
\sqrt{(M_{\tchi_2^0}^2-M_{\tell_L}^2)(M_{\tell_L}^2-M_{\lsp}^2)
\over M_{\tell_L}^2} = 447.3\,\GeV
$$
consistent with the observed end-point in Fig.~\ref{fig:pointh}. Obviously
this distribution does not distinguish $\tell_L$ and $\tell_R$.

\begin{figure}
\begin{center}
\includegraphics[width=0.4\textwidth,clip]{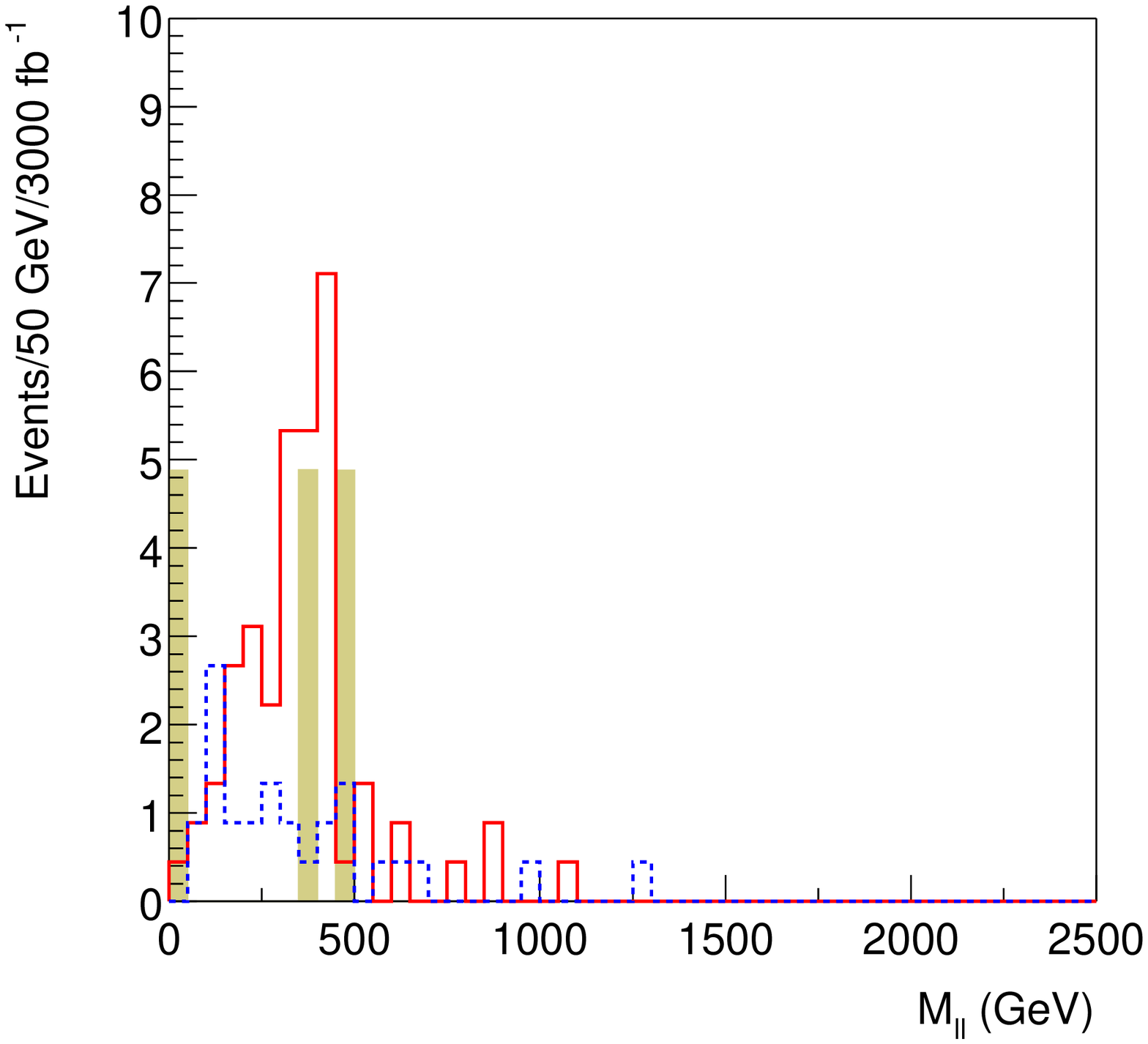} 
\includegraphics[width=0.4\textwidth,clip]{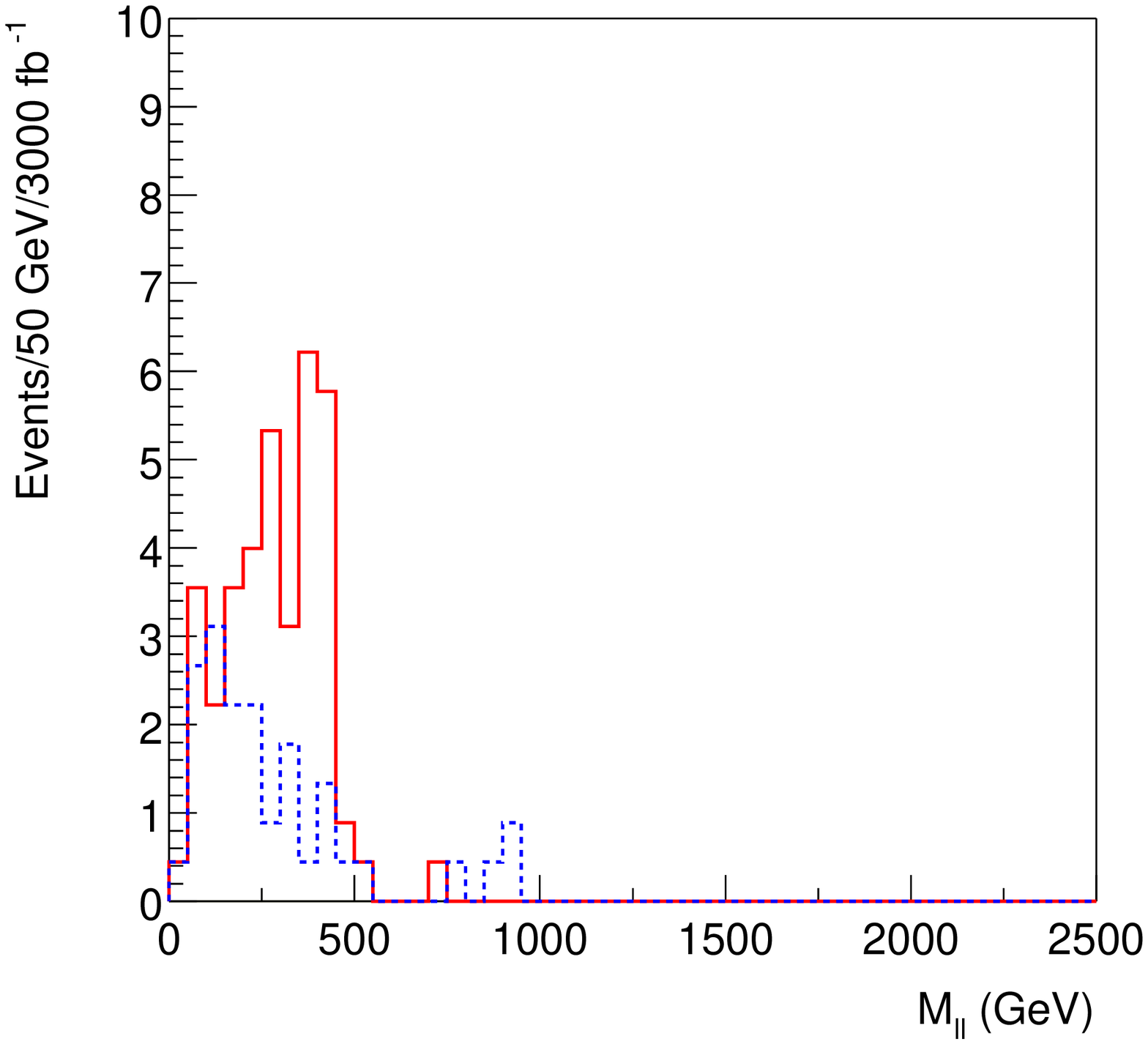} 
\end{center}
\vskip -0.4cm 
\caption{For Point~H, the $M_{\ell\ell}$ distribution for all events 
(left) and events containing a stable stau (right).  
 Solid: same-flavour lepton pairs. Dashed: different-flavour lepton
 pairs. Shaded: SM background.
\label{fig:pointh}}
\end{figure}

 If the stable stau signature is used, then the situation improves considerably.
The di-lepton mass for events containing a $\ttau_1$ with a time delay 
in the ATLAS Muon Spectrometer in the range $7 < \Delta t < 21.5\,{\rm ns}$ 
 is shown in Fig.~\ref{fig:pointh}.
 Since $\Delta t>10\ \sigma$, the Standard Model  background is expected  to be
negligible. A measurement of the end-point should 
be possible at the SLHC.

\subsection{New gauge bosons}
\label{sec:zprime}
The potential of the SLHC for the discovery of additional heavy
gauge bosons has been studied by considering a $Z'$ with $Z$-like
couplings to leptons and quarks.  The $Z'$ production cross-section
times branching ratio into electron pairs $\sigma (Z'\rightarrow e^{+}
e^{-})$ and the $Z'$ width are shown in Table~\ref{tab:zprime} as a
function of mass.

\begin{table}
\begin{center}
\caption{$Z'$ production cross-section at the LHC 
 times branching ratio into electron pairs 
 and $Z'$ width, as a function of mass.}
\label{tab:zprime}
\vspace*{0.1cm}  
\begin{tabular}{|c|c|c|c|c|c|c|} \hline
 $Z'$ mass (TeV) & 1 & 2 & 3 & 4 & 5 & 6 \\ 
\hline
$\sigma (Z' \rightarrow e ^{+} e^{-}) (fb)$   & 512  & 23.9 & 2.5  & 0.38 & 0.08 & 0.026 \\ \hline
$\Gamma _{Z'}$ (GeV) & 30.6 & 62.4 & 94.2 & 126.1& 158.0& 190.0 \\ \hline
\end{tabular}
\end{center}
\end{table}

The study is based on the CMS detector performance and the result is
then extrapolated to the case of two experiments.  The analysis takes
into account acceptance, reconstruction efficiency and resolution for
muons and electrons. The expected pile-up noise at \slhclum\ is 
also included, as well as saturation effects in the CMS
crystal calorimeter.

The $Z'$ mass is reconstructed conservatively without correcting
for internal photon radiation. This leads to some tails in the mass
spectra.  The expected number of signal and background events is
calculated from the gaussian part of these spectra. The background is
found to be small, i.e. $\simeq$ 2\% from Drell-Yan production and
less than 1\% from $\ttbar$.
 The expected number of signal events for two experiments is shown in
Fig.~\ref{fig:zprime}. Assuming that a discovery can be claimed if the
number of observed events is at least ten, the LHC discovery reach
improves from $\sim$~5.3~TeV (standard LHC, 600 fb$^{-1}$) to
$\sim$~6.5~TeV (SLHC, 6000 fb$^{-1}$).  For comparison, a machine with
$\sqrt{s}$=~28~TeV running at \lhclum\ would
extend the mass reach of the standard LHC by 50\%~\cite{atlas-note}.

\begin{figure}
\begin{center}
\includegraphics[width=0.65\textwidth,clip]{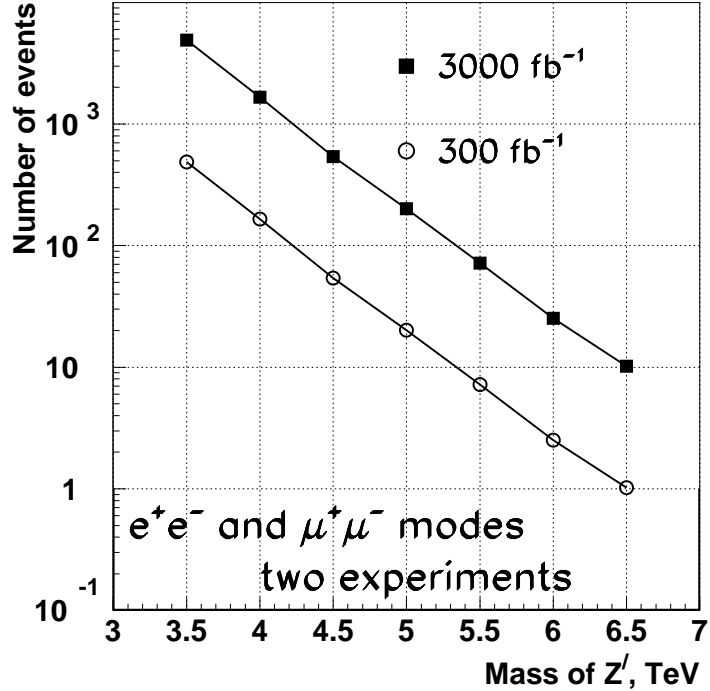} 
\end{center}
\vskip -0.4cm 
\caption{Expected number of $Z'\to\mu^+\mu^-, e^+e^-$ events in
 both experiments for integrated luminosities 
 of 300 \ifb\  per experiment 
 and 3000 \ifb\  per experiment.\label{fig:zprime}}
\end{figure}

It should also be noted that discovery will be mainly based on the
electron final state, which provides the best mass resolution, whereas
measurements of couplings and asymmetries will use both lepton
channels.

\subsection{Extra-dimensions}
\label{sec:extradim}
 Theories with large extra-dimensions, which aim at solving the
hierarchy problem by allowing the gravity scale to be close to the
electroweak scale, have recently raised a lot of interest.  They
predict new phenomena in the TeV energy range, which can therefore be
tested at present and future colliders.
   Several models  and signatures have
been considered in the study presented here. They are discussed below.

\subsubsection{Direct graviton production in ADD models}

 In these models~\cite{ref:ADD}, the extra-dimensions are compactified
to the sub-millimiter size and 
 only gravity is allowed to propagate
in them, whereas the SM fields are confined to a
4-dimensional world. Gravitons in the extra-dimensions 
occupy energy/mass levels
which are separated by very small splittings, and therefore give rise to a
continuous tower of massive particles (`Kaluza-Klein (KK) excitations').
The presence of additional dimensions can therefore produce new
phenomena involving gravitons, such as direct graviton production at
high energy colliders.\\  
The most sensitive channel at the LHC should
be the associated production of KK gravitons with a quark or
a gluon. The resulting signature is an energetic jet plus missing
transverse energy, since the gravitons escape detection.  The
cross-section depends on two parameters, the gravity scale $M_D$ and
the number of extra-dimensions $\delta$, and decreases with increasing
values of both $M_D$ and $\delta$.  The background is dominated by the
final state $Z(\to \nu\nu)+jets$.
  
 The discovery potentials of the LHC and SLHC are compared in 
 Fig.~\ref{fig:extradim}.    
\begin{figure}
\begin{center}
\includegraphics[width=0.65\textwidth,clip]{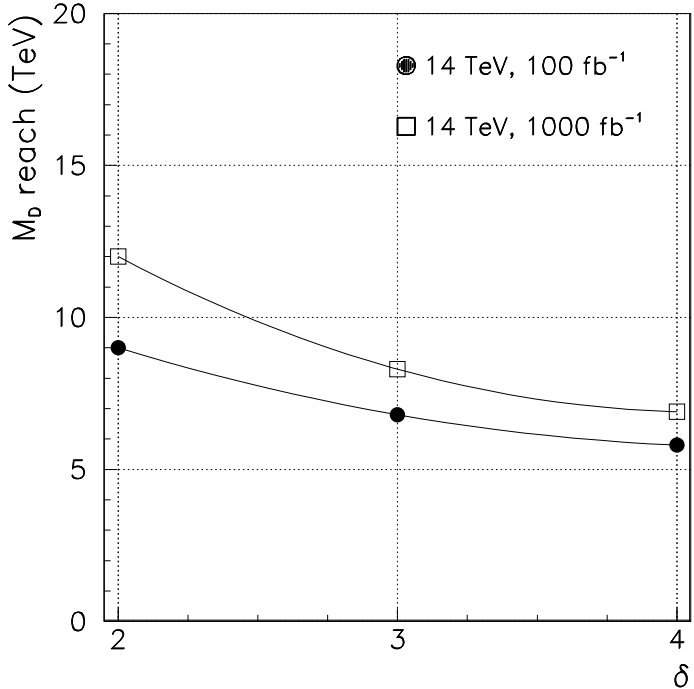} 
\end{center}
\vskip -0.4cm 
\caption{Expected $5\sigma$ discovery reach on the gravity scale 
 $M_D$ as a function of the number of extra-dimensions $\delta$ 
 in ATLAS in the framework of ADD models, for 100~\ifb\  
 (LHC) and 1000~\ifb\  (SLHC).\label{fig:extradim}}
\end{figure}   
 It can be seen that a factor of ten in luminosity would improve the
LHC mass reach by typically 30\%.  Major detector upgrades 
are not crucial for this physics, since the
search is based on events with jets and missing energy in the TeV
range. For comparison, doubling the LHC
energy but keeping the instantaneous luminosity of \lhclum\ would
approximately double the reach in $M_D$ for any value of
$\delta$~\cite{atlas-note}.

\subsubsection{Virtual graviton exchange in ADD models} 

Virtual KK gravitons can also be exchanged between incoming
and outgoing SM particles in high-energy collisions, thereby leading
to modifications of the cross-section and angular distributions
compared to the SM expectations.  Since graviton effects are enhanced
at high energy, due to the large number of accessible Kaluza-Klein
excitations, such manifestations of Extra-dimensions are expected at
large invariant mass and $p_T$ of the particles in the final state.
Drell-Yan and two-photon production are among the most sensitive
channels at high-energy colliders.  Using these channels, it was found
that the reach in the gravity scale $M_D$ for $\delta=3$ 
 increases from $\sim$8~TeV (100 \ifb, standard LHC) to
11.7~TeV (3000 \ifb, SLHC).

\subsubsection{Resonance production in Randall-Sundrum models}

In the Extra-dimension scenario proposed by Randall and
Sundrum~\cite{randalfS} the hierarchy between the Planck and the
electroweak scales is generated by an exponential function called
``warp factor".  This model predicts KK graviton resonances
with both weak scale masses and couplings to matter.  In its simplest
form, with only one extra-dimension, two distinct branes (the TeV
brane and the Planck brane), and with all of the SM fields living on
the TeV brane, the Randall-Sundrum model has only two fundamental
parameters: the mass of the first KK state $m_1$ and the
parameter $c = k/M_{\overline{PL}}$, where $k$ is related to the
curvature of the 5-dimensional space and $M_{\overline{PL}}$ is the
effective Planck scale. The parameter $c$ governs the width of the
resonances, and is expected to be not far from unity.

Direct production of Randall-Sundrum resonances ($pp\to \ G$) can lead
to spectacular signals, for instance in the clean di-lepton decay mode
($G\to \ \ell\ell$).  They should be observable already in the first
years of LHC running if $m_1$ is in the range 1-3 TeV. In addition,
their properties (e.g. their spin-2 nature) can be measured
and should distinguish them
from e.g. $Z'$ production~\cite{Allanach}.
 Figure~\ref{fig:rs} summarises 95\% C.L. exclusion limits in the plane
 $m_1$ versus $c$-parameter. It shows the present constraints from LEP
and Tevatron, as well as the potentials of the LHC and SLHC.  The
region labelled as ``allowed" is theoretically favoured~\cite{rizzo},
although the rest of the plane is not strictly forbidden.  It can be
seen that the SLHC should extend the LHC reach by almost 1~TeV.

\begin{figure}
\begin{center}
\includegraphics[width=0.45\textwidth,angle=90,clip]{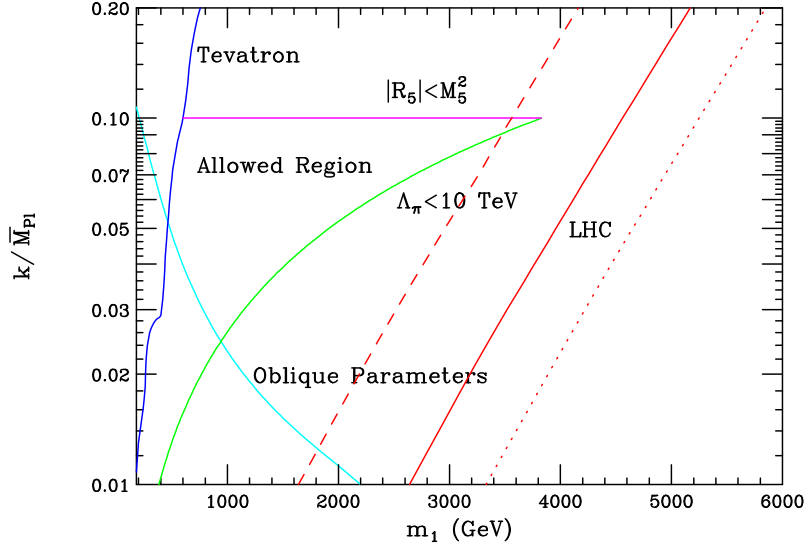}
\end{center}
\vskip -0.4cm 
\caption{95\% C.L. exclusion limits in the plane $m_1$ versus $c$
  (see text) for Randall-Sundrum graviton resonances decaying into
  electron or muon pairs. The triangular region labelled as `allowed'
  represents the theoretically favoured domain. Here $M_5$ is the
  5-dim Planck scale, $R_5$ is the 5-dim curvature invariant and
  $\Lambda_\pi$ is the inverse coupling strength of the KK gravitons.
  The bound $\vert R_5\vert < M_5^2$ is applied so that quantum
  gravity loop effects are small and we can treat the RS model
  classically.  The dashed line and the full line show the LHC
  potential for integrated luminosities of 10~fb$^{-1}$ and
  100~fb$^{-1}$ respectively, the dotted line shows the potential of
  the SLHC with 1000~fb$^{-1}$. The exclusion domains lie to the left
  of the lines. The present Tevatron limit is also indicated, as well
  as the region excluded by precision measurements of the electroweak
  oblique parameters $S,\ T,\ U$ at LEP.\label{fig:rs}}
\end{figure}

\subsubsection{Resonance production in TeV$^{-1}$ scale Extra-dimensions}

In these models~\cite{MOD1} only the
fermion fields are confined to a 4-dimensional brane, whereas the 
SM gauge fields are allowed to 
 propagate in a number of additional small
extra-dimensions (compactification radius $\sim 1~$TeV$^{-1}$),
 orthogonal to the brane.
 The most important phenomenological consequence 
is the predicted existence of KK excitations of the SM 
gauge bosons, $\gamma, Z, W$ and $g$.
 For simplicity, only the case of one extra-dimension is considered here. 
 The model is completely 
 specified by a single parameter $M_c$, the compactification scale,
 from which the masses $M_n$ of the KK excitations of the gauge bosons can
 be derived using the relation $M^2_n= (nM_c)^2+M_0^2$, where $M_0$ is 
the mass of the SM gauge boson.
 The couplings are the same as the corresponding SM couplings, 
scaled by a factor $\sqrt{2}$. 
 Constraints  from precision electroweak 
measurements give an approximate lower limit $M_c~>~4$~TeV~\cite{Rizzo1}. 

The possibility of detecting the leptonic decays of the $KK$
excitations of the $\gamma$ and $Z$ bosons at the LHC and SLHC has
been studied. The production of the gauge boson excitations, including
interference terms and angular information, was performed using the
full Breit-Wigner shape for the first two excited states~\cite{RT},
and a resummed expression for the higher states.  The matrix elements
were interfaced to PYTHIA, and the produced events were passed through
a fast simulation of the ATLAS detector~\cite{ATLFAST}.

 The main experimental issue in the study of these 
models is the measurement of leptons with very high transverse momenta.
 As an example, the ATLAS detector has been designed to measure leptons 
with $p_T$ up to $\sim$2-3 TeV. 
 For electrons the main issue is the saturation of the dynamic range of the 
calorimeter electronics, which can be possibly compensated by 
a modification in the gain of the readout electronics. 
 For muons, the momentum is obtained from the track curvature 
 in the external Muon Spectrometer, and this measurement is
 very poor for transverse momenta above 4~TeV.

Events were selected by
 requiring two opposite-sign isolated leptons ($\ell=e,\mu$) with 
 $p_T~>$~20~GeV, inside the rapidity range $|\eta|<2.5$ and  
 with invariant mass
$m({\ell^+\ell^-})>~1000$~GeV.
 In the absence of New Physics, approximately 18000 events 
survive these cuts for each lepton flavour and for an integrated
luminosity of 3000~fb$^{-1}$.
 Figure~\ref{fig:tevscale} shows 
 the expected two-lepton invariant mass spectrum for an Extra-dimension
 signal with $M_c$~=~5~TeV on top of the SM background. 
\begin{figure}
\begin{center}
\includegraphics[width=0.45\textwidth,clip]{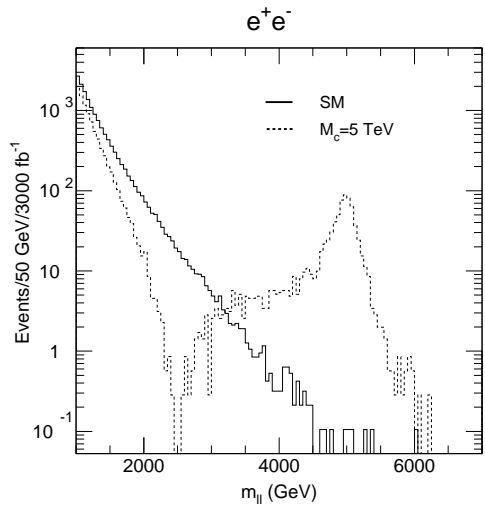}
\end{center}
\vskip -0.4cm 
\caption{Invariant mass distribution of $e^+e^-$ pairs as expected from
 the Standard Model (full line) and from a gauge excitation model with 
$M_c$~=~5~TeV (dashed line).
The histograms are normalised to 3000~fb$^{-1}$.\label{fig:tevscale}}
\end{figure}
 One can notice two structures. 
 A peak centered around $M_c$, corresponding to the superposition
 of the first $\gamma$ and $Z$ resonances, 
 and a suppression of the cross-section with respect to 
the SM expectation for masses below the resonance. This suppression is due
to the negative interference between the SM gauge bosons and
the whole tower of KK excitations, and is sizeable even for compactification
scales well above those accessible form a direct detection of the mass peak.

 The reach for the observation of a resonant peak 
in the ${\ell^+\ell^-}$ invariant mass distribution
can be obtained from Table~\ref{tab:tevscale},
which  gives the expected numbers of signal and 
background events in the electron and muon channels
separately, for an integrated luminosity of 3000 fb$^{-1}$ and
for different values of $M_c$. 
 Using as a discovery criterion that at least ten 
events (summed over both lepton species and both experiments) be detected
 with two-lepton invariant mass above a given value, and
 that the signal statistical significance be $S/\sqrt{B}>5$,  the reach  
should be $\sim$6~TeV for 300~fb$^{-1}$ per experiment
 and $\sim$7.7~TeV for 3000~fb$^{-1}$ per experiment
(both experiments combined). It should be noticed that 
 to achieve this result a good knowledge of the rate
of background events at high masses, which could be affected
by mismeasurements of the lepton momenta, is crucial.
  For  $M_c~=4$~TeV a few events 
 from the second resonance at 8~TeV could be observed, thereby hinting at the 
periodic structure of the mass spectrum of the resonances. 
 For higher compactification
scales, only the first KK excitation will most likely be accessible.

\begin{table}
\begin{center}
\caption{For an integrated luminosity of 3000~fb$^{-1}$ and one experiment,
 expected number of signal events in the peak region for different values 
 of the mass of the lowest KK excitation $M_c$, 
  and expected number of SM background events. 
  The peak region is defined by a cut on the minimum  $\ell^+\ell^-$ invariant mass 
 given in the second column. The results for electrons and muons are shown separately.}
\label{tab:tevscale}
\vspace*{0.1cm}  
\begin{tabular}{|r|r|r|r|r|r|}
\hline
 $M_{c}$(GeV)  & Cut (GeV) & ${\rm Signal}\ (e)$ & ${\rm Signal}\ (\mu)$ & Background ($e$)& Background ($\mu$)\\
\hline
 4000 & 3000 & 5160   & 4680   & 44   & 54 \\
 5000 & 4000 & 690    &  600   & 4.5  & 6.6 \\
 5500 & 4000 & 270    &  240   & 4.5  & 6.6 \\
 6000 & 4500 & 99     &  84    & 1.5  & 3   \\
 7000 & 5000 & 13.5   &  11.4  & 0.45 & 0.5 \\
 8000 & 6000 & 1.3    &  1.6   & 0.045 & 0.36 \\
\hline
\end{tabular}
\end{center}
\end{table}

 Sensitivity to a signal can also be obtained from the observation of  
 off-resonance (negative) interference effects in the Drell-Yan mass spectrum 
(see Fig.~\ref{fig:tevscale}). A detailed estimation requires
a likelihood fit to the shape of the distribution. 
  A simpler method was used to obtain the results reported here, which
 consists of evaluating the decrease in the number of events 
 (compared to the SM expectation) inside a given $m ({\ell^+\ell^-})$ range,  
 as a function of $M_c$. 
 The statistical significance of the cross-section suppression 
is approximately given by $(N (tot)-N (SM))/\sqrt{N (SM)}$,
 where $N (tot)$ is the total number of observed events and $N(SM)$ the 
 expected number of events from the Standard Model.
  There are however uncertainties in the knowledge of the shape 
 of the two-lepton invariant mass distribution, 
 due to both instrumental effects (absolute energy scale, linearity), and 
 theoretical effects (structure functions, higher 
 order corrections). A precise estimation of these uncertainties requires
 a dedicated study, which is beyond the scope of this paper. 
  Therefore here we have simply assumed that a signal can only
  be claimed if $R\equiv \vert N (tot) /N (SM)-1 \vert >5\%$ or 
  $R >10\%$.  
   For an integrated luminosity of 3000~fb$^{-1}$ per experiment, 
   combining both experiments and both lepton flavours, and looking   
  at the mass range $1500~<~m({\ell^+\ell^-})~<3500$~GeV,  
  the reach should be $M_c=14$~TeV ($10\sigma$ 
 statistical significance) requiring $R>10\%$ and 
$M_c=20$~TeV ($5\sigma$ statistical significance) requiring 
$R>5\%$.  
 As a comparison, in one year at the nominal LHC luminostiy, 
 i.e. with 100~fb$^{-1}$, and considering the interval
$1000~<m ({\ell^+\ell^-})~<3500$~GeV, the 
 5$\sigma$ reach should be $\sim$10~TeV,
 corresponding to a 14\% deviation from the Standard Model cross-section.

 In conclusion, 
 with an integrated luminosity of 3000~fb$^{-1}$ per experiment, 
the LHC experiments should be able to detect directly KK excitations 
of the $\gamma$ and $Z$ gauge bosons in their leptonic decay modes
for compactification scales of up to 7.7~TeV. 
 By studying the deviation from the Standard Model expectation 
 of the non-resonant part of the two-lepton invariant mass distribution, this reach 
can be extended to higher compactification scales.
 The sensitivity in this case 
is limited by the assumed uncertainty on the knowledge of 
the SM Drell-Yan spectrum at high mass, and is of order 15-20~TeV 
for a systematic uncertainty of 5--10\%.

\subsection{Quark substructure}
\label{sec:compo}

A tenfold increase in the LHC luminosity should give access to jets
of up to $E_T\sim 4.5$~TeV (see Fig.~\ref{fig:jets}), thereby
extending the machine kinematic reach for QCD studies by more
than 0.5-1~TeV.
\begin{figure}
\begin{center}
\includegraphics[width=0.65\textwidth,clip]{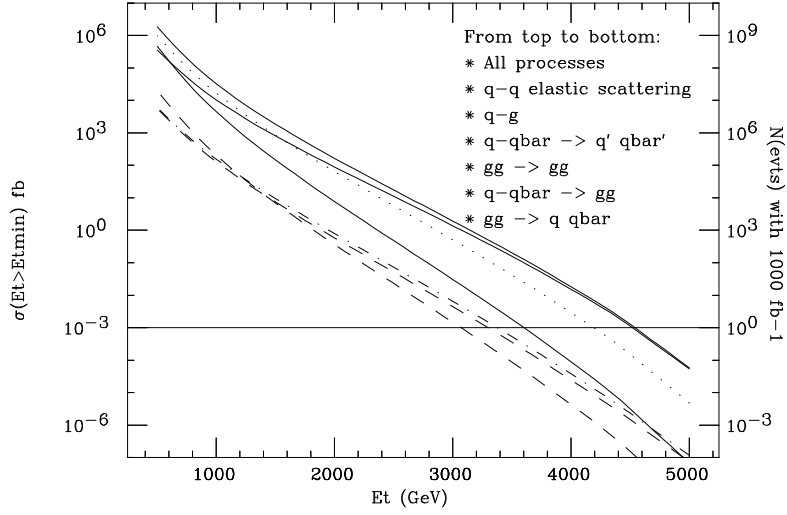}
\end{center}
\vskip -0.4cm
\caption{Integrated production cross-section and 
 rates for inclusive central ($\vert \eta
  \vert < 2.5$) jets. The different curves label the various
  contributions to the total the cross-section.
 \label{fig:jets}}
\end{figure}
 This improved sensitivity should have an impact also on the search for
quark sub-structures.  Indeed, signals for quark compositeness should
reveal themselves in deviations of the high energy part of the jet
cross-section from the QCD expectation. The angular distribution of
di-jet pairs of large invariant mass provides an independent signature
and is less sensitive to systematic effects like possible
non-linearities in the calorimeter response.  This method was
therefore used in this study.
 
Figure~\ref{fig:compo} shows the expected 
 deviations from the SM prediction
for two values of the compositeness scale $\Lambda$, as a function of
the variable $\chi$ defined as
$\chi=(1+|\cos\theta|)/(1-|\cos\theta|)$. Here $\theta$ is the angle
between a jet and the beam in the centre-of-mass frame of the di-jet
system.  The invariant mass of the di-jet system was required to be
larger than 6~TeV.  The effect of compositeness shows up as an
increase in the event rate at small values of $\chi$.
\begin{figure}
\begin{center}
\includegraphics[width=0.55\textwidth,clip]{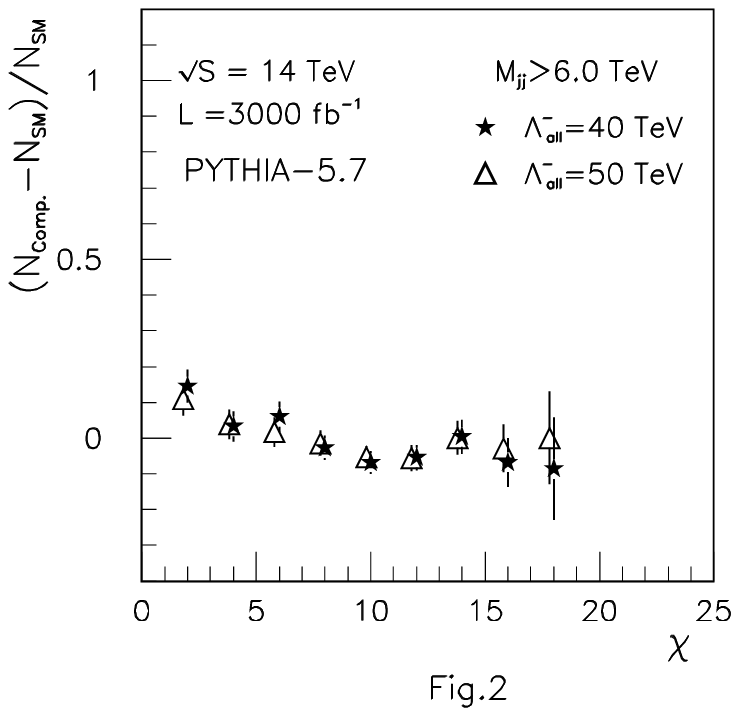}
\end{center}
\vskip -0.4cm 
\caption{Expected deviations from the Standard Model
  predictions for the angular distribution of di-jet pairs at the SLHC
  (ATLAS experiment, integrated luminosity of 3000~\ifb), for two
  values of the compositeness scale $\Lambda$.  Di-jet pairs are
  required to have invariant mass greater than
  6~TeV.\label{fig:compo}}
\end{figure}
The compositeness scales that can be probed in this way at the LHC and
SLHC are summarised in Table~\ref{tab:compo}. For comparison, the
potential of a 28~TeV machine is also shown. It can be seen that a tenfold
luminosity increase would have a significant impact for this physics.

\begin{table}
\begin{center}
\caption{The 95\% C.L. lower limits
  that can be obtained in ATLAS on the compositeness scale $\Lambda$
  by using di-jet angular distributions and for various
  energy/luminosity scenarios. }
\label{tab:compo}
\vspace*{0.1cm}  
\begin{tabular}{|c|c|c|c|c|}\hline
Scenario         &    14 TeV 300 fb$^{-1}$&  14 TeV 3000 fb$^{-1}$
&28 TeV 300 fb$^{-1}$ &28 TeV 3000 fb$^{-1}$ \cr\hline
$\Lambda$(TeV)  &     40  &     60 &     60   &    85\cr 
\hline
\end{tabular}
\end{center}
\end{table}
  
As these measurements involve only the calorimeters and jets in the
TeV range, they can be performed at the SLHC without major detector
upgrades.  Ability to extend the heavy-flavour tagging to the very
high \et\ region could however help disentangling the flavour
composition of a possible cross-section excess. We evaluated that only a
fraction smaller than 1\% of the jets with $\et>2$~TeV should contain
bottom or charm quarks, therefore any indication of a long lifetime
component in these jets beyond this level would signal New Physics.

\section{THE EXPERIMENTAL CHALLENGES AND THE DETECTOR R\&D}
\label{sec:experiments}
The main motivation for a luminosity upgrade is to extend the physics
reach of the experiments by providing more statistics. For a full
exploitation of this upgrade it is imperative that the general
detector performance remains at the same level as at the nominal LHC.
In order to face the challenge of operation at an order of magnitude
higher luminosity than foreseen in the original LHC
design\footnote{For further reference the neutron flux at \slhclum\ 
  and the dose for an integrated luminosity of 2500~\ifb\ are shown in
  Fig.~\ref{fig:expfig1}.}, we have therefore deliberately chosen to
pose first the question of whether the currently planned detectors
could survive and operate at luminosities of \slhclum. If they cannot
then the question is posed of possible replacements of detectors or
technologies.

   Development of new particle detectors takes a long time and goes
   through many phases starting from the idea or concept, progressing
   through intensive R\&D, prototyping, systems integration,
   installation and commissioning and finally data taking. This can be
   illustrated using any one of the many detector technologies in the
   LHC experiments. Typical time-scales stretch over more than a
   decade. In order to create a R\&D roadmap we therefore assume that the
   upgraded detectors should be installed and commissioned by around
   2012/2013, and try to answer the following questions:

\begin{itemize}
\item What R\&D is necessary?
\item What priority should be assigned to the R\&D?
\item When should the R\&D start, taking account of the manufacturing phase?
\item What resources would be required (financial and manpower)?
\item How would the R\&D interface with that carried out elsewhere?
\end{itemize}
In this Section, we shall address the above points for 
each sub-detector in turn, as well as for the trigger, data acquisition
and electronics. While, as discussed in
Section~\ref{sec:machine}, several options can be envisaged for the bunch
structure at high luminosiry, we limited our studies to the 
case of 12.5~ns bunch spacing.

\begin{figure}
\begin{center}
\includegraphics[width=0.85\textwidth,clip]{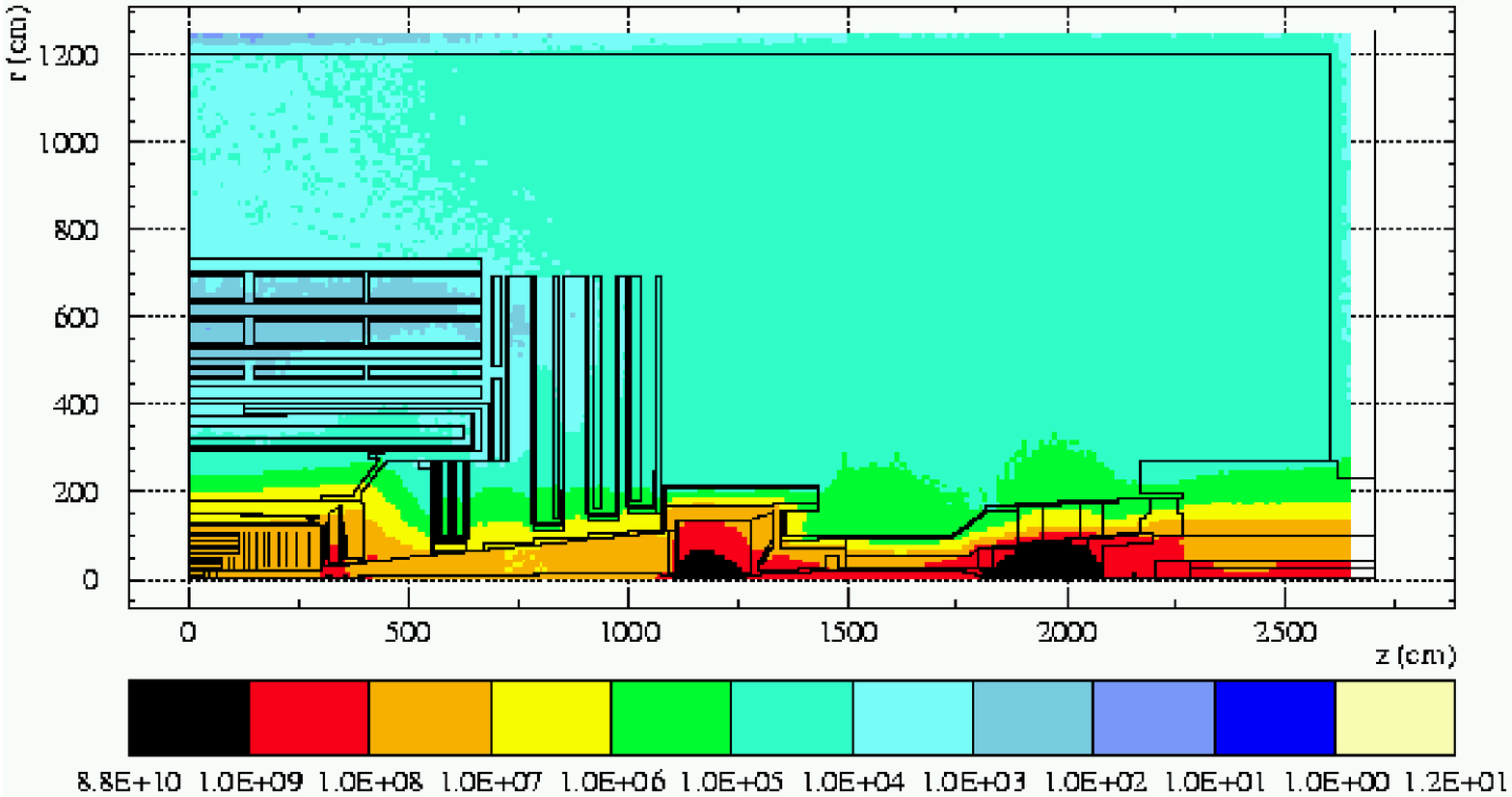} \\
\includegraphics[width=0.85\textwidth,clip]{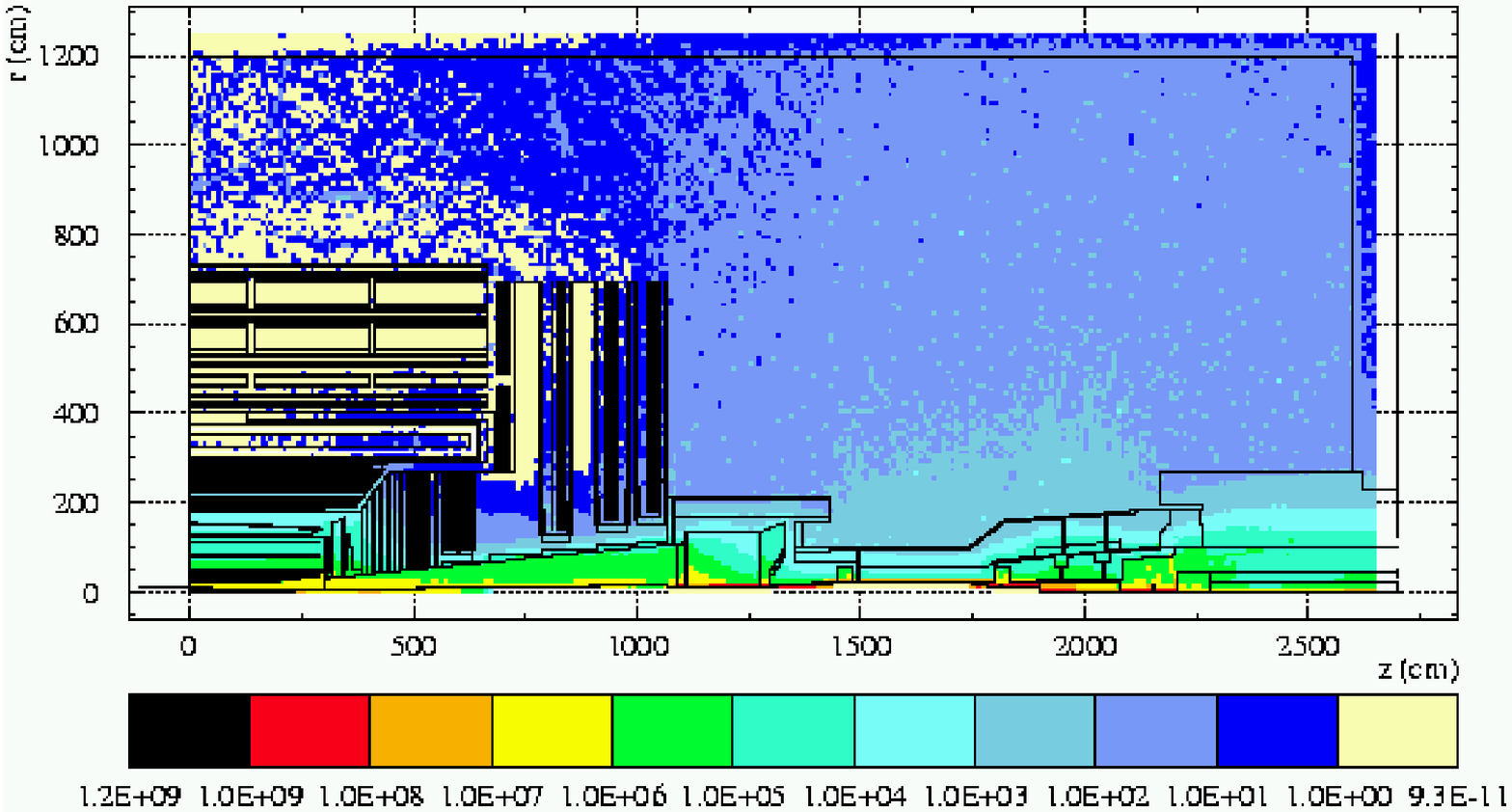} 
\end{center}
\caption{Upper: the neutron flux ($\rm{cm}^{-2}s^{-1}$) 
at an instantaneous luminosity of
\slhclum. Lower: dose (Gy) for an integrated luminosity
         of 2500 fb$^{-1}$ \label{fig:expfig1}}
\end{figure}
 
\subsection{Inner Tracking Detector}
\label{sec:IT}
The performance of the tracker is characterised by momentum
resolution, track reconstruction efficiency and b-tagging efficiency
and purity.  In order to keep the occupancy and two-track resolution
at the standard LHC levels, the cell sizes have to be decreased in
order to compensate for the increased track density at SLHC. Cell
sizes have to be decreased by a factor 10, though simulations will be
needed to optimise the granularities required at a given radius. The
total cost should not much exceed that of the currently planned
trackers, implying that the cost per channel has to be decreased by a
factor of 10. Cost reduction should therefore be a driving feature of
any planned R\&D.

In what follows we assume that the tracker will be equipped with
electronics that is fast enough to distinguish individual crossings.
Otherwise, at \slhclum, the detectors will have to 
deal with about 200 collisions per 25 ns, producing about 1200 charged tracks 
per unit of pseudo-rapidity.

The limiting factor for the lifetime of the detectors will be radiation
damage, which for cooled silicon detectors is mainly a function of the
integrated luminosity. For the latter we assume 2500 $fb^{-1}$, which is 5
times more than the assumption used e.g. in the CMS Technical Design
Report for the design luminosity of LHC. The integrated fluence and dose
for the CMS Tracker are given in Table~\ref{tab:exptab1}. Because the 
radiation environment is dominated by the pp secondaries, these values are of
rather generic nature and not strongly CMS specific.

\begin{table}
\begin{center}
\caption{Hadron fluence and radiation dose in different radial layers of
the CMS Tracker (barrel part) for an integrated luminosity of 
2500~\ifb.}
\label{tab:exptab1}
\vspace*{0.1cm}  
\begin{tabular}{|c|c|c|c|}\hline
 Radius (cm)    & Fluence of fast            & Dose (kGy) & Charged Particle \\
            & hadrons ($10^{14}$cm$^{-2}$)&  & Flux (cm$^{-2}$s$^{-1}$) \\  
\hline
4    &   160  &   4200  &  5 $\times$ 10$^{8}$   \cr\hline
11   &   23   &   940   &  10$^{8}$              \cr\hline
22   &   8    &   350   &  3 $\times$ 10$^{7}$   \cr\hline
75   &   1.5  &   35    &  3.5 $\times$ 10$^{6}$ \cr\hline 
115  &   1    &   9.3   &  1.5 $\times$ 10$^{6}$ \cr\hline
\hline
\end{tabular}
\end{center}
\end{table}
The currently planned silicon systems (designed for integrated
luminosities of approximately 500\ipb) would probably be nearing the end
of their lifetime at the start of a potential SLHC programme.  They
would not be able to handle the fluences at SLHC. The silicon sensors,
both strip and pixel systems, would suffer substantial radiation damage,
increased noise and thermal runaway as a consequence of the increased
leakage currents. Since the current sensor design would not support the
bias voltage required for full depletion, the sensors would operate
under-depleted, decreasing the signal and increasing the noise, and
hence degrading the performance.

For the electronics the situation is somewhat more favourable. The
indications are that deep sub micron (DSM) electronics can probably
withstand higher doses, but is not fully characterised at the doses and
fluences under discussion. Mitigation of single event effects will also
have to be carefully investigated.
At low radius it is unlikely that the current
DSM chips could survive. We also expect the opto-electronic components
to be affected, and most of the materials, glues, and cooling fluids are
not tested to the doses discussed above. 

The innermost strip layers (at a radius of around 25cm) will have occupancies above
10 \%. The TRT in ATLAS will face occupancy approaching 100 \% and cannot be
used.  So even if the detectors could tolerate the radiation damage, they
will not be suitable for the SLHC.

{\bf We conclude that the only viable solution is to completely rebuild the 
Inner Detector systems of ATLAS and CMS.}

Simple extrapolations based on the available experimental data show
that current detector technologies, with some new developments, could
work at a factor 3 larger radius, i.e. strips at R $>$ 60~cm, pixels
at R $>$ 20~cm.  This can also be seen from Table~\ref{tab:exptab1},
where the fluences at these radii correspond to the currently foreseen
fluences for strip and pixel detectors at the LHC. As a result of these
rather straightforward considerations the tracker volume can be split
into three radial regions:

\begin{itemize}
\item R $>$ 60 cm: where detectors can be built by further pushing existing
silicon strip technology. 
\item 20 $<$ R $<$ 60 cm: where further developed hybrid pixel technology can
work
\item R $<$ 20 cm: where most likely new approaches and concepts are required. 
\end{itemize}
{\bf The R\&D programs should be focussed towards upgrades of ATLAS and CMS.} 
For both experiments it is important to establish common guidelines
and requirements for the relevant studies from the beginning. This
implies good overall co-ordination of the R\&D and good contact between
the various R\&D groups. There are in particular two areas where common
understanding and ground rules must be established with dedicated
studies:

\begin{itemize}
\item {Common software and physics studies}: pattern recognition,
  fluences, activation, occupancy, cell size optimisation,
  segmentation (e.g. macro pixels, mini strips), radius of the
  innermost layer (driven by b-jets), material effects on the key
  physics channels to be studied at SLHC.
\item {Tracker system studies}: understand the timing,
  synchronisation, readout/trigger rates and limitations, engineering
  issues (see Section~\ref{sec:elec}).
\end{itemize}
This focus is necessary to make appropriate choices and set priorities
relatively early in the R\&D phase.
In the following Sections we discuss these three regions and propose and
motivate various R\&D programs for suitable tracking detectors.

\subsubsection{Tracking at a radius greater than 60 cm.}

Silicon micro-strip detectors can be used to instrument the outer
tracker (60 $<$ R $<$ 110 cm) in a layout similar to that chosen by CMS.

Six layers of silicon micro-strip detectors with read-out pitch ranging
between 80 and 160 $\mu$m would be good enough to cope both with the
occupancy expected for operation at \slhclum\ and have a point
resolution needed for good transverse momentum resolution. By
extrapolating estimates of occupancy in the CMS Si tracker at nominal 
LHC the global occupancy is expected to vary from less than 1 \% in the outer
layers to a maximum of a few \%. In this region a 50 \% increase in number
of channels would probably be needed compared to the current CMS strip 
tracker.

The fluence of fast hadrons in this region for an integrated luminosity
of 2500\ifb\ is estimated to be between 1 and 
$3 \times 10^{14}$cm$^{-2}$. 
Extensive
studies have been done in preparation of the current generation of LHC
inner trackers to validate the use of silicon detectors up to integrated
fluences of $1.5 \times 10^{14}$cm$^{-2}$  \cite{expref1, expref2}. In 
particular it has been demonstrated
that the current high breakdown technology ($V_{break} > 500$ V) guarantees
safe operation of these devices for the entire LHC lifetime.

For SLHC many of the characteristics of what are considered to be
standard ``radiation resistant'' silicon micro-strip detectors will be
maintained:  p$^{+}$n technology, integrated AC coupling, poly-silicon bias
circuit, $\langle 100 \rangle$ 
crystal orientation, standard 0.20-0.25 w/p ratio and
metal overhang on the strips. A careful study will be needed on the
charge collection efficiency in heavily irradiated devices read-out by
very fast shaping time electronics. A focused R\&D program will be needed
to optimise the Si-strip detector characteristics, performance and cost. 

A factor of 2 higher fluence in SLHC would require further improvement
in breakdown voltage in industrially produced devices though use of
lower resistivity silicon and thinner devices could lower the bias
voltage at full depletion. The choice of wafer resistivity and thickness
will have an impact on system aspects (detector module layout and strip
length, read-out granularity, noise performance of the front-end
electronics, cooling needs etc) and on the total cost (number of
channels, number of detectors, technology and cost of processing).

The microelectronics industry is rapidly migrating to 8'' (and then to
12'') processing lines. It is likely that the currently used 6'' lines
will not be available in 6-8 years from now.  As a consequence a part of
the R\&D program should be devoted to the exploration of the feasibility
of processing of detector grade Si wafers with a larger diameter, and
to transfer the existing detector processing technology to the new 
lines. 

\subsubsection{Tracking between a radius of 20 cm and 60 cm.}
  
Today's pixel technology is expected to work at radii
above 20 cm. Questions related to the infrastructure, services, power and
cost are likely to be decisive in determining the minimum radius at
which the pixels can be deployed at the SLHC. 

It is possible to marry the current pixel architecture (both sensor and
electronics) for cell-sizes that are ten times bigger than the current
pixels but ten times smaller than the size  of the current microstrips
in this region. Hence a critical goal of the necessary R\&D would be to
achieve a cost/pixel that is between 10 and 100 times smaller than the
current cost/microstrip. Such devices can be labelled 'macro-pixel'
devices. Many issues such as routing will need to be addressed. Such
devices could already figure in the upgrades of innermost silicon strip
detectors of current LHC trackers. 

Defect engineered silicon (section 5.1.4.a) is already used in the current
LHC trackers, in particular to reduce damage due to charged hadron
irradiation at low radius. Defect engineered silicon has the potential
of reducing the radius at which current technologies can be
used at SLHC.

\subsubsection{Tracking at a radius smaller than 20 cm.}

At the LHC design luminosity the innermost pixel detectors are expected to 
be placed at a radius of 7 cm from the beamline. The occupancy at 7
cm is estimated to be $\simeq$ $3 \times 10^{-4}$. At SLHC, in order to 
preserve the occupancy at a tolerable level at such radii, the pixels area 
should be decreased by at least a factor of 5. The b-tagging 
performance should then not be degraded.
  
At SLHC, and at a radius of $\simeq$ 7 cm, the dose and fluence of hadrons is
expected to be 100 kGy and $5 \times 10^{15}$ hadron/cm$^{2}$ respectively. 
Hence this is an extremely harsh region. Short of changing the hybrid pixel 
detectors annually, or perhaps even more frequently, there are currently 
few solid and demonstrated possibilities. Hence fundamental research rather
than only development is needed. New concepts, geometries and materials are 
probably needed to attain the required speed and radiation hardness. 

The pixel systems in the current trackers will possibly be changed
during the LHC period and some of the R\&D mentioned above will also be
very relevant for these upgrades.

\subsubsection{Subjects for R\&D}
{\bf a) Use of defect-engineered silicon}
\\
The term ``defect-engineering'' stands for the deliberate incorporation of
impurities or defects into the silicon bulk material before or during
the processing of the detector. The aim is to suppress the formation of
microscopic defects that have a detrimental effect on the macroscopic
detector parameters during or after irradiation. In this sense defect
engineering tries to cope with the problem of radiation damage at its
root. 

One example of a defect-engineered material is the ``Diffusion Oxygenated
Float Zone'' (DOFZ) silicon developed by the ROSE (CERN RD48)
collaboration \cite{expref3}. It was shown that this oxygen-enriched material
exhibits an improved radiation tolerance with respect to charged hadron
irradiation. The increase of depletion voltage after type inversion is
reduced by a factor of three and the so-called ``reverse annealing''
saturates at fluences above about $2 \times 10^{14}$cm$^{-2}$ 
(24 GeV/c protons).
Furthermore, the reverse annealing is slowed down allowing for longer
warm up (maintenance) periods. These properties and the relatively
simple and cheap implementation into detector processing make DOFZ an
ideal material for the detectors located closest to the interaction
points. After thorough testing the ATLAS collaboration is now using it
for the pixel layers. It is anticipated that the CMS pixel collaboration
will follow the same path. 

It is expected that further optimisation studies of the oxygenation
process will lead to better results with respect to radiation tolerance.
Furthermore it is worth exploring other promising possibilities for
defect-engineering \cite{expref4}.
\\[0.3cm]
{\bf b) 3D detectors and new biasing schemes}
\\
The main characteristic of '3D' detector concept is to place the
electrodes (p,n) throughout the bulk in the form of narrow columns
instead of being deposited parallel to the detector surface. In a
conventional silicon sensor the depletion and charge collection across
the full wafer thickness (usually 300 $\mu$m) requires high voltages and
becomes incomplete after intense irradiation. The main advantage of the
3D approach is the short distance between the collecting electrodes.
This allows very low depletion bias voltage ($\simeq$ 10V) as well as very 
fast collection times and low noise. The principle has been successfully
proven on small prototypes operated after irradiation with protons up to
$2 \times 10^{15}$n$_{1MeV~eq}/cm^{2}$ \cite{expref5}. Other approaches are 
also under study \cite{expref6}.

The 3D concept is a very new approach. Considering that above $10^{15}$
particle/cm$^{2}$ only electrons contribute to the generated signal, due to
the rapid trapping of the holes and the reduced effective drift length
of the electrons, '3D' concepts potentially are probably 'rad-hard' up
to fluences of $5-10 \times 10^{15}$ particles/cm$^{2}$ if collection is 
performed at the n$^{+}$-electrode. Moreover, if '3D' concept would be 
combined with defect engineered material and operated in forward bias at 
cryogenic
temperatures, radiation tolerance might be achieved for even higher
particle fluences. Size and uniformity of the sensors are important
issues to be addressed together with optimisation of the manufacturing
techniques.
\\[0.3cm]
{\bf c) New sensor materials}
\\
CVD diamond is almost an ideal material for radiation hard charged
particle detectors. Its outstanding radiation tolerance, fast charge
collection, low dielectric constant, and low leakage current make it a
good candidate for high luminosity colliders.  

The RD42 collaboration has demonstrated that small area strip and pixel
detectors based on diamonds can be fabricated; they can collect signals
corresponding to about 8,000 e$^{-}$ and work up to a proton fluence of
$5 \times 10^{15}$ p/cm$^{2}$ \cite{expref7}.
 
A complete qualification of the technology would require a strong
partnership with the producers of the raw material to increase the
collected charge and to perform basic research on defects and
impurities, both before and after irradiation. In terms of processing
techniques an optimization of the electrical contacts is needed. Further
studies would be needed on the characterization of complete detector
modules. 
\\[0.3cm]
{\bf d) Cryogenic Silicon Tracker development}
\\
The main motivations for silicon detectors operated at 130~K are:
\begin{itemize}
\item a factor of 10 increase in radiation hardness due to the 'Lazarus'
effect \cite{expref8};
\item a factor 5 higher mobility for both carriers which leads to very fast
sensors;
\item negligible bulk and surface current generation rate at high voltage,
even after substantial radiation damage.
\end{itemize}
In addition the factor 3 increase in the thermal conductivity of silicon
between 130~K and 300~K would facilitate significantly the engineering
design of the detector modules as far as evacuation of heat is
concerned.

Results of the RD39 collaboration show that prototype modules can be
fabricated with simplified techniques and successfully operated at very
high SPS lead ion fluences of $(5 \pm 2) \times 10^{14}$ ions/cm$^{2}$ 
yielding almost 100 Grad energy deposit \cite{expref9}. It was also shown 
that the development of fast
low-noise cryogenic read-out electronics is feasible, and that the basic
engineering issues of operating complex systems of low temperature
detectors can probably be addressed. 

The basic advantage of this approach is that, once proven to be feasible
and reliable on the large scale required for the SLHC trackers, it could
be used for all the three radial regions. The need for further
replacement will then be mainly motivated by the limits of radiation
resistance of the DSM electronics.

From the point of view of the engineering design, it would be attractive
to close the ends of the cavity that would house the entire tracker in
cryogenic vacuum. Microtubes integrated in the module design would
evacuate the heat produced locally by the electronics. The clean vacuum
would help in avoiding contamination of the detector components, and a
very thin-walled beam pipe would then be feasible.

While the cryogenic micro-strip sensors and modules are already being
developed with promising initial results, there is much work still to be
done in optimisation and testing of the pixel devices and DSM front-end
electronics for low-temperature operation. At present the
current-injected or forward-biased cryogenic sensors can be operated up
to the fluence of $2 \times 10^{15}$cm$^{-2}$ in the strip segmentation, 
and perhaps 5
times higher in the pixel devices. These results were obtained, however,
using simplified devices; it is clear that further R\&D is required for
large area detectors to be produced with high yield and then assembled
in an automated manner. The device structures and current injection
schemes should be further studied and optimised. Low-temperature studies
of the bulk silicon under high damage should also be pursued, with view
on possible optimisation of the defect neutralization and charge carrier
trapping.

The engineering studies of the cryogenic tracker would also be a
critical issue for the R\&D.
\\[0.3cm]
{\bf e) Monolithic Pixel Detectors}
\\
In a monolithic pixel detector a two-dimensional array of detecting
diodes and the associated miniaturized readout electronics are
integrated on the same silicon substrate. Compared to the hybrid
approach, in which the arrays of sensing elements and readout cells are
manufactured on different wafers, which are then bump-bonded together,
the monolithic approach presents clear advantages in terms of detector
assembling and handling. 

Another important advantage of the monolithic technology is the
reduction of the amount of matter to be traversed by the particles.
Moreover, the very small capacitance (down to few fF) presented by the
detecting element at the input of the close-by front-end transistor
result in reduced noise.

Monolithic pixel detectors will only be attractive if standard
technologies can be used to keep costs affordable. A key issue is to
understand whether fast switching front-end electronics needed for the
SLHC can be integrated on the same piece of silicon with the detecting
element. Recent tests of such sensors are promising though it is not
clear whether the required level of radiation hardness will be
achievable.

\subsubsection{Engineering aspects} 

For the support structures a highly modular approach based on carbon
fibre composite elements seems to be appropriate to cope with the
increased radiation levels of SLHC. Moreover a large part of the
existing outer supporting structures could probably be copied. 

An increased number of channels and higher radiation levels would
increase cooling needs. Since the requirements will be driven by the
inner parts of the tracker, and a unique operating temperature for the
entire tracking volume is advisable, the most likely operating scenario
foresees an overall temperature T $\simeq~ -15/-20^{0}$ C. We presume that 
the currently foreseen cooling techniques could be used.

The power requirements for the front-end electronics should be defined
early to enable implementation of radiation-hard local voltage
regulators. Their use is mandatory in order to reduce the mass of power
cables which otherwise would be one of the dominant contributors to the
material budget.

Extensive use of high density interconnections, low mass hybrids and
flexible circuitry will be necessary to reduce the amount of material
for the read-out and ancillary electronics. 

It will be important to maintain some level of accessibility for
maintenance of critical detector components. 

\subsubsection{Electronics} 
See Section 5.5

\subsection{Calorimetry}

The chosen calorimeter technologies of ATLAS and CMS \cite{expref10,
expref11, expref12, expref13} are designed to 
withstand an integrated luminosity in excess of500 pb$^{-1}$. 
ATLAS has chosen liquid
argon sampling calorimetry for the electromagnetic, the endcap hadronic
and the forward calorimeters. The barrel hadronic calorimeter comprises
iron/plastic scintillator sandwich with wavelength shifting fibres. CMS
is using scintillating crystals for the electromagnetic, and brass/plastic
scintillator sandwich with wavelength shifting fibres for the hadronic 
calorimeter. The forward calorimeter uses quartz fibres embedded in grooves 
in iron plates. 
The neutron fluence and radiation dose at shower maximum and at
different pseudorapidities for an integrated dose of 2500 fb$^{-1}$ are given
in Table~\ref{tab:exptab2}.

\begin{table}
\begin{center}
\caption{The neutron fluence and radiation dose at shower maximum at 
different pseudorapidities for an integrated luminosity of 2500 fb$^{-1}$.}
\label{tab:exptab2}
\vspace*{0.1cm}  
\begin{tabular}{|c|c|c|c|}\hline
 Pseudorapidity  & ECAL Dose  & HCAL Dose & ECAL Dose Rate \\
 $\eta$         & (kGy)      & (kGy)     & (Gy/h)         \\  
\hline
0 - 1.5 &   15 &      1  &      2.5 \cr\hline
2.0     &  100 &     20  &      14  \cr\hline
2.9     & 1000 &    200  &     140  \cr\hline
3.5     &    - &    500  &      -   \cr\hline
5       &    - &   5000  &      -   \cr\hline
\hline
\end{tabular}
\end{center}
\end{table}
Below we consider the possible limitations of these techniques and
calorimeters for a 10-fold increase in instantaneous luminosity, and a 
5-fold increase in integrated luminosity.

\subsubsection{Liquid Argon Calorimeter}

The ATLAS Calorimeter was optimised for the nominal LHC luminosity of
\lhclum\ and a centre of mass energy of 14 TeV. A factor of 10
increase in this luminosity would raise concerns that are
considered below.
\\[0.3cm]
{\bf a) Space charge effects.}
\\
 During steady operation of the calorimeter, an equilibrium is reached
between ion creation by the passage of charged particles and ion
collection by the electric field in the gaps. When the positive ion
space density integrated over the gap becomes comparable to the charge
density on the electrodes due to the HV polarisation, the field is
distorted, and a distortion of the response may occur. Fortunately, no
practical change in the response occurs until a region near the anode
with zero field appears.  The onset of such a regime goes like
$V^{2}/d^{4} \mu$, where V is the operating voltage, d the gap and 
$\mu$ the Ar$^{+}$ mobility. This
last quantity is not very well measured experimentally, and so far we
have had more confidence in checking directly the onset of some
saturation. Measurements in test beam using prototypes of the ATLAS EM
calorimeter \cite{expref10} show that losses at the level of 1 \%
occur for an energy flux of about $5 \times 10^{6}$ GeV cm$^{-2}$s$^{-1}$. 

Table~\ref{tab:exptab3} compares the ``critical density'' with the energy 
density in
various parts of ATLAS liquid Argon calorimeter for two values of the
luminosity. The critical density in the various areas is scaled from the
measured number, taking into account the actual geometry and the sampling
fraction, as well as the shower extension in length. A major step occurs
for the FCAL with gaps as narrow as 0.25 mm in a dense tungsten matrix

\begin{table}
\begin{center}
\caption{Comparison of the critical density with the energy density for 
         ATLAS liquid argon calorimeters}
\label{tab:exptab3}
\vspace*{0.1cm}  
\begin{tabular}{|c|c|c|c|}\hline
                & Critical density  & ATLAS 10$^{34}$ & ATLAS 10$^{35}$ \\
\hline
Barrel EM, $\eta$=0    & $5 \times 10^{6}$     & $0.5 \times 10^{5}$ & 
$5 \times 10^{5}$ \cr\hline
Barrel EM, $\eta$=1.3  & $4 \times 10^{6}$     & $1.2 \times 10^{5}$ & 
$1.2 \times 10^{6}$ \cr\hline
End-cap EM  $\eta$=1.4  & $3 \times 10^{6}$     & $1.3 \times 10^{5}$ & 
$1.3 \times 10^{6}$ \cr\hline
End-cap EM  $\eta$=3.2 & $5 \times 10^{6}$     & $2.5 \times 10^{6}$ &
$25 \times 10^{6}$  \cr\hline
FCAL $\eta$=3.2        &  $1500 \times 10^{6}$ & $2.5 \times 10^{6}$ & 
$25 \times 10^{6}$  \cr\hline
FCAL $\eta$=4.5        &                       & $130 \times 10^{6}$ & 
$1300 \times 10^{6}$ \cr\hline
\hline
\end{tabular}
\end{center}
\end{table}
The numbers in the Table indicate a comfortable margin in the barrel, while the
inner parts of the EM endcap, and of the FCAL may be affected.
Ways to stay away from limits set by the space charge effects have to be
investigated. These could involve different liquids (perhaps liquid
krypton),or a cold dense gas under pressure (compatible with what the
cryostats can withstand).
\\[0.3cm]
{\bf b) Voltage drop in the HV distribution}
\\
The current induced by the drift of electrons and ions in the gap,
circulates in the HV polarisation chain, which incorporates resistors to
isolate from each other channels hooked to the same HV supply. The value
of these resistances should not be too low in order to avoid cross-talk,
and not too high such that the induced drop would require a rate
dependent correction. 

\begin{table}
\begin{center}
\caption{The voltage drops expected in ATLAS liquid argon calorimeters}
\label{tab:exptab4}
\vspace*{0.1cm}  
\begin{tabular}{|c|c|c|c|c|}\hline
 & Resistance/0.05 & Current at 10$^{34}$ & Voltage drop & Voltage drop \\
 &                 &                      &   $10^{34}$  &  $10^{35}$   \\ 
\hline
Barrel EM,  $\eta$=0   & $\sim$ 1 Mohm &    80 nA &  0.08 V &      \cr\hline
Barrel EM,  $\eta$=1.3 &               &   200 nA &   0.2 V &  2 V \cr\hline 
End-cap EM, $\eta$=2.4 &               &   400 nA &   0.4 V &  4 V \cr\hline
End-cap EM, $\eta$=2.5 &               &  4000 nA &   4.0 V & 40 V \cr\hline
End-cap EM, $\eta$=3.2 &               &  8000 nA &   8.0 V & 80 V \cr\hline
\hline
\end{tabular}
\end{center}
\end{table}

The value of the resistances is about 10 times larger at LAr temperature
than at room temperature. This factor $\simeq$ 10 has large fluctuations from
pad to pad (rms/average $\simeq$ 0.3) which precludes using measurements made 
at room temperature  to correct for the drop.
 
The expected voltage drops are given in Table~\ref{tab:exptab4}. As for 
the ion
build-up there is a comfortable margin for the barrel. The ``small wheel''
of the EM endcap, because of its smaller number of electrodes (256
against 768) sees a large current per sector. Significant effects are
expected in this area, which would not allow precision measurements.

In order to get rid of the limitation induced by this effect, a
different liquid should be explored, with less charge deposited per GeV
or one that is at higher temperature. A cold dense gas should also be
evaluated in this respect.
\\[0.3cm]
{\bf c) Activation}
\\
While this affects the ``logistics'' and not the detector response itself,
activation may become a very serious limitation for the practical use of
the detector. In particular for ATLAS one would have to reconsider the
level of Ar$^{40}$ that can be released into the atmosphere in case of
evacuation of the liquid due to a fault in the cryogenic system.
\\[0.3cm]
{\bf d) Radiation damage to the detector}
\\
There is a comfortable margin in this respect. The only measurable
effect found so far is a small increase of the polarisation resistances,
that are silk screened on electrodes (see section b), under neutron
irradiation. 
\\[0.3cm]
{\bf e) Radiation damage to the electronics}
\\
Apart from the HEC cold preamplifiers, all active circuits are in warm
environment, and therefore accessible and replaceable if needed. By
design all ASICs (including the HEC GaAs preamplifiers) were
manufactured in a technology offering a large safety margin w.r.t. the
expected radiation level at the nominal high luminosity of LHC.

Nevertheless a dedicated analysis would be needed to evaluate the
potential problems for 10 times more radiation. Particularly critical
may be the case of the few COTS  (commercial components) used in the
front-end crates.
\\[0.3cm]
{\bf f) Sequencing of the readout}
\\
The effect of shorter bunch spacing, such as 12.5 ns, has to be
evaluated though the current scheme may well be adequate. In ``standard''
ATLAS conditions calorimeter pulses are sampled every 25 ns, and the
energy and time are calculated using ``optimal filtering'' with data from
5 consecutive samples. Timing is arranged such that the third sample
falls at the maximum of the pulse.

In the test beam, which is asynchronous, all events in a window of $\pm$
12.5 ns around the optimum timing are used, and give results
indistinguishable from those sampled at the peak.  In the case of 12.5
ns bunch spacing, one could therefore continue to clock the readout at
40 MHz, requiring only that the LVL1 identifies whether the triggered event
is synchronous with this clock or advanced by 12.5 ns.
\\[0.3cm]
{\bf g) Formation of LVL1 signals}
\\
The bandwidth of the present LVL1 authorises `` BCID'' for all signals
above a given threshold. To obtain a similar ``BCID'' at 80 MHz is likely
to require an upgrading of this system. Since everything is accessible,
such a change would cost money but should not be an {\it a priori}
show-stopper
\\[0.3cm]
{\bf h) Level of pileup}
\\
With no change to the circuits nor to the signal treatment, a factor of
10 in luminosity translates into a factor of 3 in pile-up, while
electronics noise in unchanged. An optimal filtering could allow to
``speed-up'' the equivalent response, for a better balance between
electronics and pile-up noise, leading to a smaller degradation.

The use of a different liquid, or cold dense gas, would indeed require
a full re-evaluation of the signal to noise ratio, and in general of the
calorimeter performance.

In conclusion, an increase of luminosity would certainly create space
charge and voltage drop problems in the large $\eta$ region of the Endcap
electromagnetic calorimeter ($\eta>$2.5)  and the forward calorimeter (FCAL).
R\&D will be needed to understand better these effects, and to
investigate the use of other liquids or of a cold dense gas.

\subsubsection{CMS Crystal ECAL} 
{\bf a) Crystals}
\\
In case of an order of magnitude luminosity increase, the dose rate in
the Barrel (at shower max) will increase from 0.15Gy/h to 1.5 Gy/h,
which corresponds to the current nominal situation at $\eta$=2.4. There should
therefore be few problems. For the endcaps, the dose rate reaches 30
Gy/h at $\eta$=2.6 and 75 Gy/h at $\eta$=3. This is close to the ``saturation'' 
irradiation conditions actually used at the Geneva Hopital Cantonal (250
Gy/h). It is known that the light attenuation in this condition is $>$ 1
meter and therefore the light loss is less than 25 \%. 

However, for a complete understanding, a programme of irradiations under
these conditions should be performed including

\begin{itemize} 
\item  long-term irradiation  (days) 
\item  irradiations with high fluxes of hadrons, for comparison with gammas.
\item  calibration studies 
\end{itemize}
{\bf b) Photosensors}
\\
The leakage current in the APDs used in the Barrel will increase by
approx 20 $\mu$A per year at SLHC. This should translate
into a large increase in electronics noise, reaching 100 MeV per crystal
after a few years. Improved recovery mechanisms (e.g. higher temperature
during shut down) could be investigated.
 
For Endcap VPTs the glass window has to be tested at the expected very
high radiation doses. The behaviour of the tube with a strong steady
current should also be studied.
\\[0.3cm] 
{\bf c) On-detector Electronics}
\\
{\it Radiation hardness.}
\\
The electronics is built using radiation-hard processes that are
qualified for the nominal situation in the endcap (5Mrad, 
$2 \times 10^{14}~n/cm^{2}$
for an integrated luminosity of 500 fb$^{-1}$). One can therefore conclude
that the electronics in the Barrel will survive an order of magnitude
luminosity increase.
The situation for the endcaps is much more critical. The actual
electronics has already been recessed from the beam axis to limit the
maximal fluence to $2 \times 10^{14}~n/cm^{2}$.  An increase by a factor 10 
requires either a move from the actual position to the periphery -which is
probably impossible - or a replacement after a vigorous R\&D to find more
radiation hard technologies. This affects all the components of the
Front-End cards (preamplifier, ADC, optolinks)
\\
{\it Faster bunch crossing}.
\\
Running the actual electronics at 80MHz is impossible.  As in the case
of the ATLAS liquid Argon calorimeter, the first thoughts are  that one
could cope with a doubling of the frequency (12.5 ns bunch crossing)  by
still sampling at the original 40 MHz frequency. The excellent time
resolution obtained with the multi-sample electronics would allow a
corect assignment of the bunch crossing The LV1 trigger primitives are
created by filters using 5 consecutive samples. The consequences on the
actual CMS Trigger system have to be studied. 
\\[0.3cm] 
{\bf d) Pileup}
\\
Pileup effects have to be assessed, both from the physics point of view
and also for the electronics (for example the variation of the base line
for the Floating Point Preamplifier). The last point may be an issue in
the high $\eta$ range.
\\[0.3cm] 
{\bf e) Activation}
\\
One of the biggest worries is the activation level that would be reached
in the endcaps,; certainly several mSv/h after a run at a luminosity of
\slhclum. This means that interventions will be very 
difficult
(the integrated allowed yearly dose being reached  in a few hours).  One
should not envisage regular replacements of hardware in these
conditions.

The results of the radiation hardness tests suggested above for VPTs and
for electronics are therefore crucial.  One should note that already the
replacement campaign for the endcaps electronics before the upgraded 
Luminosity period  will require careful planning. 

In summary the CMS ECAL Barrel could probably be used with an increase
of luminosity, even if the performance may be somewhat  degraded due  to
an increase in noise and pileup. The situation for the Endcap is more
difficult to assess: R\&D is required to verify the behaviour of the
crystals and photosensors under high dose rates. The Front End
electronics will have to be replaced in difficult activation
conditions.  

\subsubsection{Plastic Scintillator Based Hadron Calorimeters}
CMS employs a brass/plastic scintillator sampling hadronic calorimeter up
to $|\eta |<3$. ATLAS uses a similar technique in the barrel region. The
plastic scintillator used in these calorimeters loses half of its light
output after a dose of about 50 kGy. Hence, from Table~\ref{tab:exptab2}, 
it can be seen that the ATLAS and CMS hadron calorimeters in the barrel region 
($|\eta |<1.5$) should not need changing.
 
The situation is more difficult for the CMS endcap hadron calorimeter.
The deleterious effects of radiation can be substantially mitigated by
individully reading out the scintillator layers in the first 3-4 
interaction lengths. The signals from these layers can then be weighted to
compensate for the loss of light. Periodic replacement of the scintillator,
albeit difficult, could also be envisaged. A programme of R\&D should be
undertaken to search for an alternative active medium and for a more 
radiation tolerant scintillator. 

\subsubsection{CMS Very Forward Calorimeter}

The radiation dose in the forward region (3 $<~|\eta |~<$ 5) changes rapidly.
The CMS iron/quartz fibre calorimeter uses plastic-clad quartz fibres.
Quartz -clad quartz fibres are more radiation resistant but are much
more expensive. Replacing plastic-clad fibres by quartz-clad fibres,
will allow the use of the same technique up to integrated luminosity of
2500 fb$^{-1}$. 
  
Novel technologies that can operate at ultra high radiation levels
should be searched and developed. A possibility is to detect the
Cerenkov light emitted in pressurised gas contained in 2 mm diameter
steel tubes whose  reflectivity on the inner surface at grazing
incidence is very high. The idea has been tested in CERN-H4 beam
\cite{expref15}

\subsection{Muon Systems}
\subsubsection{Intensity considerations}
{\it Background radiation}
\\
The ATLAS and CMS muon systems \cite{expref16, expref17} have been designed 
according to
conservative assumptions in the background rates (factor 3 to 5 safety
margin above estimates from simulations). The real safety margin can
only be established from measurements once LHC operates. It is possible
that in some regions measured rates will be substantially higher than
current expectations and this information, available only five years hence, 
will influence the modifications necessary for SLHC. 

The particle fluence scales with luminosity and is an even stronger
function of $\eta$ and r (factor of 100 variation). It is predominantly
composed of low energy ($<$ 100 MeV) neutrons, high energy neutrons and
photons (with typical energy $<$ 10 MeV). At SLHC, the fluence of each
species ranges up to some $10^{5}$cm$^{-2}s^{-1}$ in the highest 
$\eta$ region. The detection efficiency for these particles is in the range 
of 0.1-1 \%.  They
dominate the observed hit rate at low $\eta$ ($\leq$ 2.2). The rate for 
charged particles (hadrons, muons, isolated electrons) is typically much 
lower than the rate of detected neutral ones, except in the high $\eta$ 
region (2.2-2.7), where it progressively becomes dominant.
 
When considering technologies for use at SLHC, it is worth pointing out
that the strong geometric dependence implies that {detector types which
function at high-$\eta$ in LHC will certainly work quite adequately at 
low-$\eta$ in SLHC}.
\\[0.3cm]
{\it Shielding and muon tracking in the high-$\eta$ region}
\\
{The simplest viable modification for SLHC is to increase shielding
around the beam-pipe at high-$\eta$}, 
which reduces the overall background
rate and is particularly effective at low-$\eta$, where neutrals dominate 
the observed hit-rate in the muon detectors. The penalty is a cut in the
high-$\eta$ acceptance. However, the rate from charged particle background
(20-40 \% muons), effectively irreducible by shielding, may in any case
limit the deployment of detectors in the forward regions, and force a
reduction in the acceptance at  SLHC. 
\begin{figure}
\begin{center}
\includegraphics[width=0.85\textwidth,clip]{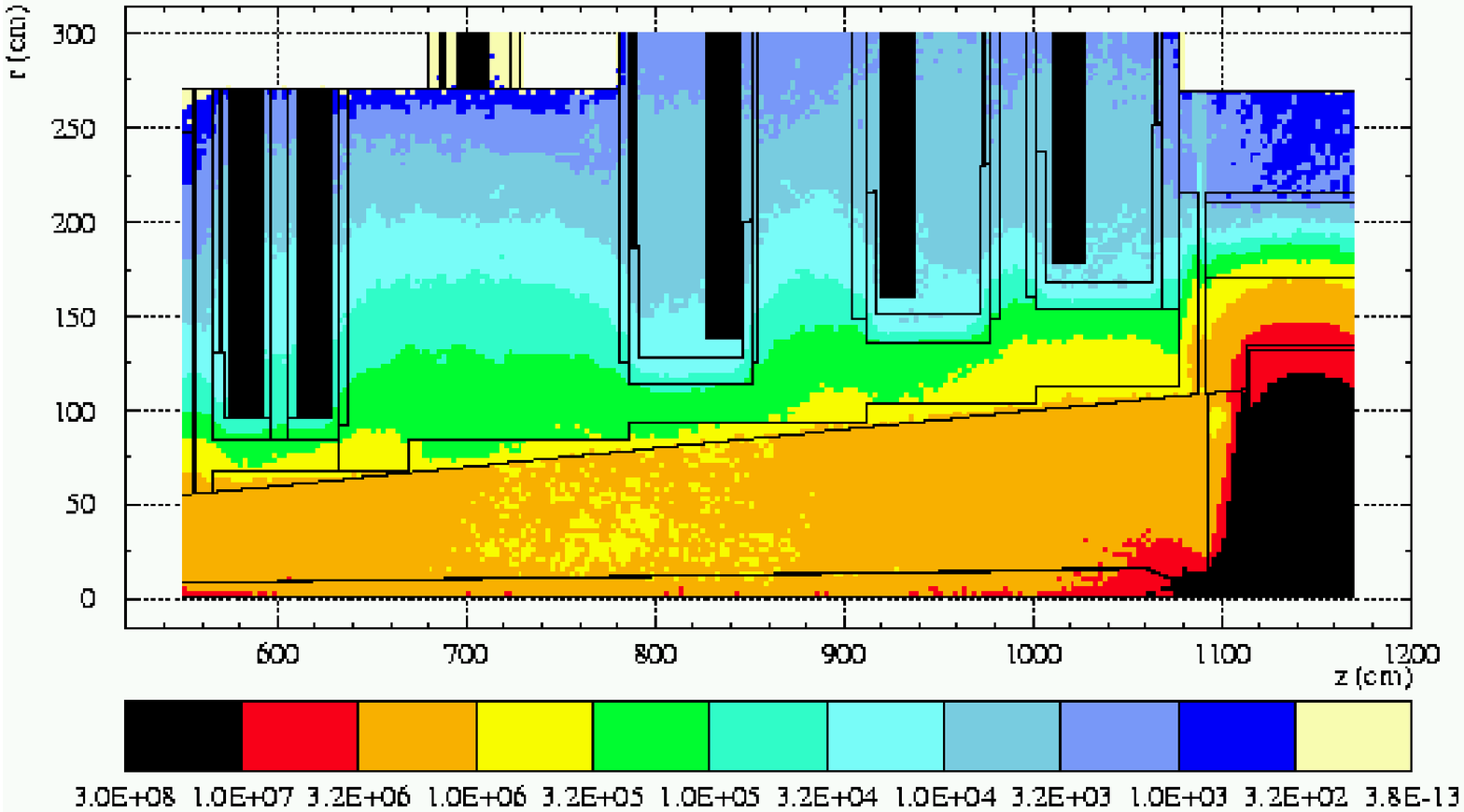} \\[0.3cm]
\includegraphics[width=0.85\textwidth,clip]{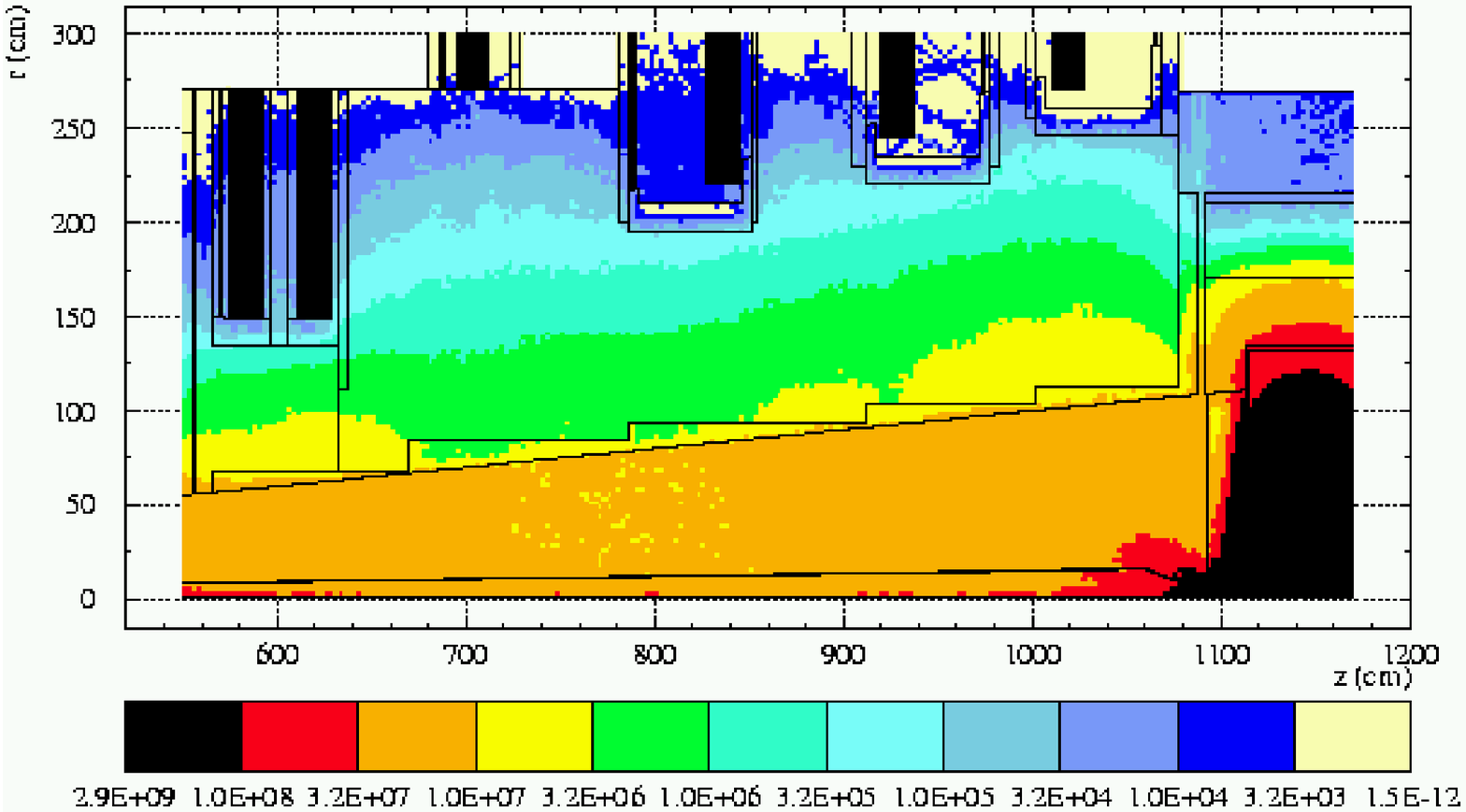} 
\end{center}
\caption{Upper: the neutron fluxes ($\rm{cm}^{-2}s^{-1}$) 
in the low radius, high $\eta$-region of the
CMS endcap muon detector for $\eta<2.4$ and present LHC shielding
\@ \lhclum. Lower: same, for for $\eta<2$ and possible shielding for
SLHC \@ \slhclum. \label{fig:expfig2}}
\end{figure}

Figure~\ref{fig:expfig2} compares the contours of fluence in the low radius, 
high-$\eta$ region of
the CMS detector, for a high-$\eta$ cut-off of 2.4 (present LHC shielding)
and 2.0 (possible shielding for SLHC).  The maximum rates to which the
cathode strip chambers are exposed are similar in the two cases, thus a
re-build of the low radius muon layers to match the new geometry could
be done using the existing technologies. The CMS resistive plate
chambers used in triggering are already limited to the $|\eta|~<~$ 2.1 region.
Depending on the exact configuration of machine
elements chosen for the SLHC low-$\eta$ insertions (see below), re-design of
the shielding around the TAS collimator, and around the beam-pipe within
the forward muon system might also imply significant engineering
re-design of the entire forward regions of the experiments, for instance
the forward toroids of ATLAS or the forward calorimetry system of CMS.
All this adds to the uncertainty in the rate estimates, and it is not
inconceivable that in the worst case the first forward muon stations
might need to be replaced by a different design concept. However, the
aim would be to use developments of technologies already proven at LHC
in a worse environment, eg those that were candidates for central
tracking.
\\[0.3cm]
{\it Tracking in the low-$\eta$ region ($\eta~<$ 2)}.
\\
Assuming that an effective forward shielding design can be maintained,
the current muon detectors in the central region $\eta<$ 2, will mostly
perform well enough. However, the insensitivity to ageing at integrated
dose levels higher than considered so far may need to be confirmed in
some cases. The expedient of substituting high-$\eta$ technologies in the
most vulnerable low-$\eta$ regions, could lead, in the CMS case, to
replacement of drift tube chambers by endcap-type cathode strip chambers
in the first and fourth barrel stations. However, the detector
performance may be otherwise affected by the increased background -
e.g., the resolution of the ATLAS MDTs may be limited by space charge
effects (which may already reduce performance at the highest $\eta$ at LHC).
Experience from LHC operation is necessary to make an informed
judgement. 

\subsubsection{Trigger}
The design of the trigger will need a significant
upgrade to cope with the increased intensity.   In particular, rejection
ability considerations will force the trigger to be driven at the
increased bunch-crossing rate of 80 MHz. Different considerations apply
to the different detector technologies:
\begin{itemize}
\item For intrinsically fast detectors  (i.e. with signal generated with
time jitter $<$ 12.5 ns), where the trigger decision is taken
asynchronously (or over-sampling above 40 MHz) (e.g.: the ATLAS RPCs),
part of the existing trigger might be usable at SLHC.
\item For fast detectors with trigger logic effectively driven by the LHC 40
MHz bunch-crossing rate (e.g. ATLAS TGCs, CMS Drift Tubes); the trigger
logic would need to be redesigned to operate at 80 MHz.
\item For slow detectors, with signal time jitter comparable or above 12.5
ns  (e.g. the ATLAS TGCs in the largest $\eta$ region: for tracks at large
incidence angle); a different detector technology, or at minimum a more
sophisticated electronics, might have to be considered in order to
operate in these regions.
\end{itemize}

\subsubsection{Read out}
The detector read out would have to cope with increased bandwidth, due
to higher background rates, and possibly to higher trigger rate.  The
actual read out speed may or may not need to be upgraded (e.g., the
drift tubes could presumably be read out with the same speed and memory
depth). 

There may be sufficient impact on  power dissipation in certain
detectors to require an upgraded cooling system. Even if most existing
detector elements are usable at luminosities around \slhclum, the time
taken to replace on-board electronics and cooling  may be a major
contribution to the shutdown length needed to re-configure experiments
for SLHC.

\subsubsection{Beam optics and radio-activation}
The increase in luminosity would be achieved in part by doubling the
bunch-crossing rate, in part by an increase in the bunch intensity and
in part by decreasing the beam cross section ($\eta ^{\ast}$). A new design 
of the
beam optics might require moving the quadrupole triplet closer to the
interaction point or installing a D1,3Q,D2 configuration with the last
dipole very close to the interaction point. Either may imply a major
re-design of the collimators/shielding, and would affect the detector
layout at the large $\eta$ limit. For ATLAS, this would add reasons for
reducing the angular acceptance, and for modifying the design of the
forward quadrupoles. For CMS, this would probably force integration of
the forward calorimeter within the forward muon region and might force a
further reduction in the acceptance.

Activation of shielding and supports might limit the access-time to the
detectors, placing constraints on installation and maintenance
scenarios. 

\subsubsection{Conclusion on the muon systems}
The modifications needed for the SLHC to the muon systems of ATLAS and CMS
should be determined by a cost-benefit analysis
depending on the perceived physics potential in the light of results
from LHC at \lhclum. Benchmarking of background simulations from actual LHC
experience is also important in deciding how to proceed. In general the
choice to be made is between maintaining high-$\eta$ acceptance using new,
super-robust, low maintenance detectors, or accepting a reduced high-$\eta$
acceptance limit due to additional shielding that permits the existing
LHC muon detector technologies to survive in most locations. The forced
re-design of other sub-systems may in any case, ultimately determine the
effective high-$\eta$ region accessible for muon detection.

Depending on the exact configuration (particularly high-$\eta$ cut-off)
changes in technology might be needed in certain specific regions (eg
first forward stations), but the aim should be to use technologies
already developed and applied for high-rate tracking at LHC.

To cope with the 80Mhz bunch-crossing rate, much of the on-board trigger
and readout electronics and cooling will have to be replaced, even
without change in detector technology. This task may be a major
contribution to the re-fit time for SLHC.

\subsection{Trigger and Data Acquisition}
\label{sec:TDAQ}
The consequences are considered for trigger and data-acquisition
(TDAQ) systems if LHC is upgraded to have higher luminosity (in the
following we assume \slhclum) and a reduced
bunch-crossing (BC) period to 12.5 ns. After discussing some issues
related to the detectors that also affect TDAQ, we outline the
programme of R\&D on TDAQ systems that would be needed for the
successful exploitation of such a machine. Finally, we give an
indication of the thresholds that might be possible with a first-level
(level-1) trigger at SLHC.

\subsubsection{Higher luminosity}

The obvious consequences of raising the luminosity of LHC are higher
detector occupancy, increased trigger rates at fixed transverse-momentum
thresholds (or higher thresholds for fixed rates), and larger levels of
radiation that could damage or perturb the detectors and the on-detector
electronics.
\\[0.3cm]
{\it a) Occupancy}
\\
Increased occupancy has two important consequences for the TDAQ system:
degraded performance of trigger algorithms due to the increase in
pile-up, and a larger event size to be read out. Examples of the
degradation of the trigger performance include reduced rejection at
fixed efficiency from isolation requirements on electron/photon
candidates, and increased muon-trigger background rates arising from
accidental coincidences between radiation-induced ``noise'' hits in the
muon detectors. 
\\[0.3cm]
{\it b) Trigger rates}
\\
The increased event size reduces the maximum allowed level-1 rate for
fixed readout bandwidth. This suggests that one should perhaps try at
least to avoid increasing level-1 rate beyond the maximum of 100 kHz
presently envisaged in ATLAS and CMS. Such a strategy appears to be
possible, as discussed later, but implies raising the
transverse-momentum thresholds on candidate electrons, photons, muons,
etc., and using less inclusive triggers. The increase in the thresholds
has to compensate for the larger interaction rate, and also for the
degradation in algorithm performance due to the higher occupancy (less
rejection for fixed efficiency).
\\[0.3cm]
{\it c) Radiation damage}
\\
The increased levels of radiation at SLHC could cause problems in terms
of damage to detectors and to the on-detector electronics (either
permanent damage or single-event-upset effects). Note that part of the
level-1 trigger electronics in both ATLAS and CMS is mounted on the
detectors. The radiation tolerance of this electronics, as well as the
front-end electronics of the detector systems, would need to be assessed
in view of increased radiation levels at SLHC. However, it should be
kept in mind that the actual radiation levels are presently uncertain
(``safety factors'' are applied in qualifying the electronics), and will
only be known with precision after LHC starts operation.

\subsubsection{Reduced BC period (12.5 ns)}
{A reduction of the BC period below its present value of 25 ns has
important consequences for the level-1 trigger and the detector
front-end electronics}. The present trigger systems are pipelined
processors driven by the 25-ns period (40 MHz) LHC machine clock; they
select events indicating exactly which BC produced the interaction of
interest. In the following we assume a BC period of 12.5 ns (80 MHz
frequency). Frequencies higher than this would amount to almost
continuous beam given the rise times of signals from the detectors and
the timing resolution achievable in the detector and trigger
electronics. {The present strategies for timing-in and time monitoring of
the experiments rely on using the bunch structure of the machine. This
may simply not be possible for BC intervals of less than about 12.5 ns}.

{From the point of view of performance, it would be advantageous to
rebuild the level-1 processor systems to work with data sampled at 80
MHz} (internally some data movement and/or processing are already done at
80 MHz and above in both the ATLAS and CMS systems). This would optimise
the performance of the algorithms (rate versus efficiency) by limiting
the effects of pile-up that become more important at SLHC. This provides
the best chance to hold the level-1 output rate below 100 kHz, leading
to cost savings elsewhere by avoiding replacement of the front-end and
readout electronics, and avoiding an increase in bandwidth and
processing power in the high-level trigger and DAQ systems. It is the
only way to identify with 12.5 ns precision the BC that caused the
trigger. 

In some cases the so-called ``trigger primitive'' information from the
detectors (e.g. energy in calorimeter trigger towers) could still be
derived from existing detector front-end systems. For example, the
trigger primitive generation electronics that prepares the trigger tower
data from the CMS digital calorimeter front-end electronics could
possibly be modified to calculate the energy deposited in each 12.5 ns
BC period from the time sequence of measurements made at 40 MHz
frequency.

An alternative would be to keep some of the level-1 trigger processor
electronics clocked at 25 ns. Here the trigger-primitive information
from the detectors (e.g. energy in calorimeter trigger towers) from
pairs of BCs would be assigned to 25 ns intervals. This would still
require modifications to the front-end part of the level-1 trigger, but
part of the processing chain could be retained as is. The trigger would
then identify pairs of BCs to be read out (and where appropriate with
data from surrounding BCs) forming so-called time-frames. The data from
these time-frames could then be used to reconstruct the hit time in the
higher-level triggers and offline - note that for many detectors the
resolution is better than 12.5 ns. Drawbacks of this approach are
increased pile-up (since the activity from pairs of BCs is combined at
least at the level of the trigger processing), and a larger event size
(since the size of time-frame has to be enlarged to allow for the
ambiguity in the BC that caused the trigger).

\subsubsection{Comments on detectors (for 12.5 ns BC interval)}
The inner-tracking detectors will have to be replaced for operation at
SLHC along with their front-end electronics (radiation damage,
occupancy). The new detectors and associated electronics with 12.5 ns
sampling period can benefit from a level-1 trigger that identifies the
BC with 12.5 ns precision.

Although it may be possible to retain the existing calorimeters, it may
be necessary to re-optimize the shaping time of the analogue
electronics. It might be possible to retain the existing front-end and
readout electronics, provided the analysis can be done with data sampled
at 25 ns period. The time of deposition of the energy in the calorimeter
can be reconstructed with high precision using data from a series of
measurements in time. With the digital calorimeter readout in CMS, the
digital processing that prepares the level-1 trigger tower data would
have to be modified to calculate the energy in each 12.5 ns BC period.
ATLAS has a separate system of ADCs for the trigger. Of course, the
survival and operability of the on-detector electronics in the higher
radiation environment would need to be checked.

For the muon spectrometers, it may be possible to retain (some of) the
detectors and associated front-end and readout electronics. In some
cases (e.g. ATLAS TGCs), the time resolution of the detectors may be
marginal to trigger unambiguously on bunch crossings separated by 12.5
ns. As for the calorimeters, the survival and operability of the
on-detector electronics in the higher radiation environment would need
to be checked. The rate of spurious triggers induced by radiation in the
cavern would also need to be checked.

\subsubsection{Trigger menu}

Three types of triggers will most likely be needed at a SLHC:
\begin{itemize}
\item Triggers for very high-$p_{T}$ discovery physics. 
These do not cause big
rate problems since thresholds can be as high as several hundreds of GeV.
\item Triggers to complete the LHC physics programme, e.g. precise
measurements of the Higgs sector. These require thresholds on
leptons/photons/jets as low as those used at the LHC. However, since the
final states to be studied are known, one can use exclusive menus (e.g.
one lepton plus two b-jets plus missing energy) targeted to the final
states that need to be studied. 
\item Control/calibration triggers with low thresholds, selecting, e.g., W,
Z and top events. These can be pre-scaled. 
\end{itemize}
A first, very preliminary, study has been made to determine the expected
rate of some basic inclusive triggers at level-1. An illustrative set of
selection criteria is as follows:    
\begin{itemize}
\item inclusive single muon $p_{T}>30$ GeV (rate $\sim$ 25 kHz);
\item inclusive isolated $e/\gamma$ $E_{T}>55$ GeV (rate $\sim$ 20 kHz);
\item isolated $e/\gamma$ pair $E_{T}>30$ GeV (rate $\sim$ 5
  kHz)\footnote{The isolated $e/\gamma$ pair trigger with an \et\ threshold
    of 30~GeV on each cluster could be replaced by a trigger with two
    different thresholds, e.g. 40 GeV and 25 GeV, which could be more
    efficient for studies involving channels such as $H->\gamma \gamma$.}
\item muon pair $p_{T}>20$ GeV (rate $\sim$ few kHz?);
\item jet $E_{T}>150$ GeV .AND.$E_{T}^{miss}>80$ GeV required in 
coincidence (rate $\sim$ 1-2 kHz);
\item inclusive jet trigger $E_{T}>350$ GeV (rate $\sim$ 1 kHz);
\item inclusive $E_{T}^{miss}>150$ GeV (rate $\lsim$ 1 kHz);
\item a multi-jet trigger with thresholds determined by the affordable rate
(still to be evaluated).
\end{itemize}
The rates are very preliminary estimates based on scaling rates from
the ATLAS and CMS level-1 trigger TDRs, not allowing for the
degradation in performance of isolation, etc. due to the higher level
of pile-up at SLHC. This is particularly true for the muon rates,
which do not take into account the degradation in performance of the
trigger with $p_{T}$ (i.e. the threshold is less sharp at higher
$p_{T}$), or the possibility of large rates from the increased
background due to radiation in the cavern (which could be a serious
problem for the inclusive muon trigger in ATLAS). We guess there might
be some chance to lower the dimuon threshold, but we err on the safe
side for now.

There will certainly be triggers in addition to the above, for example
a trigger requiring a muon-electron pair with a $p_{T}$ threshold of
about 20-30 GeV for each lepton is likely to have an affordable rate.

\subsubsection{Data Acquisition}

In spite of the continuous and extraordinary evolution of the computing
and communication technologies, a research and development programme is
necessary in the following domains:  

{\bf a) Readout network}: implementation has to follow the LHC machine 
luminosity thus exploiting the parallel evolution of technologies 
The main building block of any LHC data acquisition
system is the network interconnecting the data sources (detector
digitizers) to the processing nodes (event filters). While processor
farms are becoming off-the-shelf commercial components, the same is not
yet true for the interconnection technologies whose progress, even if
impressive, started more recently than the one in the field of
computing. For example today a full commercial network system with the
performance required to build a LHC data acquisition network is not yet
available in the market (i.e. a switch with thousand ports,
non-blocking, 1 Terabit/s aggregate bandwidth etc.). Therefore
implementations of the event builders at LHC will be made via subsequent
upgrades following both the machine luminosity and (we hope in phase)
the evolution of the communication technologies.
 
{The network technologies should be tracked}. The new implementations
should be applied in the running data acquisition readout systems. In
particular the integration of the 10 Gb/s Ethernet and the emerging
Infiniband technologies should be tested to interconnect large farms of
processors (e.g. the farms foreseen for the LHC computing and the Grid
projects can provide suitable test beds). 

{\bf b) Complexity handling}, critical at the start of the experiment, because
the management of such a large system is a real new problem ('opening a
new airport syndrome' e.g. Malpensa)
The online computing systems will most likely have
more than 10000 CPUs. In addition to the hardware boxes (CPUs) there
will be millions of software boxes (jobs) to be managed and controlled.
These numbers are sufficiently large that the designers will have to
confront reliability problems not seen in any previous laboratory setup
or slow control system. Moreover the experiment control and information
systems will need to be accessible to multiple users with different
profiles, expertise and therefore access privileges. The hardware and
software management of such a large complex is similar to that found in
the present Internet Service Provider centers (e.g. 6000 CPUs in
google.com search engine).

{Modern technologies should be studied to control distributed computing
and exploit the Web tools, currently used in the e-commerce world, to
implement the experiment high level controls and user interfaces.} Under
the new run control the handling of future experiments will not be very
different from that of an e-commerce company (same problems: security,
remote access, databases, world wide access, knowledge data bases,
on-line orders etc.). The exploitation of the immense developments
ongoing in the domain of e-commerce will open new ways to operate large
collaborations and large set of distributed processors.

\subsubsection{Main R\&D issues}

The main R\&D issues for the level-1 trigger relate to the reduction in
the BC period. Data movement is probably the biggest issue for
processing at 80-MHz sampling rate. Interconnection issues (links,
back-planes, etc.) already drive the design of the level-1 processors in
ATLAS and CMS with 40-MHz BC rate. Triggers for SLHC would need higher
bandwidth and/or more use of zero suppression, data compression, etc.
Processing at higher frequencies and with higher input/output data rates
to the processing elements also needs to be investigated, although
technological advances (FPGAs, etc.) will help here.

Synchronisation (using the TTC system, etc.) becomes an issue for short
BC periods. Present strategies for timing-in and time monitoring of the
experiments that rely on using the bunch structure of the machine may
have to be reviewed. Finally, some detectors currently used in the
trigger may be too slow for 12.5 ns timing precision, requiring R\&D on
alternatives.

Concerning the high-level triggers and DAQ, the main issue is how to
handle the larger bandwidth (rate times event size) at SLHC. Bandwidth is an
issue both for readout and for event building. Processing power is
likely to be less of an issue assuming continued growth in the
performance/price ratio.

\subsection{Electronics for SLHC}
\label{sec:elec}
Electronics Technology has consistently developed at a rate described
by an empirical relationship, known as 'Moores Law'.  This relationship
predicts that the minimum feature size in silicon microelectronics
circuits will decrease by a factor of two every five years.  This trend
has proved to be true since the early 1970s.  The smallest feature size
commercially available today is 0.13 $\mu$ and Intel has recently
demonstrated a 0.03 $\mu$ transistor in the research laboratory.  The
significance of this continuing trend is that the number of usable logic
gates in a microelectronics chip increases by a factor of 4 every 5
years with a corresponding increase in speed and decrease in power
dissipation.  In addition the time taken for research developments to
become viable commercial products is ten years.  Hence we can predict
with certainty that the trend will continue for at least a further ten
years, but that new technology developments will be required to maintain
the growth in the electronics industry in the period from ten to twenty
years.

The Particle Physics experiments presently in construction for the CERN
LHC would not have been possible without the extensive use of
microelectronics technology.  This technology was developed for the
worldwide computer and telecommunications industries.  {The successful
application of these technologies to the requirements of Particle
Physics experiments was the result of an intensive R\&D programme
approved by the CERN DRDC. Without this programme the present
generation of experiments would not have been possible.}

For SLHC it will be necessary to build on the expertise and
infrastructure that has been established for LHC, both at CERN and at
the network of collaborating Institutions throughout the international
Particle Physics research community.  

\subsubsection{Proposed R\&D Projects}

One of the major successes of the LHC development programme was the
demonstration that commercially available 0.25 $\mu$ CMOS technology can 
be radiation hard \cite{expref19}.  
This technology is now used extensively in LHC
applications with very significant performance and financial gains.  

Many radiation-hard circuits for the LHC experiments have been, or are
being, converted into DSM technology. Radiation effects can be divided
into two categories: total dose effects and single event effects. 

Total dose effects in CMOS are mainly associated with charging-up of
oxides. During irradiation electrons and holes are generated in the
oxides. While the electrons are evacuated rapidly (within ns) holes
accumulate in traps leading to transistor threshold shifts. As the
oxides get thinner the charging decreases in proportion to the volume of
the oxide. Hence thinner oxides are inherently more radiation hard. As
the oxide thickness falls below 10 nm the reduction in the radiation
induced threshold voltage becomes even more pronounced \cite{expref20}.
Therefore for deep sub-micron processes (a 0.25 $\mu$m process has a
gate oxide thickness of 5 nm) radiation induced threshold voltage shift
becomes negligible even at very high radiation doses. However there is
still the possibility of leakage paths from drain to source and from one
transistor to another that have to be eliminated by special layout
techniques \cite{expref21}. 

Single event effects will perhaps be the ones that cause the most
difficulty for the tracker electronics at SLHC. The effects comprise:
\begin{itemize}
\item single event gate rupture that only manifests itself above a critical
threshold electric field and should not be an issue for deep sub-micron
CMOS circuits. 
\item single event latch-up that can probably be avoided by the use guard
rings that are also used to limit total dose effects.
\item single event upsets that cause the logical level of the node to switch
state. This effect occurs above a threshold LET. The threshold LET tends
to decrease for smaller feature sizes and is a real concern for deep
sub-micron circuits. 
\end{itemize}
{The more recently available technologies (0.13 $\mu$ and beyond) will 
require characterisation for SLHC applications and the development of the new
design techniques and the required libraries}. Understanding the limits,
and applicability, of DSM electronics should be the subject of vigorous
R\&D for SLHC and for the upgrades of LHC experiments.

Data rates in SHLC detectors will scale with luminosity. This raises the
issue of whether to process data at the detector and reduce data volumes
before transfer to Off-Detector electronics, or whether to invest in
advanced data link technology to minimise the risk to electronics on the
detector.

The development of intelligent architectures to reduce the volume of
data transferred off the detector; especially in the case of high
granularity detectors (Tracker and Pixel Systems) will require common
development projects. 

{The development of very high-speed data links for Particle Physics
applications, based on the commercial developments is a common
development project}.  Commercial developments are not optimised for
extreme environments found in LHC experiments.  

Another consequence of the very high data rates anticipated at Super LHC
is the power dissipated in the CMOS electronics in the detectors.  There
are alternative technologies that could be considered (for example
Si-Ge), but the LHC research community has little experience of
designing in these technologies.  In addition, the advantages of working
with a modern commercial process may be lost.

The understanding of the issues involved in using alternative
technologies will require substantial work.  In addition, more work on
power removal techniques in the environment found at LHC will be
required.  

In the development of electronics for LHC the most pressing problems
were those of developing electronics that would operate inside the LHC
detectors.  As a result of the DRDC programme, these problems were
solved, but in many cases the overall systems design issues did not
receive the required attention.  

A new R\&D initiative should recognise this shortcoming and encourage the
research community to focus on the systems design issues from the
outset.  The ultimate performance of a detector system is often limited
by the noise that is generated by non-optimal grounding systems. 
Understanding all the systems issues is the focus of another R\&D project
that will develop common solutions where possible.  

\subsubsection{Organisational Issues}

When the initial DRDC projects were approved in 1989, most of the
microelectronics were produced on either 4 or 6 inch wafers.  In
addition most of the Particle Physics community could obtain access to
the best available design software tools through the Europractice
programme which made these tools available to the European Teaching and
Research community at very low cost. 
 
The 0.25 $\mu$ CMOS technology used in many LHC experiments is now produced
on 8 inch wafers and the next technology to be used (0.13 $\mu$ CMOS) will
soon be produced on 12 inch wafers.  In addition the number of
interconnection planes is also increasing to give the designer more
freedom in connecting the elements within the circuit.

The result of these developments is much more efficient chip designs and
much cheaper chips for the very large users (Computer and
Telecommunication Industries) who have been driving the development of
the technology.  Not only will the wafers be more expensive, but the
number of masks used in the processing will also increase, which in turn
will increase the Non-Recurrent Engineering (NRE) costs.  The potential
complexity of the designs will also increase together with the
complexity of the techniques required to layout and simulate the
behaviour of the circuits.  The commercial cost of the design software
will be very high ($\gg$~1M dollars), and hence it is crucial to maintain 
access to the Europractice programme.  

The number of research organisations, world-wide, that will be able to
access these technologies will be very small and will probably be led by
the major Particle Physics Laboratories, where CERN has a leading role. 
{Without the establishment of a world-wide network, involving both the
Particle Physics research community and commercial partners, to develop
the next generation of electronics, future Particle Physics experiments
will not be possible}.

Experience with the development required for the LHC implies that the
time required to develop the electronics for Super LHC will be $\sim$ 8-9
years. 

In the mid 1980s the CERN structure was changed to recognise the
importance of co-locating a critical mass of the best electronics
engineers to develop the electronics required for LHC.  The developments
required for SLHC will also require an equivalent change in structure at
CERN to provide not only critical mass of engineers and the required
infrastructure, but also the focus for the required world-wide network
within the research community.  

If successful, not only will CERN provide the leadership of the
international Particle Physics community, but it will become the
international focus for the development of multi-disciplinary advanced
instrumentation. 

\subsection{Conclusions: Experimental Challenges and the Detector R\&D} 

A luminosity upgrade of the LHC to \slhclum\ will require
significant detectors R\&D especially for the inner tracking systems
including that for radiation-hard front-end electronics and optical
links. CERN should launch a new R\&D programme as soon as resources allow.
This should be modeled on the Detector R\&D Committee
 programme of the 1990's, but initially
with most of the R\&D targeted to the needs for the SLHC.

In the immediate future the highest priority should be given to 
the completion of
the current detectors and only a very limited R\&D effort should be
considered at CERN. However there are several reasons to continue with
minimal effort; outside CERN numerous groups are performing generic R\&D.
The (S)LHC community will benefit from having a good contact with these
groups; the SLHC challenges can provide guidelines for the R\&D
effort; CERN can provide test beam and irradiation facilities for these 
groups and
can be an important reference for these groups when they define their
national projects; finally this type of work attracts instrumentation
students in general and also experts outside the traditional HEP
community.    

A low-level of human and financial resources should be made available
from CERN mainly for co-ordination of several R\&D programs financed by
member states on a national basis. A significant increase of activity and
therefore resources should be planned for the years 2006 and beyond to
give an appropriate impetus for focused activities in view of an SLHC
running in the early part of the next decade.
If the new R\&D programme is successful, not only will CERN provide the
leadership of the international Particle Physics community, but it will
become the international focus for the development of multi-disciplinary
advanced instrumentation.

Below we draw the conclusions for each of sub-detectors considered above.

\subsubsection{Inner Tracking}

The current ATLAS and CMS trackers have to be completely rebuilt for
SLHC in order to withstand a factor 10 higher luminosities. The general 
approach suggested is to: 

a) further develop with industry the current silicon strip technology
for use at radii $>$ 60 cm. 

b) further develop the current pixel technology that is expected to work
at radii between 20 cm and 60 cm. 

c) for the vertex region (R $<$ 20) new concepts and new materials are
required to attain the necessary speed and radiation hardness. 

Furthermore, there is a need for engineering studies related to
materials, power distribution, cooling and development of radiation hard
electronics together with a full readout scheme 

\subsubsection{Calorimetry}

The calorimetry in the barrel regions of ATLAS and CMS should be able to
withstand the ten times higher luminosities. However careful attention
has to be paid to the endcap and forward regions.

For liquid argon calorimetry the issues to be investigated are space
charge effects and current induced voltage drops amongst others.
Operation using different liquids, such as krypton, or even dense cold
gases should be evaluated.

For the CMS lead tungstate crystal calorimeter a programme of
irradiations emulating SLHC conditions has to be carried to evaluate the
performance of the crystals, photodetectors and the front-end
electronics.

For the CMS endcap HCAL, short of using a novel technique, methods
should be developed for mitigating the effects of higher radiation
levels. These could include a combination of a) individual readout of
scintillator layers, b) periodic replacement of scintillators, and c)
development of a more radiation-tolerant scintillator. For the CMS
forward calorimeter replacement of the plastic-clad quartz fibres by
quartz-clad quartz fibres has to be envisaged. Development of new
techniques should also be pursued.

\subsubsection{Muon Systems} 

The LHC experiment muon systems have been designed according to
conservative assumptions in the background rates (factor 3 to 5 safety
margin above estimates from simulations). The real safety margin can
only be established from measurements once LHC operates. The simplest
viable modification for SLHC is to increase shielding around the
beam-pipe at high-$\eta$. The penalty is a cut in the high-$\eta$ acceptance.
The modifications to the ATLAS \& CMS muon systems needed for SLHC should
be determined by a cost-benefit analysis depending on the perceived
physics potential in the light of results from LHC at \lhclum. In general
the choice to be made is between maintaining high-$\eta$ acceptance using
new, robust, low maintenance detectors, or accepting a reduced
high-$\eta$ acceptance limit due to additional shielding that permits the
existing LHC muon detector technologies to survive in most locations. 
In the worst case the first forward muon stations might need to be
replaced by a different design concept. However, technologies can be
used that are already proven at LHC in a worse environment, eg those
that were candidates for central tracking.

\subsubsection{Trigger and Data Acquistion}

A reduction of the BC period below its present value of 25 ns has
important consequences for the level-1 trigger and the detector
front-end electronics. From the point of view of performance, it would
be advantageous to rebuild the level-1 processor systems to work with
data sampled at 80 MHz. 
Concerning the high-level triggers and DAQ, the main issue is how to
handle the larger bandwidth (rate and event size) at SLHC. Bandwidth is an
issue both for readout and for event building. In spite of the
continuous and extraordinary evolution of the computing and
communication technologies, a research and development programme is
necessary in the domains of readout network and complexity handling. The
network technologies should be tracked. Modern technologies should be
studied to control distributed computing and exploit the Web tools.

\subsubsection{Electronics}

The successful application of microelectronics technologies to the
requirements of Particle Physics experiments was the result of an
intensive R\&D programme approved by the CERN DRDC. One of the major
successes of the LHC development programme was the demonstration that
commercially available 0.25 $\mu$ CMOS (DSM) technology can be radiation hard. 
This technology is now used extensively in LHC applications with very
significant performance and financial gains. The more recently available
technologies (0.13 $\mu$ and beyond) will require characterisation for SLHC
applications and the development of the new design techniques and the
required libraries. The 0.13 $\mu$m CMOS will be produced on 12 inch wafers. 
In addition the number of interconnection planes is also increasing to
give the designer more freedom in connecting the elements within the
circuit. The Non-Recurrent Engineering (NRE) and the costs of the design
software will be very high, and hence it is crucial to
maintain access to the Europractice programme. 
Understanding the limits, and applicability, of DSM electronics should
be the subject of vigorous R\&D for SLHC and for the upgrades of LHC
experiments. 
Another vital subject for R\&D is the development of very high-speed data
links for Particle Physics applications, based on the commercial
developments.
 
All of the new R\&D initiatives should encourage the research community to
focus on the systems design issues from the outset.

\section{CONCLUSIONS}
\label{sec:concl}
  The physics potential of an upgraded LHC running at a 
 luminosity of \slhclum\ can be summarised as follows:
 
\begin{itemize} 
\item The measurement of some of the TGC's will reach an accuracy 
  comparable with the size of EW, and possibly SUSY, virtual
  corrections.
  
\item New rare decay modes of the SM Higgs boson will become accessible,
   e.g. $H\to \mu^+\mu^-$ and $H\to Z\gamma$. The determination of
  the Higgs couplings to  bottom and top quarks, as well
  as to  EW gauge bosons, will reach precisions of 10\% or better,
  over a good fraction of the $\mh<200$~GeV mass range.  In the 
   MSSM, the region of SUSY parameter space where at least two
  Higgs bosons will be observed is significantly enlarged relative to the
  LHC reach.
  
\item The first observation of SM Higgs pair production may be
  possible in the $170<\mh<200$~GeV mass range, with a determination of
  the Higgs self-coupling
  $\lambdahhh$ at a level of 19\% (25\% ) for $\mh=170$~GeV
  ($\mh=200$~GeV), after background subtraction.
  The precise size of the backgrounds has however large
  theoretical uncertainties. The use of data control
  samples  will be necessary to  fully pin down these uncertainties
  and to strenghten the estimates of the significance of
  the $HH$ signal.
  
\item In the absence of a Higgs signal, studies of 
     resonant and non-resonant scattering of electroweak
     vector boson pairs at high mass 
     will benefit from the larger statistics, which should give access
     to a larger variety of channels and in general to more
     convincing signals than at the LHC. These conclusions, however, 
     depend upon the possibility of maintaining adequate forward
     jet tagging performances.

\item The FCNC decay modes of the top quark $t\to \gamma/Z q$ may be
  accessible if their BR is of order $10^{-6}$. This range is
  of relevance for some theories beyond the Standard Model.  
  
\item The mass reach for squarks and gluinos will be extended from
 $\sim$~2.5~TeV (standard LHC) to $\sim$~3~TeV (SLHC). 
   In addition, some exclusive SUSY channels 
  which are rate-limited at the standard LHC 
  could be studied in detail 
  with a tenfold increase in statistics, thereby providing
  additional information about the underlying theory. 
  
\item The mass reach for new gauge bosons, or for signatures of
  Extra-dimension models, will be extended by approximately 30\%
  relative to the LHC; in the case of compositeness, the sensitivity
  to deviations from
  the expected behaviour of quarks in the SM will be extended from 
  a scale $\Lambda~=$~40~TeV  to $\Lambda~=$~60~TeV.
  
\end{itemize}
All of the above can be obtained at a moderate extra cost relative to
the overall initial LHC investment, extending the lifetime of the LHC
complex, completing its physics potential, and bridging the time gap
with future activities.

With the exception of final states containing very energetic objects
(e.g. jets, photons or muons with transverse energies in the TeV
range), the feasibility of the above physics programme requires
detector upgrades able to maintain the performances expected at the
standard \lhclum\ luminosity.  In many of the examples discussed in
this document, the performance of the LHC detectors is affected not
only by the high-luminosity environment, but also by the intrinsic
detector limitations in terms of detection efficiency and measurement
accuracies. Future studies should therefore aim at identifying an
optimal technological and financial balance between luminosity upgrade
and detector upgrades, with the goal of maximising the overall physics
performance.

The foreseen detector upgrades will require significant detector R\&D,
especially for the inner tracking systems (including radiation-hard
front-end electronics and optical links). CERN should launch a new
R\&D programme as soon as resources allow. This should be modeled on
the Detector R\&D Committee programme of the 1990's. We believe that a
vigorous R\&D activity for the SLHC will entail general and significant
progresses in the area of particle detector developments, and
therefore will ultimately have impacts on future machines (e.g. a
VLHC) and on particle physics in general.


\subsection*{Acknowledgements}
We thank M. Battaglia, C. Da Via, E. Heijne, R. Horisberger, P.
Jarron, B.~McElrath, M.  Moll, T. Rizzo, P. Weilhammer and R. Wunstorf
for their contributions to this document.


\begin{thebibliography}{99}
\bibitem{RT} We warmly thank Tom Rizzo for providing us with the FORTRAN 
code for $Z/\gamma$ excitation production and for the 
help in implementing it.
\end{thebibliography}
\end{document}